\documentclass[%
 reprint,
nofootinbib,
 amsmath,amssymb,
 aps,
prd,
]{revtex4-1}

\usepackage{graphicx}
\usepackage{dcolumn}
\usepackage{bm}
\usepackage{geometry}
\usepackage{color}	
\usepackage[left]{lineno}
\usepackage{amssymb}
\usepackage{amsmath}
\usepackage{amssymb}
\usepackage{graphicx}
\usepackage{hyperref}
\usepackage{subfigure}
\usepackage{float}
\usepackage{fullpage}
\usepackage{braket}
\usepackage{url}
\usepackage{upgreek}
\usepackage{outlines}
\usepackage{braket}
\usepackage{appendix}
\usepackage{xspace}
\usepackage{tikz}
\usepackage{placeins}
\usepackage{multirow}

\usetikzlibrary{matrix,arrows,decorations.pathmorphing}
\usetikzlibrary{patterns}
\usetikzlibrary{trees}
\usetikzlibrary{decorations.pathmorphing}
\usetikzlibrary{decorations.markings}
\usetikzlibrary{positioning,arrows}
\usetikzlibrary{decorations.pathreplacing}
\usetikzlibrary{decorations.pathmorphing}
\usetikzlibrary{decorations.markings}

\newcommand\ignore[1]{}

\newcommand\be{\begin{equation}}
\newcommand\ee{\end{equation}}
\newcommand\bea{\begin{eqnarray}}
\newcommand\eea{\end{eqnarray}}

\newcommand{\bdm}{\begin{displaymath}}
\newcommand{\edm}{\end{displaymath}}
\newcommand\nn{ \nonumber\\}

\renewcommand{\epsilon}{\varepsilon}
\renewcommand{\phi}{\varphi}

\numberwithin{equation}{section}
\numberwithin{figure}{section}

\graphicspath{{fig/}{../}}

\begin{document}


\title{Minkowski Conformal Blocks and the Regge Limit for SYK-like Models}

\author{Timothy G. Raben}
 \email{Timothy.Raben@ku.edu}
 \affiliation{University of Kansas, Department of Physics \& Astronomy\\
1082 Malott,1251 Wescoe Hall Dr. Lawrence, KS 66045 
}%
\author{Chung-I Tan}%
 \email{Chung-I\_Tan@brown.edu}
\affiliation{%
Department of Physics, Brown University\\
Box 1843 182 Hope Street Providence, RI 02912
}%

\date{\today}

\begin{abstract}
We discuss scattering in a CFT via the conformal partial-wave analysis and the Regge limit. The focus of this paper is on understanding an OPE with Minkowski conformal blocks. Starting with a t-channel OPE, it leads to an expansion for an s-channel scattering amplitude in terms of t-channel exchanges. By contrasting with Euclidean conformal blocks we see a precise relationship between conformal blocks in the two limits without preforming an explicit analytic continuation. We discuss a generic feature for a CFT correlation function  having singular $ F^{(M)}(u,v)\sim   {u}^{-\delta}\,$,  $\delta>0$, in the limit  $u \rightarrow 0$ and $v\rightarrow 1$.  Here, $\delta=(\ell_{eff}-1)/2$, with $\ell_{eff}$ serving as an effective spin and it can be determined through an OPE. In particular, it is bounded from above, $\ell_{eff} \leq 2$, for all CFTs with a gravity dual, and it can be associated with string modes interpolating the graviton in AdS. This singularity is historically referred to as the Pomeron. This bound is nearly saturated by SYK-like effective $d=1$ CFT, and its stringy and thermal corrections have piqued current interests. Our analysis has been facilitated by dealing with Wightman functions. We provide a direct treatment in diagonalizing dynamical equations via harmonic analysis over physical scattering regions. As an example these methods are applied to the SYK model. 
\end{abstract}

\pacs{Valid PACS appear here}
\maketitle

\tableofcontents

\section{Introduction}\label{sec:intro}

Most current studies in conformal field theories (CFT) are carried out in the Euclidean limit.  This is particularly true when using {\bf Euclidean conformal blocks} (ECB) in exploiting the consequences of conformal invariance~\cite{Dolan:2004iy,Dolan:2003hv,Dolan:2011dv}.  Conversely, a scattering process is intrinsically Minkowski~\cite{Paulos:2016fap,Paulos:2016but,Pasterski:2016qvg,Pasterski:2017kqt,Lam:2017ofc,Cheung:2016iub,Hartman:2015lfa,Brower:2006ea,Brower:2007qh,Brower:2007xg,Cornalba:2007fs,Cornalba:2008qf,Cornalba:2009ax,Costa:2012cb,Cornalba:2006xm,Brower:2014wha,Banks:2009bj}.  Earlier studies in CFT scattering, first discussed for CFT with gravity dual~\cite{Brower:2006ea,Brower:2007qh,Brower:2007xg},  mostly adopted an Euclidean treatment and an analytic continuation is then performed~\cite{Cornalba:2006xm,Cornalba:2007fs,Cornalba:2008qf,Cornalba:2009ax,Costa:2012cb}. Recent interest in CFT in a Minkowski setting has increased to warrant a more systematic and direct treatment~\footnote{In the context of holography there is a history of directly investigating Lorentz correlators of the gravity theory to learn about thermal properties of the strongly coupled CFT. We are not concerned with this approach in this paper, but rather focus on the CFT directly regardless of a gravity dual.}.  Such an approach provides a framework where one can directly  treat scattering problems, for example, inclusive and exclusive high energy near-forward scattering~\footnote{In literature this is often referred to as the ``eikonal limit" or the ``Regge limit"}, among others.  Many phenomenological applications to high energy physics at the LHC and HERA have been carried out with encouraging successes~\footnote{Many holographic models have been used to successfully model collider physics. For a brief list of some applications that directly investigate conformal properties see~\cite{Nally:2017nsp,Brower:2014sxa,Brower:2014wha, Brower:2015hja,Brower:2012mk,Costa:2012fw,Costa:2013uia,Brower:2010wf,Ballon-Bayona2016,Cornalba:2008qf,Cornalba:2009ax,Costa:2012cb}.}. In this paper we demonstrate a new method for directly computing {\bf Minkowski conformal blocks} (MCB) as well as elucidating details about the Minkowski conformal block expansion relevant for arbitrary dimension.


\begin{figure}
\begin{center}
\includegraphics[width = 2in]{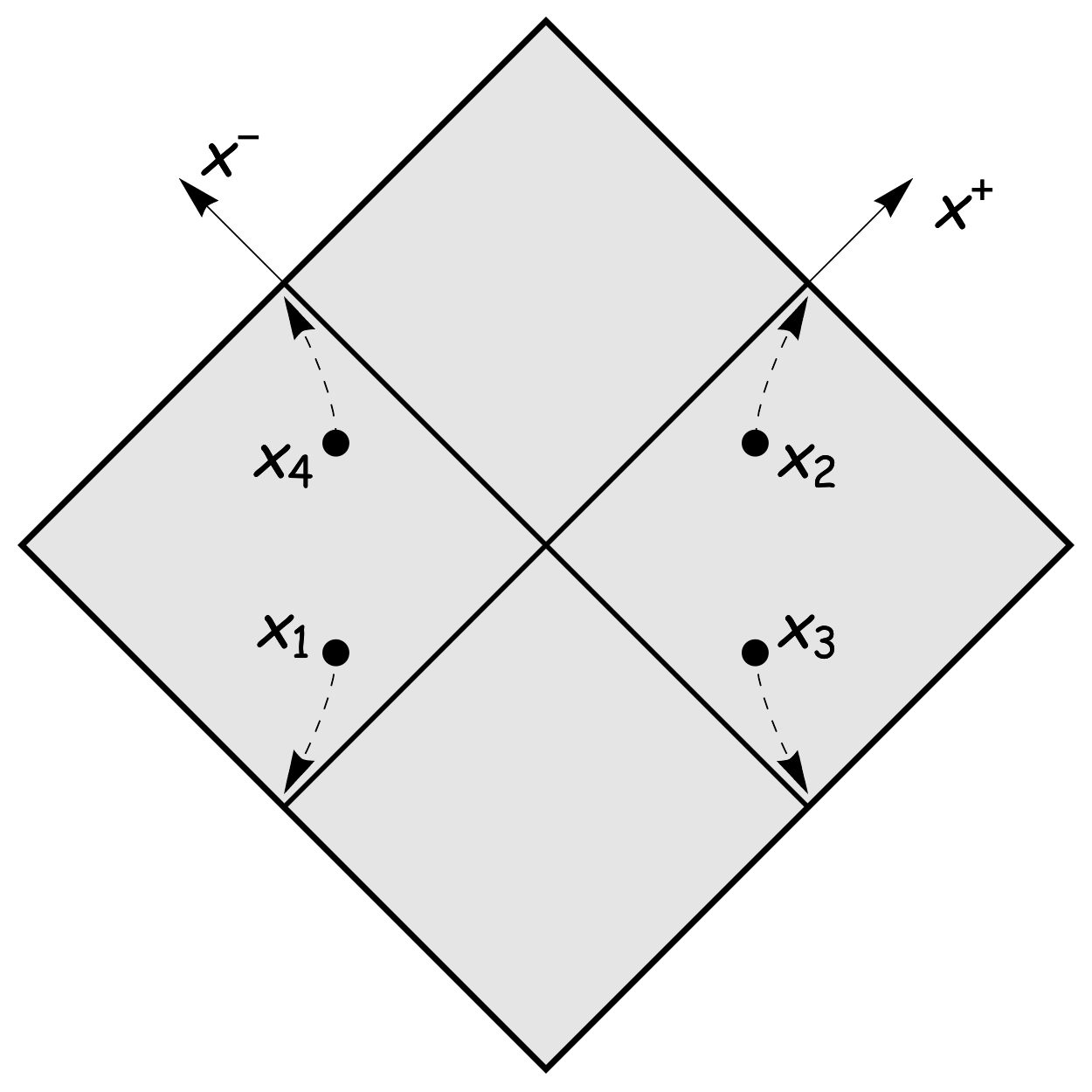}
\caption{Conformal compactification of the Minkowski light-cone showing points taken to null infinity in the Regge limit. In light-cone coordinates $(x^+,x^-,x_\perp)$ we take  $- x^+_1 \sim x^+_2\to\infty$  and $- x^-_3 \sim x^-_4\to\infty$, keeping $x^i_\perp$ fixed.  Here $x_1^0<x_3^0<0<x_4^0<x_2^0$.  With $(-x_1,x_2)$ approaching the  forward light-cone and $(-x_3,x_4)$ the backward light-cone, 
this will be referred to as a ``double light-cone limit". }
\label{fig:ReggeLimit}
\end{center}
\end{figure}
\subsection{Overview}
This paper deals with the intersection of three sometimes disparate subjects: (1) conformal field theory, (2) analytic scattering amplitudes, and (3) string-gauge duality. Because these three subjects often discuss similar methods, for example conformal block expansion vs partial wave expansion, using different formalism and notation-we outline here our approach to conformal scattering processes that best illustrates these intersections.

A conformal approach to scattering processes was initially developed through the AdS/CFT correspondence~\cite{Brower:2006ea,Brower:2007qh,Brower:2007xg}, but it can be presented entirely in a CFT  language~\cite{Cornalba:2006xm,Cornalba:2007fs,Cornalba:2008qf,Cornalba:2009ax,Costa:2012cb}.   Both approaches are equivalent and each offer  separate intuitive frameworks. In this paper, we closely follow the CFT approach~\footnote{We summarize briefly the AdS/CFT perspective in Appendix \ref{sec:holography}.}, but describe physical insights it can tell us about interpreting the gravity dual. A typical example of conformal scattering is the off-shell photon process
\be
\gamma^*(1) + \gamma^* (3) \rightarrow \gamma^* (2) + \gamma^* (4), \label{eq:4gamma}
\ee
which does not involve asymptotic states. The amplitude is related to a time-ordered ($T$) four-point current correlator,  $\langle 0| T ({\cal J}_1 {\cal J}_2{\cal J}_3{\cal J}_4)|0\rangle $~\footnote{In more intuitive notation that conforms with other literature, we will sometimes write $\langle 0| T ({\cal J}_1 {\cal J}_2{\cal J}_3{\cal J}_4)|0\rangle $ as $\langle T ({R}(x_1) {R}(x_2) {L}(x_3) {L}(x_4) )\rangle$ or $\langle T ({R}(1) {R}(2) {L}(3) {L}(4) )\rangle$.}.   Our convention refers to scattering from  $(1+3)$ to $(2+4)$ as the s-channel.  The t-channel OPE, ${\cal J}_1{\cal J}_2= \sum_\alpha c_{12,\alpha} {\cal O}_\alpha$, can be expressed in terms of MCB, $G^{(M)}_{(\Delta,\ell)}$, as in Eq. (\ref{eq:MOPE}). This ultimately leads to an expansion for s-channel scattering amplitudes in terms t-channel exchanges~\footnote{The word ``channel" can refer to a scattering process or an OPE. See Appendix~\ref{sec:channels} for our conventions.}. 

Recently, high energy scattering in CFT has become important for holographic models with black holes~\cite{Shenker:2013pqa,Shenker:2014cwa,Roberts:2014ifa} and the related SYK model~\cite{kitaev1,kitaev2,Sachdev:1992fk,Polchinski:2016xgd,Jevicki:2016bwu,Maldacena:2016hyu,Murugan:2017eto}: understanding the bounds of chaotic behavior and elucidating the flow of information via thermodynamics. It has been explained in \cite{Shenker:2014cwa} that understanding this behavior is equivalent to examining high energy behavior of near-forward scattering through the AdS/CFT correspondence following the formalism introduced in \cite{Brower:2006ea,Brower:2007qh,Brower:2007xg}. High energy scattering, depicted in Fig. \ref{fig:ReggeLimit}, involving a time-ordered four-point correlator, can address stringy and thermal corrections to scrambling times by calculating ``out-of-time-ordered" thermo-correlation functions, $\langle W(t) V (0)W(t) V(0)\rangle_{{\beta}}$, with $\beta$ the inverse temperature.

Near-forward scattering for a process like Eq. (\ref{eq:4gamma}) involves a small momentum transfer between  1 and 2, with a small deviation in their directions of travel~\footnote{Small with respect to the center of mass energy: $|t|<<s$.}. The process can best be illustrated by Fig. \ref{fig:ReggeLimit} where (1,2) (right-movers) move near the forward light-cone and (3,4) (left-movers) close to the backward light-cone. For simplicity, consider conformal scalars, with pairwise equal conformal dimensions $\Delta_2=\Delta_1$ and  $\Delta_3=\Delta_4$.  Due to conformal invariance, we have 
\bea
\langle T ({R}(1) {R}(2) {L}(3) {L}(4) )\rangle=\qquad\qquad\nonumber \\
\qquad=  \frac{1}{(x^2_{12})^{\Delta_1}(x^2_{34})^{\Delta_3}} 
  \,F^{(M)}(u,v)\,, \label{eq:A(X)}
\eea
where  $F^{(M)}$ depends only on invariant cross ratios, here chosen~\footnote{An alternative choice is  $u'= u/v$ and $v'=1/v$, corresponding to $1\leftrightarrow 2$ or $3\leftrightarrow 4$ interchange. We will return to this point in Sec. \ref{sec:Crossing}.} to be 
\be
u= \frac{x_{12}^2 x_{34}^2}{x_{13}^2 x_{24}^2}\, , \quad v= \frac{x_{23}^2 x_{14}^2}{x_{13}^2 x_{24}^2}\, ,\label{eq:cross-ratios}
\ee
with  $x_{ij}=x_i-x_j$ and $x_{ij}^2$ defined with {\bf Lorentzian signature}.    

As is well-known~\cite{Dolan:2004iy,Dolan:2003hv,Dolan:2011dv}, one can  express the invariant function $F^{(M)}(u,v)$  via a conformal block expansion
\begin{equation}\label{eq:MOPE}
F^{(M)}(u,v)= \sum_{\alpha} a^{(12\;34)}_{\alpha} G^{(M)}_{\alpha}(u,v) \, ,  
\end{equation}
where ${G}_\alpha^{(M)}(u,v)$ are MCB, each associated with a conformal primary  ${\cal O}_\alpha$ entering into the {\bf t-channel OPE}. Eq. (\ref{eq:MOPE}) defines the {\bf t-channel Minkowski conformal block expansion}. Most of our results will apply generically to CFT's of arbitrary dimension, but many techniques are motivated by previous analyses of${\cal N}=4$ SYM. Specifically for  ${\cal N}=4$ SYM, we focus on contributions from single-trace conformal primaries. In general, conformal primaries can be organized according to their twists $\tau_0$. The dimension $\Delta$ and spin $\ell$, are related by the relation 
\be
\Delta_\alpha(\ell) =\ell + \gamma_\alpha(\ell) + \tau_0, \label{eq:twist}
\ee
 with $\gamma_\alpha(\ell)$  the anomalous dimension. In the absence of interactions, $\gamma(\ell)=0$. In this representation, the dynamics lies in  knowing all the participating conformal primaries, ${\cal O}_\alpha$, and the  associated ``partial-wave coefficients", $a^{(12;34)}_{\alpha}$. The partial-wave coefficient is real and factorizable, $a^{(12;34)}_{\alpha}\sim c_{12,\alpha}c_{34,\alpha}$. An important focus of this paper is to demonstrate how the formal sum, Eq. (\ref{eq:MOPE}), can be interpreted, through the use of a Sommerfeld-Watson transform, as the principal  series for an unitary representation of non-compact groups, Eq. (\ref{eq:newgroupexpansionDisc}), contrasting Minkowski and Euclidean behavior. 

We treat CFTs where $F^{(M)}(u,v)$ can diverge at  $u\rightarrow 0$ but is polynomially bounded. Since  $G^{(M)}$ is constructed to be real, it follows that  the contribution to Eq. (\ref{eq:MOPE}) from each conformal primary is also real.  However, as a scattering amplitude $F^{(M)}(u,v)$ is in general complex.  A complex phase can emerge as a consequence of summing over higher spins~\footnote{Complex phases can also be generated through summing over multiple trace primaries of low spins. This can lead to eikonalization. See \cite{Brower:2007qh,Cornalba:2006xm,Fitzpatrick:2015qma} for a discussion about eikonalization in CFTs.}.  Therefore, as a scattering problem, it is equally important in addressing the issue of  re-summation for OPEs in a Minkowski setting~\footnote{In CFT bootstrap program (reviewed in \cite{Simmons-Duffin:2016gjk}), the OPE sums are typically truncated in all channels, thus the issue of re-summation does not arise. However, this is a \emph{separate} issue from defining the region of convergence for conformal blocks via a series expansion. As noted in \cite{Dolan:2011dv}, the region of convergence for standard euclidean conformal blocks is restricted to $\sqrt u + \sqrt v)\leq 1$. Continuation to the Minkowski region necessarily requires going beyond the region of convergence for euclidean OPE. Our treatment here avoids this cumbersome step. }. In this vein, Conformal invariance has historically also been applied to simplify the analysis for ladder-type integral equations as is commonly done in resummations leading to high energy Regge behavior~\cite{MPM-3,MPM-2,MPM,CGL,Bali:1967zz,Balitsky:1978ic,Kuraev:1977fs,Fadin1975,Gribov1972,Altarelli1977,Dokshitzer1977}. By working with Wightman functions, the absorptive part of forward scattering amplitudes~\footnote{To be precise, the absorptive part is a discontinuity. In a coordinate space treatment this corresponds to a vacuum expectation value of a double commutator, for example $\bra{0} [R(2), R(1)] [L(4) ,L(3)] \ket{0}$, appropriate for Eq. (\ref{eq:A(X)}) and non-zero only in the physical region. An explicit example is that for deep-inelastic scattering (DIS) which involves an OPE of currents. See \cite{Polchinski:2002jw} for a review and  Appendix \ref{sec:application-II} for connection to this work.}, our treatment  leads to a simpler diagonalization procedure for dynamical equations via the appropriate harmonic analysis.

We show in this paper that a generic feature for CFT correlation functions $F^{(M)}(u,v)$ is its singular growth
 \be
 F^{(M)}(u,v)\sim   {u}^{-\delta}\,, \label{eq:Pomeron}
 \ee
in the limit (DLC)~\footnote{This limit can be thought of as \emph{a} Regge limit, which is normally formulated in the momentum space. With the momentum space description comes a long history of phenomena associated with Regge behavior. As illustrated in Fig. \ref{fig:ReggeLimit}, this limit can also be treated as a double light-cone limit in position space. However, we caution that some authors use this label to refer to \emph{broader} limits, for example \cite{Banks:2009bj}.  Our statement here will be made more precise in Sec.~\ref{sec:MCB} and also in Appendix \ref{sec:kinematics} via a Rindler-like parametrization~\cite{Lam:2017ofc,Cheung:2016iub}.}
\be
u\rightarrow 0, \quad v\rightarrow 1,  \label{eq:MinReggeLimit}
\ee
with  $\,\, 1/2>\delta>0$. In Sec. \ref{sec:MCB}, this limiting behavior is shown to allow MCB to be directly calculated by considering the corresponding boundary conditions for solutions to a conformal casimir. The limit will be defined more precisely in Sec. \ref{sec:DLC} where it is formulated as a {\bf double light-cone limit}~\footnote{A similar limit to ours is explored in \cite{Alday:2015eya,Li:2015rfa,Costa:2011dw}.}. There are various ways to map conformal cross ratios on to the light-cone. Some involve single light-cone limits--like $u \rightarrow 0$ in \cite{Fitzpatrick:2014vua,Fitzpatrick:2015qma}, or various double light-cone limits correspoding to different physical regimes \cite{Komargodski2013,Alday:2015ota,Kaviraj:2015cxa,Vos:2014pqa,Alday:2015ewa}. For example, the conformal bootstrap program examines, among other limits, crossing relations after taking a double light-cone limit where  $u\rightarrow 0, v \rightarrow 0$ asymmetrically as in \cite{Alday:2015ota}. 

\begin{figure}[ht]
\begin{center}
\includegraphics[width = 3in]{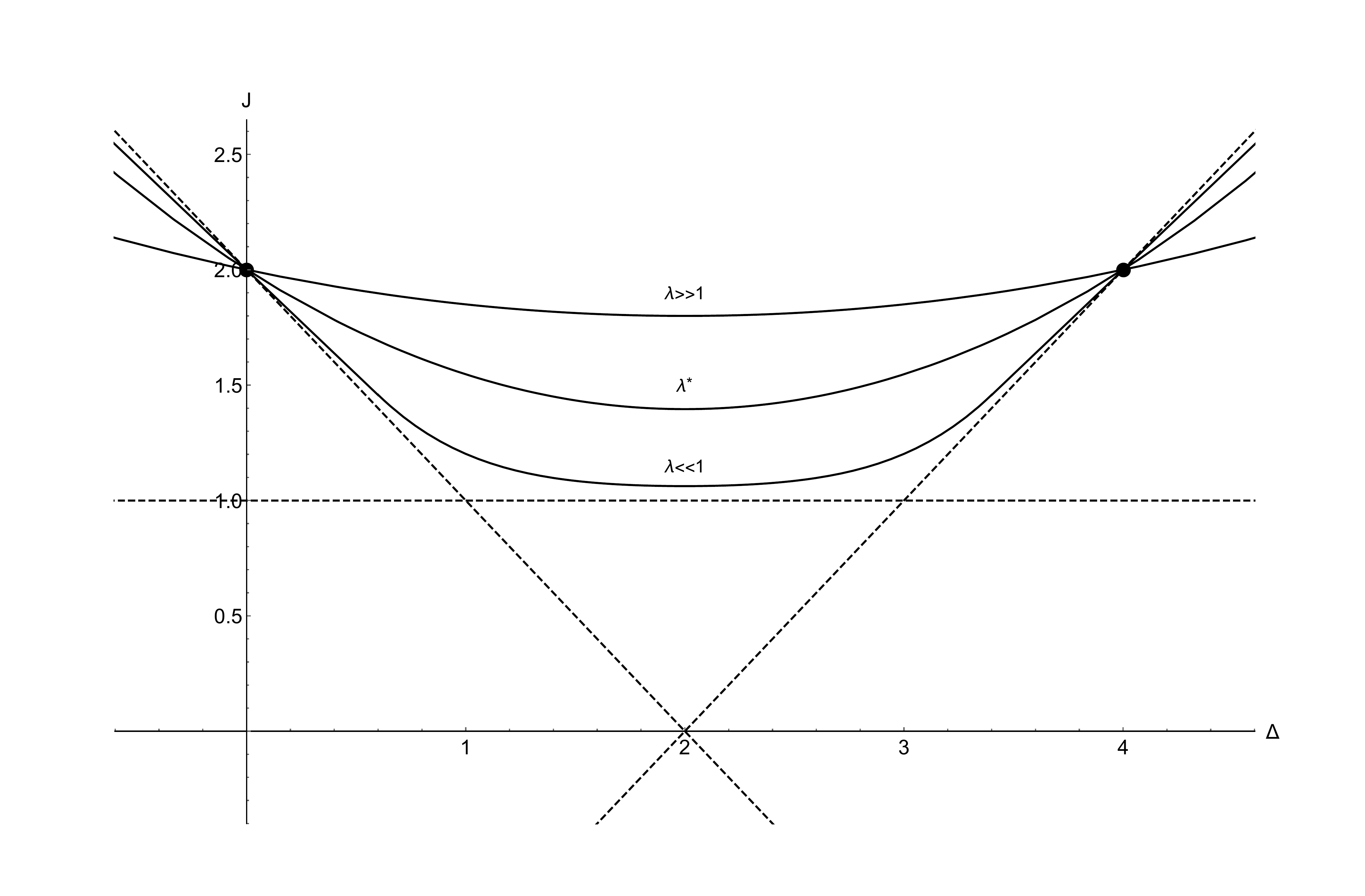}
\caption{Schematic form of the $\Delta-\ell$ relation at $d=4$ for twist-2, ($\tau_0=2$),  at  weak ($\lambda\ll 1$) and strong coupling
($\lambda\gg 1$). This figure is similar to that from \cite{Brower:2006ea} where it was first introduced.}
\label{fig:BFKLDGLAP}
\end{center}
\end{figure}

By working directly in a Minkowski limit, we show $\delta = (\ell_{eff}-1)/2$ with $\ell_{eff}$ serving as an effective spin. For all CFTs with a gravity dual, this effective spin obeys an upper bound~\cite{Brower:2006ea}, 
 \be
 \ell_{eff}\leq 2,
 \ee
In the case of ${\cal N}=4$ SYM it can be associated with string modes interpolating the graviton in AdS~\cite{Brower:2006ea,Brower:2010wf}.  (See Fig. \ref{fig:BFKLDGLAP}.)  This singularity is historically referred to as the Pomeron in both the context of QCD and for strongly coupled gauge theories.  For SYK-like 1-d effective CFT~\cite{kitaev1,kitaev2,Sachdev:1992fk,Polchinski:2016xgd,Jevicki:2016bwu,Maldacena:2016hyu},  $(\ell_{eff}-1)$ drives the Lyapunov behavior for thermo-correlators, $\langle W(t) V (0)W(t) V(0)\rangle_{{\beta}}$, with Eq. (\ref{eq:Pomeron}) becoming $ e^{-2\pi (\ell_{eff}-1) t/\beta}$ at $t$ large. These models are nearly maximally chaotic with the stringy and thermal deviations being driven by this Regge limit effective spin.

\subsection{Outline}
This paper, including the major results, is organized  as follows:

In Sec.~\ref{sec:DLC} we discuss the kinematics of near forward scattering in a CFT where the relevant regime can be described as a DLC. Understanding this limit and the physical scattering regions are essential to the analysis in Secs. \ref{sec:MCB}-\ref{sec:CFT-1}. The OPE in an Euclidean setting exploits dilatation invariance leading to a \emph{single-scale} scaling, which specifies the relevant boundary conditions for ECB. In contrast, in a Minkowski setting there can be \emph{two scaling limits}. It is useful to adopt a new parameterization where this scaling is easily expressed as
\bea
{\rm Dilatation}: \,\, &&\sigma\rightarrow \infty, \label{eq:dilatation}\\
{\rm Boost}: \,\, &&  w\rightarrow \infty. \label{eq:boost}
\eea
The dilatation limit is characterized by a scaling parameter $\sigma$, Eq. (\ref{eq:sigma}), which singles out the conformal primary of leading twist.  The second scaling parameter $w$, defined  by Eq. (\ref{eq:w}),  relates to a Lorentz boost, specified by a conformal rapidity $y$, as in Eq. (\ref{eq:boost2}),  where
\be
w\simeq 2 /\sqrt u\sim e^{2y}.
\ee
From the s-channel scattering perspective, t-channel spin, $\ell$, is conjugate to the rapidity~\cite{Brower:2007xg}. In the scaling limit of large rapidity,  conformal symmetry manifests itself through an effective spin as in Eq. (\ref{eq:Pomeron}). Eq. (\ref{eq:Pomeron}) is a generic feature for the CFT correlation function $F^{(M)}(w,\sigma)$, in the limit  $w  \rightarrow \infty$ with $\sigma $ fixed. 

In Secs. \ref{sec:Crossing}-\ref{sec:graviton-exchange} we discuss crossing. To contrast Minkowski \emph{t-channel OPE} with \emph{s-channel OPE} we use the $w$, $\sigma$ parameterization where the s-channel physical region corresponds to $1<w<\infty$ and u-channel in $-\infty<w<-1$. Of particular importance, we clarify why the contribution from the stress-energy tensor in a t-channel OPE, as well as its stringy correction in a dual description, serves as the dominant contribution in the DLC limit.

In Sec.~\ref{sec:MCB} we  discuss conformal blocks themselves, directly deriving MCB, and looking at relations to ECB and their asymptotic behavior. More technical details are left to Appendix \ref{sec:MCFB-append}. For a t-channel OPE, MCB obey boundary conditions
\begin{align}
G^{(M)}_{(\Delta,\ell)} (u,v)  \sim \sqrt u^{1-\ell} \, \Big( \frac{1-v}{\sqrt u}\Big)^{1-\Delta}\, .\label{eq:Mbdry}
\end{align}
The limit $u\rightarrow 0$ is to be taken first before $v\rightarrow 1$,  with  $ 1<\frac{1-v}{2\sqrt u}< \infty$.
 In contrast, for the corresponding limit of $u\rightarrow 0$ and $ v\rightarrow 1$ in an Euclidean OPE,   conformal blocks obey boundary conditions
\be
G^{(E)}_{(\Delta,\ell)} (u,v)  \sim\sqrt u^{\Delta} \, (1-v)^{\ell} \, . \label{eq:Ebdry}
\ee
This direct approach shows that the $G^{(M)}_{(\Delta,\ell)}(u,v)$ are  {\bf related to, but not directly given by} the analytic continuation of $G^{(E)}_{(\Delta,\ell)}(u,v)$~\footnote{The literature often refers to the fact that the Lorentzian version is proportional to the analytic continuation, but here we spell out an exact relation.}.

While the conventional asymmetrical limits of taking $u\rightarrow 0$ first before $v\rightarrow 1$ is useful to differentiate MCB from ECB, we show in Sec.~\ref{sec:symmetric} that a more symmetric treatment, in terms of variable $(w,\sigma)$, allows us to generalize our approach in treating conformal blocks for  general dimension, d. It also helps elucidate their crossing properties, and allows a more explicit demonstration the connection in the high energy limit to Euclidean $AdS_{d-1}$ bulk propagators. These generalizations allow a smooth transition to the interesting case of $d=1$.
 
In Sec. \ref{sec:application-I}, we provide a more precise treatment on how t-channel OPE, for s-channel scattering, should be interpreted. This involves identifying the principal series representation for a non-compact group via a standard harmonic analysis. Through a Sommerfeld-Watson resummation, it is shown that $F^{(M)}(u,v)$ takes on a Mellin-like representation, Eq. (\ref{eq:newgroupexpansion}). It follows that, in the physical region,  its imaginary part is given simply by
\begin{align}
{\rm Im} \, F^{(M)}(u,v)=&  \sum_\alpha     \int\limits_{L_0-i\infty}^{L_0+i\infty}  \frac{d \ell}{2 i}   \;  
	   a(\ell, \Delta_\alpha(\ell))\nonumber \\
      &\times  {G}^{(M)} (\ell,\Delta_\alpha(\ell) ; u,v) \, .
\label{eq:newgroupexpansionDisc}
\end{align}
By pushing the integration contour in Eq. (\ref{eq:newgroupexpansionDisc}) to the left, contributions  from singularities in the complex-$\ell$ plane become dominant in the high energy limit. The leading contributions come from the family of conformal primaries which interpolate the stress-energy tensor, $\Delta_P(\ell)$, with a branch-point singularity at $\ell_{eff}\lesssim 2$. For holographic theories, the deviation from $\ell=2$ can be understood in terms of stringy corrections for integrable theories and receives additional temperature corrections in thermal theories.  We emphasize that ${\rm Im} \, F^{(M)}(u,v)$ has non-vanishing support only in the s-channel and u-channel physical region, $1<|w|$. In the region $-1<w<1$, the contour can be closed  to the right, leading to vanishing ${\rm Im} \, F^{(M)}(u,v)$.

As a special application of this new approach, we turn in Sec. \ref{sec:CFT-1} to CFT scattering in $d=1$ and SYK-like models.  We use the above Mellin representation to formulate the relevant 4-pt correlator in Sec. \ref{sec:SYK1} and discuss the role of effective spin for SYK-like models in Sec.~\ref{sec:SYK}.  By taking advantage of reparametrization invariance, an integral equation for  ${\rm Im} \, \Gamma(w)$~\footnote{Although $\Gamma(w)$ is a CFT 4-point function we have changed notation to make comparison with the $d=1$ literature simpler.} is constructed in Eq. (\ref{eq:SD-ImW-2}). We stress that we formulate the model  directly as  a Minkowski scattering problem, leading to an equation involving ${\rm Im} \, \Gamma(w)$ in the physical region only. This integral equation can be diagonalized readily by exploiting the conformal boost invariance. Using this approach, the integral equations involved can be used to derive simpler algebraic relations as described in Sec. \ref{sec:SYK2}.  For the case of SYK models,  the partial-wave amplitude is given as an integral over a Legendre function of the second kind, Eq. (\ref{eq:inversion-2}). The effective spin shows up as the right-most singularity of the partial-wave amplitude, $A(\ell) $, Eq. (\ref{eq:inversion-2}), a pole at $\ell=2$ and  is analytic to the right.

We end in Sec. \ref{sec:discuss} with a short summary and adding further discussion on the role of {\bf spectral curves}.    (See further discussion in Sec. \ref{sec:propagator}.) For the canonical AdS/CFT correspondence, conformal invariance leads to spectral curves, $\Delta (\ell)$, that are symmetric under
 \be
 \Delta(\ell) \leftrightarrow d-\Delta(\ell). \label{eq:Delta-j-symmetry}
 \ee
Its importance for high energy scattering is discussed in Sec. \ref{sec:SpectralCurve} for $d=4$ CFTs. This property can be seen in Fig. \ref{fig:BFKLDGLAP}, and plays an important role in determining the effective spin. The effective spin can be obtained by solving an equation involving anomalous dimension,~\cite{Brower:2006ea,Brower:2007xg}
\be
\gamma(\ell_{eff}) + \ell_{eff}=0 \label{eq:PomeronIntercept}
\ee
where $\gamma(\ell)$ is the analytically continued anomalous dimension. More generally, non-thermal deviation from $\ell=2$  can be attributed to stringy corrections. Thermal theories receive additional temperature dependent corrections. For the graviton, Eq. (\ref{eq:Delta-j-symmetry}) can be thought of as coming from the AdS mass condition~\footnote{It has been shown that for $\mathcal{N}=4$ SYM,  $\Delta(d-\Delta)$ is Borel summable while while $\Delta$ is not~\cite{Basso:2011rs,Gromov:2012eg}. In this case, where the theory is thought to be integrable, integrability techniques can be used to determine this sum to high order.  For a review of the procedure and it's application to the Pomeron, see \cite{Brower:2014wha}. One should be careful when computing corrections to the spectral curve that the expansion is well defined.},
\be 
\Delta(d-\Delta)=m^2_{AdS}\, , \label{eq:adsmass}
\ee
with non-thermal stringy corrections \emph{respect} this symmetry. The leading correction can be interpreted as introducing a spin dependent mass $m^2_{eff}(\ell) \sim 2\sqrt{\lambda_t}\,(\ell-2)$. For ${\cal N}=4$ SYM, it can be shown that $m^2_{eff}(\ell)$ admits a systematic expansion about $\ell=2$,
\be
m^2_{eff}(\ell)=\sum_{n=1}^\infty \beta_n (\lambda_t) (\ell-2)^n. \label{eq:adsmass-2}
\ee
Each coefficient admits a strong coupling expansion in $\lambda_t^{-1/2}$, with leading behavior $ \beta_n (\lambda_t)=O(\lambda_t^{1-n/2})$~\cite{Brower:2014wha,Basso:2011rs,Kotikov:2013xu,Gromov:2014bva}.  (See Eqs. (3.14-3.15) of Ref. \cite{Brower:2014wha}.)  Including thermal corrections can introduce a new effective mass that \emph{breaks} the symmetry in Eq. (\ref{eq:Delta-j-symmetry}). Nonetheless we believe one can still apply the analysis starting from Eq. (\ref{eq:PomeronIntercept}).   

We have also included several appendices. These provide more details than is normally done since, in spite of the initial work of \cite{Brower:2006ea} more than a decade ago, scattering for CFT remain unfamiliar to most CFT practitioners. Additionally those interested in CFT scattering come from a variety of backgrounds so we have aimed to be as self contained as possible. In Appendix \ref{sec:kinematics} we set channel conventions and provide kinematic relations between invariant cross ratios and position coordinates via a Rindler-like parametrization~\cite{Lam:2017ofc,Cheung:2016iub} appropriate for the DLC limit. Appendix \ref{sec:holography} connects a conformal invariant 4-point function, $F^{(M)}(u,v)$, to an ordinary momentum-space amplitude from the perspective of the AdS/CFT conjecture. In Appendix \ref{sec:MCFB-append} we clarify in greater details the relation of MCB to the conventionally defined ECB. Appendix \ref{sec:application-II} focuses on the application of CFT scattering to DIS, focusing on exploiting the $SO(1,1)\times SO(1,1)$ symmetry and connecting DIS to conformal methods. Finally, in Appendix \ref{sec:green} a conventional Hilbert space treatment for $d=1$ CFTs is carried out and extended to the case of non-square-integrable, but  power-behaved like Eq. (\ref{eq:Pomeron}), functions. As explained in Secs. \ref{sec:application-I} and \ref{sec:CFT-1}, this illustrates that, through the Sommefeld-Watson transform via complex angular momentum, the re-summed Minkowski OPE can be interpreted as a (deformed) harmonic analysis over non-compact group. We also outline the basics of the SYK theory needed for interpreting Sec. \ref{sec:CFT-1}.

\paragraph{NOTE:}  Upon completing this work we were made aware of \cite{Simmons-Duffin:2017nub}, extending the work of Caron-Huot \cite{Caron-Huot:2017vep}, which has some overlap and similar conclusions as ours. Other related works include~\cite{Murugan:2017eto,Kravchuk:2018htv,Iliesiu:2018fao,Turiaci:2017zwd,Lam:2018pvp}. It is useful to briefly comment on the relation of this study to that of  \cite{Caron-Huot:2017vep} and \cite{Simmons-Duffin:2017nub}.  The starting point of both \cite{Caron-Huot:2017vep} and \cite{Simmons-Duffin:2017nub} is CFT in an Euclidean setting. One impetus for the study of CFT in the Lorentzian limit is related to the question of chaos bound, e.g.,  \cite{Shenker:2014cwa} and SYK model. In \cite{Caron-Huot:2017vep}, CFT scattering amplitude, ${\cal M}$, is introduced  by identifying it as a discontinuity of analytically continued Euclidean correlation function, (e.g., Eq. (2.13) in \cite{Caron-Huot:2017vep}), which in turn leads to a representation for its imaginary part, ${\rm Im}\, {\cal M}$, as a ``double-discontinuity", (Eq. (2.14) of \cite{Caron-Huot:2017vep}), or, equivalently, a double-commutator. Recognizing the importance of the constraint imposed by the Regge asymptotics, the focus has been to find a relation between the partial-wave amplitude $a(\ell, \Delta)$, analytically continued in complex $\ell$,  and ${\rm Im} \,{\cal M}$, leading to its key result, (Eq. (3.20) of \cite{Caron-Huot:2017vep}). One important feature is the asymptotic boundedness in the limit ${\rm Re} \, \ell\rightarrow \infty.$ The procedure adopted followed a traditional Regge analysis in introducing Froissart-Gribov representation.

Our direct study for  Lorentzian CFT  is motivated by that of \cite{Brower:2006ea} where conformal Regge behavior can be derived, and we discuss how a double-commutator, as the discontinuity of a CFT scattering amplitude, can be related to a t-channel OPE through a principal series representation, e.g., Eq. (\ref{eq:newgroupexpansionDisc}). In a broader context, Eq. (\ref{eq:newgroupexpansionDisc}) itself can be derived from the unitary irreducible representation of the full non-compact $SO(4,2)$, Eq. (\ref{eq:trueinvariance}). In an Euclidean setting, this leads to the principal series representation for $SO(5,1)$, Eq. (\ref{eq:groupexpansion}). The importance of this relation has also been emphasized in \cite{Simmons-Duffin:2017nub}, and also earlier in \cite{Mack:2009mi}. In a Regge context, it can be traced back to earlier work of M. Toller \cite{Toller1}. 
The key dynamical assumption in our approach is meromorphy in the complex $\Delta-\ell$ plane for the ``partial-wave amplitude", $a(\ell, \Delta)$, e.g., Eq. (\ref{eq:trueCFT}), which leads formally to a t-channel OPE via spectral curves.  

Our study here complements that of \cite{Caron-Huot:2017vep}.  Eq. (\ref{eq:newgroupexpansionDisc}) involves MCB, which can be introduced directly in a Minkowski setting, as discussed in Sec. \ref{sec:MCB}, thus avoiding the step of intricate analytic continuation.  The ability to close various complex contours require specifying boundedness of $a(\ell, \Delta_\alpha(\ell) )$ for ${\rm Re} \, \ell$ large.  The necessary assumption involved is, in the end, equivalent to the assumption of polynomial boundedness in the DLC limit, Eq. (\ref{eq:Pomeron}), which is the main focus of our study.  The close relation between these two approaches can be brought out more explicitly in the case of $d=1$, e.g. Eqs. (\ref{eq:1D-ImConformalBlock}), (\ref{eq:inversion-2}) and (\ref{eq:tw}). The issue of asymptotic boundedness can be analyzed  explicitly  in terms of an elementary  Hilbert space treatment, which is carried out in  Appendix \ref{sec:green}.

\section{The DLC Limit}\label{sec:DLC}

In this section, we spell out more precisely how dilatations and Lorentz boosts can be  related to the dependence of conformal correlators on invariant cross ratios in the limit $u\rightarrow 0$ and $v\rightarrow 1$, as in Eq. (\ref{eq:MinReggeLimit}). The causal relationship described in the introduction is depicted in Fig. \ref{fig:chan-proc} and defines the s-channel scattering region. This limit can of course be taken both  in Euclidean and Minkowski signatures. For an Euclidean OPE, this limit involves only  a single scale corresponding to a dilatation under $SO(5,1)$. The corresponding asymptotic boundary conditions for ECB are given by (\ref{eq:Ebdry}).
In a Minkowski setting,  however, because of the Lorentzian structure,  the  same limit can involve two scales, one for Lorentz boost,  and the other for dilatation. As indicated in the introduction this limit involves particles being scaled along forward and backward light-cones and we we refer to this specific double light-cone limit as the DLC. The relevant scaling limit exploits the invariance under $SO(1,1)\times SO(1,1)$, a subgroup of the full conformal symmetry $SO(4,2)$.

In Sec. \ref{sec:boost} we discuss the DLC limit more explicitly. It is useful to adopt a Rindler-like parametrization, Eqs. (\ref{eq:virtuality}-\ref{eq:boostparam}), fixing the phase space of s-channel physical region. In Sec. \ref{sec:Crossing}, we discuss the related DLC limit under crossing. In place of  $(u,v)$, we introduce new sets of independent invariants, Eqs. (\ref{eq:w}) and (\ref{eq:sigma}), which are not only more useful for the DLC limit, but also simplify the description of s-u crossing. In Sec. \ref{sec:graviton-exchange}, more explicit connection  to the near-forward scattering is discussed. We consider the contribution from   the stress-energy tensor, ${\cal T}^{\mu\nu}$, in a Minkowski setting, which in turn helps to motivate boundary conditions for MCB, (\ref{eq:Mbdry}).
 
Additional details and definitions can be found in Appendix \ref{sec:DLC-append}. 
 
 \subsection{Kinematics}\label{sec:boost}
 For both Euclidean and Minkowski  settings, Eq. (\ref{eq:MinReggeLimit}) corresponds to the limit  $x_{12}^2\rightarrow 0$ and  $x_{34}^2\rightarrow 0$ and $x_i^2\rightarrow 0$, $i=1,2,3,4$,  with other invariants between left- and right-movers fixed: $L^2\simeq  x_{13}^2\simeq  x_{23}^2\simeq x_{24}^2\simeq x_{14}^2=O(1)$. $L$ provides a scale for the relative separation between left- and right-movers, 
\be
\sqrt u \simeq   \sqrt{ x_{12}^2x_{34}^2   }/L^2\rightarrow 0 \,. \label{eq:DLC1}
\ee
Due to scale invariance, this is equivalent to increasing the left-right separation, 
\be
L^2\simeq x_{ij}^2\rightarrow \infty, \quad i=1,2, \quad {\rm and}\quad j=3,4 \, ,\label{eq:DLC2}
\ee
while keeping fixed $x_{12}^2$, $x_{34}^2$, and  $x_i^2$, $i=1,2,3,4$. For an Euclidean OPE, the limit, (\ref{eq:DLC1}) or (\ref{eq:DLC2}), involves only  a single scale, $L$, corresponding to the aforementioned dilatation under $SO(5,1)$, which specifies boundary conditions for ECB, Eq. (\ref{eq:Ebdry}).

In a Minkowski setting,  because of the Lorentzian structure,  the  same limit can involve two scales, one for Lorentz boost, and the other for dilatation. Consider  light-cone coordinates, $ x=(x^+,x^-; x_\perp)$, $x^\pm = x_0\pm x_L$, with  $x_\perp$ its  $(d-2)$-dimensional  transverse components. We shall keep all $x_i$ spacelike, $x^2=-x^+x^-+  x_{\perp}^2= (-x_{0}^2+ x_{L}^2)+  x_{\perp}^2> 0$. 

 Focus first on the case $d=2$. For each coordinate, let us   define  $r=\sqrt{-x^+x^-}>0$, which    can parametrized   by a parameter $\eta_i$~\footnote{The length scale $\mu_0$ is introduced for `visual" purpose, which can be set to unity.},
\be
 r_{i}=r(\eta_i)= \mu_0 \, e^{-\eta_i}  \, . \label{eq:virtuality}
\ee
The allowed range $0<r_i<\infty$ corresponds to $-\infty<\eta_i<\infty$. We shall refer to $r$ as ``conformal virtuality"~\footnote{We use this name as the quantity will play a similar role to conventional virtuality which is an energy like quantity defined as the off-shell energy of a particle.} and dilatation in light-cone components, $x^{\pm}\rightarrow \lambda\, x^{\pm}$,   corresponds to scaling in conformal virtuality.  In terms of  parametrization Eq. (\ref{eq:virtuality}), it corresponds to a shift, $\eta\rightarrow \eta - \log \lambda$. For $d=2$, $\eta_i\rightarrow \infty$ sends $x_i^2\rightarrow 0$.
 We can vary conformal virtualities for left-movers, (3,4), and right-movers, (1,2),  independently by performing separate scaling transformations, leading to the desired limit (\ref{eq:DLC1}) or (\ref{eq:DLC2}).
 
To identify a Lorentz boost, we next introduce rapidity variable, $0\leq y< \infty$, for each coordinate. Consider first the right-movers, $(x_1,x_2)$. For time ordering, we keep $x_1^+<0$ and $x_2^+>0$, and  parametrize their light-cone components  as
 \be
 x_i^\pm = \pm \epsilon_i r_i e^{\pm y_i} \label{eq:boostparam} 
 \ee
 with  $\epsilon_1 = -1$,  $\epsilon_2=+$. Similarly,  for the pair $(x_3,x_4)$, $x_j^\pm =\mp \epsilon_j r_j e^{\mp y_j}$, with $\epsilon_3 = -1$ and $\epsilon_4=+$, so that $x_3^-<0$ and $x^-_4>0$. Sending all rapidities $y_i\rightarrow \infty$, with  $r_i$ fixed,  leads to Eq. (\ref{eq:DLC2}). Therefore, one way  to achieve this limit is to perform a global Lorentz boost.   For our purpose, it is sufficient to fix  a single global rapidity $y$ for all four legs, with $y=y_1=y_2=y_3=y_4$. 
  
In order to connect with Eqs. (\ref{eq:DLC1}) and (\ref{eq:DLC2}) for $d\neq 2$, it is now necessary to discuss the effect of transverse coordinates. To simplify the discussion, we will  adopt a frame where $x_{i,\perp}=x_{2,\perp}$ and $x_{3,\perp}=x_{4,\perp}$ and with $b_{\perp}= x_{1,\perp}-x_{3\perp}$ as the relative separation between $(1,2)$ and $(3,4)$ in the transverse impact space~\footnote{This ansatz is unnecessary but simplify our discussion here. For related discussion, see \cite{Cornalba:2006xm,Cornalba:2007fs,Cornalba:2008qf,Cornalba:2009ax,Costa:2012cb}. }.  In terms of the global rapidity $y$ and conformal virtuality $r_i$ for each coordinate, cross ratios $u$ and $v$ take on relatively simple forms, Eqs. (\ref{eq:u-global}-\ref{eq:v-global}).  In the case of two pairs of equal conformal virtuality, $r_1=r_2$ and $r_3=r_4$,
\begin{align}
u = \frac{16}{(e^{2y} +2R(1,3) + e^{-2y})^2 } \nonumber \\
v= \frac{ (e^{2y} -2R(1,3) + e^{-2y})^2 }{ (e^{2y} + 2R(1,3) + e^{-2y})^2}\label{eq:uvrange}
\end{align}
where the transverse separation enters through 
\be
R(i,j) =\frac{r_i^2+r_j^2+b_\perp^2}{2r_ir_j}. \label{eq:R}
\ee
with $i=1$ and $j=3$.
The limit $u\rightarrow 0$ can therefore be achieved either by $y\rightarrow \infty$ or $b_\perp^2 \rightarrow \infty $ first. For near-forward scattering, or the DLC limit, we consider the first scenario of $y\rightarrow \infty$ with $b_\perp^2$ fixed.  The limit $u\rightarrow 0$ therefore exploits the scaling limit of Lorentz boost.  In this limit, with conformal virtualities fixed, one also has $v\rightarrow 1$. Together, they correspond to the DLC limit, Eq. (\ref{eq:MinReggeLimit}), as promised.

For general unequal conformal virtualities, one finds, in the limit of large rapidity,
\begin{align}
w_0^{-1}\equiv  \sqrt u /2 \simeq  \frac{(r_1+r_2)(r_3+r_4) } { 2 z_{12}   z_{34}} \,  e^{-2y} , \label{eq:boost2} \\ 
\sigma_0 \equiv   \frac{1-v}{2\sqrt u} \simeq \frac{ b_\perp^2+z_{12}^2 + z_{34}^2  }{ 2 z_{12}z_{34}}  + O(e^{-2y})\label{eq:sigma0}
\end{align}
where we have introduced a tentative set of  variables,    $w_0
$ and $\sigma_0$,  and have also introduced  joint conformal virtualities ~\footnote{For convenience, we will have occasions to switch notation with $z_{12}\leftrightarrow z$ and $z_{34}\leftrightarrow \bar z$ or $z'$, in anticipation of the ADS/CFT connection.}, 
\be
z_{12}=  \sqrt{r_1r_2}  \, , \quad {\rm and} \quad 
z_{34}=  \sqrt{r_3r_4 } .
\ee

We adopt $w_0 \rightarrow \infty$ and $\sigma_0 \rightarrow \infty$  as two independent scaling limits, global boost and dilatation, which characterizes the DLC limit for CFT. For $d\neq 2$, the limit $\sigma_0\rightarrow \infty$ generically corresponds to  
$
b_\perp^2>> z_{12} z_{34} 
$.
For both  $d=2$ and $d\neq 2$ with  $b_\perp$ fixed,  large $\sigma_0$ corresponds to the limit of small conformal virtuality, 
$
\frac{z_{12}}{ z_{34}}  \rightarrow  0$   or    $  \frac{ z_{34}}{z_{12}}  \rightarrow  0.
$
This is analogous to the  ``near massless" limit in a conventional scattering limit. 

\subsection{Physical Regions}\label{sec:Crossing} 
To understand how CFT correlators can be used for scattering, we need to comment on the kinematics of physical regions and constraints due to s-u crossing for $\langle T ({R}(1) {R}(2) {L}(3) {L}(4) )\rangle$.  To simplify the discussion, we consider correlators  for four identical  scalar conformal primaries. 

From Eqs. (\ref{eq:uvrange}) and (\ref{eq:sigma}), more generally,  from Eqs. (\ref{eq:u-global}-\ref{eq:v-global}), the s-channel physical region corresponds to  $0<u<1$ and $0<v<1$. This is the causal structure of Fig. \ref{fig:ReggeLimit}. In terms of the previous section this is $w_0$ and $\sigma_0$,  $1<w_0<\infty$ and $1<\sigma_0<\infty$. It is in this regime we can examine a t-channel OPE.

A similar t-channel OPE applies to the u-channel, Fig. \ref{fig:su-tOPE}, which can be found by interchanging either $1\leftrightarrow 2$ or  $4\leftrightarrow 3$, leading to 
\be
u\rightarrow u'=\frac{u}{v}\, , \quad {\rm and} \quad v\rightarrow v'=\frac{1}{v}.   \label{eq:su-exchange}
\ee
It is possible to adopt $(u',v')$ as an alternative choice for cross ratios. This alternative choice leaves the limit of interest (\ref{eq:MinReggeLimit}) unchanged. Under such an exchange, we see that the u-channel physical region,  $0<u'<1$ and $0<v'<1$, corresponds to $0<u<\infty$ and $1<v<\infty$.  Therefore, variables $(u,v)$ transform  asymmetrically   under s-u crossing, 

As shown in \cite{Dolan:2011dv}, it is useful to change variables from $u$ and $v$ to a  pair of independent  variables,  $(x,\bar x)$, where 
\be
u=x\bar x, \quad v=(1-x)(1-\bar x). \label{eq:xxbar}
\ee 
In a Euclidean  setting, $x$ and $\bar x$ are complex and $x^*=\bar x$.  In a Minkowski  setting, the s-channel physical region corresponds to both $x$ and $\bar x$ real and positive with  $0< x<1$ and $0< \bar x<1$. This corresponds to $0<w_0=\sqrt {x\bar x}<1$, $1<\sigma_0= (\sqrt{x/\bar x} +\sqrt{\bar x/ x} )/2<\infty$. Under s-u exchange, $x\rightarrow x'=-\frac{x}{1-x}$ and $  \bar x\rightarrow \bar x'=-\frac{\bar x}{1-\bar x}$. From the u-channel perspective, the range for $x'$ and $\bar x'$ are unbounded, $(-\infty,0)$: s-u crossing remains  asymmetrical.  

In order to take a symmetric approach, it is useful to introduce a new set of variables $(q,\bar q)$,
\be
q\equiv \frac{2-x}{x}\,, \quad {\rm and} \quad \bar q \equiv\frac{2-\bar x}{\bar x}. \label{eq:qqbar}
\ee
with the s-channel physical region corresponds to $1<q,\bar q<\infty$, and u-channel physical region corresponds to $-\infty< q,\bar q<-1$.  Under s-u crossing,  one simply has $q\rightarrow -q$ and $\bar q\rightarrow -\bar q$. In terms of these,  cross ratios are given by 
\begin{align}
u&=\frac{4}{(q+1)(\bar q+1)}\, \quad {\rm and} \nonumber \\
v&= \frac{(q-1)(\bar q-1)}{(q+1)(\bar q+1)}.
\end{align}

It will be useful to change variables one more time to $(w,\sigma)$, which we will ultimately adopt in Sec. \ref{sec:symmetric}. These parameters can be defined in terms  of $(q,\bar q)$ or more directly from $(u,v)$,  by
\begin{align}
w=&\sqrt {q\bar q} = \frac{  \sqrt{ 4-(u+2(1-v))} }{\sqrt u}\nonumber \\
&\simeq  2 \sqrt u^{-1} ,   \label{eq:w}\\
\sigma =&\frac{q+\bar q}{2\sqrt{ q\bar q}}=  \frac{1-v}{\sqrt u \sqrt{ 4-(u+2(1-v))}} \nonumber \\
&\simeq   \frac{1-v}{2\sqrt u},  \label{eq:sigma}    
\end{align}
The s-channel physical region now corresponds to   $1<w<\infty$ and $1<\sigma<\infty$. Under s-u crossing,
\be
w  \rightarrow - w\, ,  \quad {\rm and} \quad \sigma  \rightarrow \sigma.
\ee 
Therefore, the u-channel physical region corresponds to $-\infty<w<-1$ and $1<\sigma<\infty$.   These variables, $w$ and $\sigma$,  serve as the most symmetric variables for describing scattering for CFT.  
In Eqs. (\ref{eq:w}-\ref{eq:sigma}), approximate equalities hold in the DLC limit, i.e., $
w\simeq w_0$  and $\sigma \simeq \sigma_0$.

In what follows, we will use $(u,v)$, $(q, \bar q)$ and $(w,\sigma)$ as three equivalent sets of independent invariants for specifying Minkowski conformal blocks, and they can be used interchangeably.

\subsection{Eikonal Scattering}\label{sec:graviton-exchange}

It has been shown~\cite{Brower:2006ea,Brower:2007qh,Brower:2007xg,Cornalba:2006xm,Cornalba:2007fs,Cornalba:2008qf,Cornalba:2009ax,Costa:2012cb} that the connected part of the invariant function in a t-channel OPE,  $F_{conn}^{(M)}(u,v)$, can be related to the scattering      amplitude for high energy near-forward scattering at fixed virtualities~\footnote{This is the traditional momentum-space virtuality. The amplitude, $T(s,t; z_{12}, z_{34}) $, can be a CFT or an AdS amplitude for comparison.  The results here do not require an AdS dual amplitude, but our interpretation of the CFT correlation function suggests that it is a natural extension.  See Appendix \ref{sec:holography} for a more precise identification.}, $T(s,t; z_{12}, z_{34}) $. This correspondence can best be made through an eikonal phase, $\chi(s,\vec b,z_{12}, z_{34})$,
\be
\chi(s,b_\perp,z_{12}, z_{34}) \leftrightarrow F_{conn}^{(M)}(u,v)\, , \label{eq:chi-F}
\ee 
with $F_{conn}^{(M)}(u,v)$  expanded  as a sum over  t-channel  single-trace conformal primaries~\footnote{A more exact treatment, the complete eikonalization sum, requires keeping multiple trace contributions. See~\cite{Brower:2006ea,Brower:2007qh,Brower:2007xg,Cornalba:2006xm,Cornalba:2007fs,Cornalba:2008qf,Cornalba:2009ax,Costa:2012cb} for  further discussions.}. A convenient way of  introducing of this  eikonal phase is through a ``shock-wave" treatment for near-forward scattering~\cite{Cornalba:2006xm,Cornalba:2007fs,Cornalba:2008qf,Brower:2007qh}. (For early developments on this front, see  \cite{HOOFT198761,AMATI198781,PhysRev.186.1656,Kabat:1992tb,'tHooft1988,Deser1988,Deser:1993wt}.)
 At a large  impact separation, $\chi(s,\vec b,z_{12}, z_{34})$ is small and can be treated perturbatively,
\begin{align}
T&(s,t; z_{12}, z_{34}) \sim  \nonumber\\
  &\sim-i w   \int d^2\vec b\, e^{i\vec b \cdot \vec q}[e^{i\chi(s,b_\perp,z_{12}, z_{34})}-1]\\
&\sim   w \, \int d^2\vec b\, e^{i\vec b \cdot \vec q}  \chi(s,b_\perp,z_{12}, z_{34}) \, +O(\chi^2).   \label{eq:T-chi}
\end{align}
Here the eikonal is given in  a mixed coordinate and momentum representation. In performing Fourier transforms for the  light-cone components, large $s$ receives contribution from small $u$ region only, or, equivalently, large $w$ region, and the relevant Fourier integrals can be dropped. 
This representation can also be interpreted holographically as scattering in the AdS bulk~\cite{Brower:2006ea,Brower:2007qh,Brower:2007xg,Cornalba:2006xm,Cornalba:2007fs,Cornalba:2008qf,Cornalba:2009ax,Costa:2012cb}.  In (\ref{eq:chi-F}),  one has  identified the small $u$ limit with $s$ large by the reciprocal relation
\be
2 \sqrt u ^{-1} \simeq w  \Leftrightarrow   (z_{12}z_{34}s)/\mu_0^2, \label{eq:u2s}
\ee
where the scale $\mu_0^2$ introduced in (\ref{eq:virtuality}) can now be used as a global scale for scattering~\footnote{This identification can be done more formally, but it is sufficient for our present purpose to have qualitative understanding  as follows. In terms of momenta, since $s\sim p_1^+ p_3^-\simeq  p_2^+ p_4^-$, it scales  with  $(x_1^-x_3^+)$, $(x_2^-x_3^+)$, $(x_1^-x_4^+)$, $(x_2^-x_4^+)$ as $e^{-2y}$,   corresponding precisely the limit of large $s$. To fix the scale,  we  identify $s^{-1}$ with the average as $\mu_0^2(r_1+r_2)(r_3+r_4) e^{-2y}$.}. It is important to emphasize that the standard normalization we are adopting here corresponds to 
\be
T(s,t; z_{12}, z_{34})   \quad \Leftrightarrow \quad w \, F_{conn}^{(M)}(u,v)\, .
\ee
Therefore, the large $w$ behavior for $F_{conn}^{(M)}$ differs from that for $T(s,t; z_{12}, z_{34})$ by one power of $w$ or, equivalently, one power of $s$.  This extra factor of $w$ corresponds to the relativistic state normalization, leading to the conventional Optical Theorem, $ \sigma_{total} \simeq (1/s){\rm Im} \, T(s,0; z_{12}, z_{34})$~\footnote{For a careful review on the derivation of this representation, Eq. (\ref{eq:T-chi}), see Secs. 3-4 of \cite{Shenker:2014cwa} where the formalism introduced in \cite{Brower:2006ea} is also discussed. This is also summarized briefly in Appendix \ref{sec:holography}.}.  As mentioned earlier, and motivated by the AdS/CFT convention, in what follows we will switch from $(z_{12}, z_{34})$, to $(z,\bar z)$ for right- and left-movers for notational simplification.

A useful illustration is to consider the contribution from the stress-energy tensor, ${\cal T}^{\mu\nu}$, having $\Delta=d$ and $\ell=2$. Since it couples through conserved currents,  the amplitude picks up a  spin factor,  $ s^2$, from the product of two large coupling terms, involving $\partial_{x_i^-}\partial_{x^+_j}$,  $i=1,2$ and $j=3,4$. This factor reflects the effect of longitudinal Lorentz boost due to spin-2 exchange. This is consistent with the expectation $\chi \simeq T/s\sim s^{\ell-1}$.  In addition to this spin factor, at large separation, the amplitude is controlled by  a scalar propagator, 
$
\langle \phi(x)\phi(0)\rangle =1/ (x^2)^{\Delta}
$.
 Since both particles are moving near their respective light-cones,  the total amplitude, after integrating over $x^{\pm}$ components~\cite{Brower:2007qh}, leads to 
$
\chi (s,\vec b) \sim \, s^{\ell-1}\int {dx^+ dx^-}\langle \phi(x)\phi(0)\rangle \sim w^{\ell-1}\, (\frac{b^2}{2z_{12}z_{34}})^{1-\Delta},$ 
where we have scaled  the last expression on dimensional grounds, as in Eq. (\ref{eq:sigma}). Explicit conformal invariance can be achieved by expressing this  as
\be
\chi (s,\vec b) \sim  w^{\ell-1}\, \sigma^{1-\Delta}=   w \, \sigma^{-(d-1)}\label{eq:eikonal2}
\ee
 with $\ell=2$ and $\Delta=d$. This also agrees with what follows holographically for one-graviton exchange in the bulk based on AdS/CFT using Witten diagrams, as explained in Appendix \ref{sec:holography} ~\footnote{See  also \cite{Brower:2007qh}. It should be pointed out that the resulting eikonal from the stress-energy tensor  is purely real. When stringy effect is taking into account, the eikonal turns complex, with ${\rm Im}\, \chi>0$ by unitarity and with ${\rm Im}\, \chi/{\rm Re}\, \chi\simeq constant$ in the limit $w\rightarrow \infty$ first. This also necessitates in a broader discussion on the order in taking $w$ and $\sigma$ large. This issue has been discussed extensively in \cite{Brower:2007xg} and should be addressed also in application to SYK-like models. In this paper, we focus only on tree-graph contribution to the eikonal sum, and the limit is always taken with $w\rightarrow \infty$ first.}. This stress-tensor example also serves as the prototypical behavior expected in a Minkowski OPE expansion, 
\be
G_{(\Delta,\ell)}^{(M)}(u,v) \sim {w}^{\ell-1} \sigma^{1-\Delta}, \label{eq:Mbdry-1}
\ee
as $ w\rightarrow \infty$ first and then $\sigma \rightarrow \infty$.  With Eqs. (\ref{eq:w}) and (\ref{eq:sigma}), this corresponds to the boundary conditions for Minkowski conformal blocks, Eq. (\ref{eq:Mbdry}). 
 
Although this work is primarily concerned with the t-channel OPE, a similar eikonal form can be found from examining the s-channel OPE directly.  Performing an OPE in the s-channel one finds
\be
F^{(M)}(u,v) =\sum_{h,\bar h} a_s(h,\bar h) G^{(M)}_s(u,v)\, ,
\ee
with $\Delta=h+\bar h$ and $J=h-\bar h$. In the DLC limit, after removing the identity conformal primary $I$, following \cite{Costa:2012cb,Costa:2017twz} this becomes $i F^{(M)}(u,v) = -i [ e^{i\chi(s,  b_\perp, z, \bar z)}-1] \simeq \chi(s,  b_\perp, z, \bar z) + O(\chi^2)$. Although as shown above, this is most naturally interpreted as being dominated by t-channel stress-energy tensor exchange.

\section{Minkowski Conformal Blocks}\label{sec:MCB}

The conformal partial-wave expansion can be traced back to work of \cite{ferrara:72a,ferrara:72b,ferrara:74,ferrar:1975,Dobrev:76,LANG1993573,Dolan:2000ut} and has been carried out more recently  in a series of  papers by  Dolan and Osborn~\cite{Dolan:2004iy,Dolan:2003hv,Dolan:2011dv}. Application of partial-wave expansions for non-compact group has had a long history~\cite{mackey,Toller1,Toller2,Toller3,Ferrara:1973tg, Bali:1967zz, Herman}. Recent works on these expansions have been carried out exclusively in an Euclidean framework, and Minkowski results are obtained via careful analytic continuation. In this section we detail explicitly how the MCB can be obtained directly by understanding the boundary conditions of the conformal block differential Casimir.

\subsection{Definitions}
Conformal Blocks  are eigen-functions of a quadratic Casimir~\footnote{There is a quartic Casimir that can be used to construct recurrence relations between Conformal blocks in different dimensions\cite{Dolan:2011dv}. We use this below to relate results in even (respectively odd) dimensions.}, ${\cal D}$,
 \be
 {\cal D}  \; G_{\Delta,\ell}  (u,v) =  C_{\Delta,\ell}   \; G_{\Delta,\ell}  (u,v)\, , \label{eq:Casimir}
 \ee
associated with an arbitration conforaml transformation acting on a scalar four-point function.  The commutator algebra of generators involved can be realized as differential operators in terms of conformal cross ratios,
\begin{align} \label{eq:D}
{\cal D}= (1-u-v) \partial_v (v\partial_v +a + b) + u\partial_u (2u\partial_u -d) \nonumber \\
 -(1+u-v) (u\partial_u +v\partial_v +a) (u\partial_u +v\partial_v +b)\, ,
\end{align}
where $a=-\Delta_{12}/2$, $b=\Delta_{34}/2$, and $\Delta_{ij}=\Delta_i-\Delta_j$. We consider the case where $\Delta_1=\Delta_2$ and $\Delta_3=\Delta_4$ so that $a=b=0$.  Eigenvalues for the quadratic Casimir are then
\bea
C_{\Delta,\ell} &=&\Delta (\Delta -d)/2 +\ell(\ell+d-2)/2\, .\label{eq:indicial}
\eea

While most prior treatments have considered ECB, invariant under $SO(d+1,1)$, MCB, invariant under $SO(d,2)$, have been addressed recently by analytically continuing the Euclidean case. Here, we carry out a direct treatment of MCB by imposing DLC boundary conditions, Eq. (\ref{eq:Mbdry}).  We first carry out a more traditional analysis in Sec. \ref{sec:MCBexplicit}-\ref{sec:MCBexplicit}, and, in particular, point  out that  MCB as defined are not given by a direct analytic continuation of  ECB. In Sec. \ref{sec:symmetric}, an alternative, more symmetric treatment is carried out in terms of variables $w$ and $\sigma$,  (\ref{eq:w}) and (\ref{eq:sigma}). 

In the Lorentzian setting, $(\Delta,\ell)$ serve as  representation labels for $SO(d,2)$.  Since this is non-compact, $\Delta$ and $\ell$ can be continuous. However, to make contact with the OPE, we shall consider initially the situation where $\Delta$ is positive and real, and  $\ell$ is a non-negative integer. We will return to the question of restrictions on $\Delta$ and $\ell$ from the perspective of $SO(d,2)$ representation in  Sec. \ref{sec:application-I}.  

For later convenience, we  provide here two alternative expressions for $C_{\Delta,\ell} $,  with  $\epsilon = (d-2)/2$,
\begin{align}
 C_{\Delta,\ell}  &=  (\widetilde \Delta^2 +\widetilde \ell^2 )/2     - (d^2-2d+2)/4,           \label{eq:indicial1}\\
C_{\Delta,\ell}  &= \lambda_+(\lambda_+-1) \nonumber\\
&\qquad \quad +\lambda_-(\lambda_--1) + (d-2)\lambda_- ,   \label{eq:indicial2}
\end{align}
where $\widetilde \Delta =\Delta - d/2$  ,  $ \widetilde \ell = \ell + (d-2)/2$ and $\lambda_{\pm } =(\Delta \pm \ell)/2$. From Eq. (\ref{eq:indicial1}), there is a symmetry under  $\widetilde \Delta \leftrightarrow -\widetilde \Delta$, or $\Delta  \leftrightarrow d-\Delta$, and   $\widetilde \ell \leftrightarrow -\bar \ell$,  or $\ell   \leftrightarrow -\ell-d+2$. However, boundary conditions for conformal blocks break these symmetries.  Representation (\ref{eq:indicial2}) is particularly useful when treating the case $d=2$, when (\ref{eq:Casimir}) simplifies~\cite{Dolan:2011dv}. It is often useful to introduce $\epsilon=(d-2)/2$, with $\widetilde \Delta = \Delta  -(\epsilon +1)$ and $\widetilde \ell = \ell + \epsilon$.

\subsection{Indicial Analysis}
As a second order partial differential equation,  Eq. (\ref{eq:Casimir}) has removable singularities at $u=0, 1, \infty$-and similarly for $v$. Independent solutions can be specified by their behavior at these singular points. Representations (\ref{eq:indicial1}) and (\ref{eq:indicial2}) also suggest possible different variable separation procedure in solving this partial differential equation. (See Appendix \ref{sec:MCFB-append} for more details.)

As emphasized in the Introduction, and elaborated further in Sec. \ref{sec:DLC}, we are interested in the DLC limit where $u\rightarrow 0$ and $v\rightarrow 1$. A standard indicial analysis, with $G_{(\Delta,\ell)} (u,v) \sim u^p (1-v)^q$, leads to  a manifold of possible  solutions, where $p(2p-d) + q(2p+q-1)= C_{\Delta,\ell}$. From Eq. (\ref{eq:indicial1}),   the solution space is at least 4-fold degenerate. The usual OPE restricts $\ell$ to be a non-negative integer, $\Delta$ real, and $\Delta> \ell$, which is equivalent to $\lambda_+\geq \lambda_- \geq 0$.   
 A similar analysis has also be carried out in a related context in \cite{Caron-Huot:2017vep}.
 
Anticipating Eqs. (\ref{eq:Mbdry}) and (\ref{eq:Ebdry}),   we consider the following solutions:
\begin{align}
\begin{split}
(a)\,\,& p=\frac{\Delta-\ell}{2}, \quad q=\ell, \\
(b)\,\,&  p=\frac{\Delta-\ell}{2}, \quad q=-(\Delta-1), \\
(c)\,\,& p=\frac{(d-\Delta)-\ell}{2},  \quad q=\ell, \\
(d)\,\,&  p=\frac{(d-\Delta)-\ell}{2},\quad q=(\Delta-d+1).
\end{split}
\end{align}
Note that (c) and (d) are reflections of (a) and (b), under  $\Delta \leftrightarrow d-\Delta$, leading to mirror, or shadow, conformal blocks.  We will thus focus on case (a) and case (b) only. As boundary conditions, these solutions can be re-written as
\begin{align}
\begin{split} \label{eq:CB}
{(a)}& \quad  G^{(E)}_{(\Delta, \ell)} (u,v) \sim  \Big( \sqrt u\Big)^\Delta \Big( \frac {1-v}{\sqrt u} \Big)^\ell, \\
{(b)}& \quad G^{(M)}_{(\Delta, \ell)} (u,v) \sim  \Big( \sqrt u\Big)^{1-\ell}  \Big( \frac {1-v}{\sqrt u} \Big)^{1-\Delta}.
\end{split}
\end{align}
We see that case (a)  is that appropriate for Euclidean conformal blocks, Eq. (\ref{eq:Ebdry}), and case (b) appropriate for the Minkowski setting, Eq. (\ref{eq:Mbdry}). For both sets of boundary conditions,  the limit $\sqrt u\rightarrow 0$ is taken first before $v\rightarrow 1$. Equivalently,  both (a) and (b) in Eq. (\ref{eq:CB})  correspond to taking the limit $\sqrt u\rightarrow 0$ with $\frac {1-v}{\sqrt u}$ initially fixed, and then taking the limit $\frac {1-v}{\sqrt u}\rightarrow \infty$.  { \it Therefore, MCB and ECB are solutions to the same partial differential equation, defined with different boundary conditions, thus, in general, they cannot be expressed simply as  analytic  continuation of each other.}

\subsection{Explicit Construction of MCB} \label{sec:MCBexplicit}

We next focus on treating case (b) directly.  The standard procedure~\cite{Dolan:2004iy,Dolan:2003hv,Dolan:2011dv}  is  to change variables from $(u,v)$ to $(x,\bar x)$, Eq. (\ref{eq:xxbar}).  In order to maintain explicit crossing symmetry, we prefer to work first directly with new variables $(q, \bar q)$, Eq. (\ref{eq:qqbar}), where  the differential  operator for the Casimir can again be written as
 a sum of terms, (as the case for $(x,\bar x)$~\cite{Dolan:2011dv}), with 
${\cal D}(q,\bar q)  = D_0(q) + D_0(\bar q) +   D^{\epsilon}_1(q,\bar q)$, 
where $\epsilon = (d-2)/2$, with $D_0(q)$ and $D^{\epsilon}_1(q,\bar q)$ given by Eqs. (\ref{eq:D0q}) and (\ref{eq:Dmixedq}) respectively.  With $a=b=0$, (\ref{eq:D}), $D_0(q)$ takes on a simple form
\be
D_0(q)= (q^2-1) \frac{d^2}{dq^2} + 2q \frac{d}{dq},\label{eq:LegendreDO}
\ee
leading to second order ODE for Legendre functions. More importantly, $D^{\epsilon}_1=0 $ at $d=2$.  (Relation to the more traditional use of  $(x,\bar x)$ pair~\cite{Dolan:2011dv} is also provided in  Appendix \ref{sec:MCFB-append}.)

Boundary conditions in the DLC  limit are specified by taking  $u\rightarrow 0$ first before $v\rightarrow 1$. In terms of $q$ and $\bar q$, two possible approaches can be adopted: (1) an asymmetrical limit taking either $q\rightarrow \infty$ before $\bar q \rightarrow \infty$ or the opposite, and (2) a symmetrical treatment of $w=\sqrt{q\bar q} \rightarrow \infty$ before taking $\sigma =(q+\bar q)/2\sqrt{ q\bar q}\rightarrow \infty$. We will follow the asymmetrical approach here and postpone the symmetrical treatment to Sec. \ref{sec:symmetric}.

In an asymmetrical limit, the  boundary condition for MCB, Eq. (\ref{eq:CB}) (b),  becomes
\be
G^{(M)}_{(\Delta,\ell)} (u,v) \sim q_<^{(1-\lambda_+)} q_>^{\lambda_-} \label{eq:Mbdry-2}
 \ee
where $q_<$ is the smaller of the pair $(q,\bar q)$ and $q_>$ the other,  as   both  $q,\bar q \rightarrow \infty$. 
To proceed, we will first treat the case of $d=2$ and $d=4$ before commenting on   general $d$. 

\begin{itemize}
\item{}
{\bf $d=2$:} Here $\epsilon=0$ and the differential equations for $q$ and $\bar q$ decouple. From Eq. (\ref{eq:indicial2}),   we choose
\be
D_0(q) \, g(q)   = \lambda(\lambda-1)\, g(q)\, ,  \label{eq:fx}
\ee
with $\lambda =\lambda_\pm$, and similarly for $D_0(\bar q)$.   Each can be reduced to standard  hypergeometric      differential equations: the solution can be expressed as, for $1<q<\infty$,
 \be
g(q) =a\,  \widetilde k_{2\lambda} (q)  + b\, \widetilde k_{2(1-\lambda)} (q)    \label{eq:q-solution}
\ee
where 
\bea
\widetilde k_{2\lambda} (q)& =&  q^{-\lambda}\,  _2F_1(\lambda/2+1/2,\lambda/2; \lambda +3/2; q^{-2})\nn
&=& 2^{\lambda}\, \frac{\Gamma(\lambda+1/2)}{\pi^{1/2}\Gamma(\lambda)} \, Q_{\lambda-1}(q) \, .\label{eq:q-function}
\eea
In the second line,  $Q_\nu(q)$ is simply the standard Legendre functions of  the second kind. Our normalization corresponds to $\widetilde k_{2\lambda} (q)\simeq q^{-\lambda}$ for $q \rightarrow \infty$.

It remains to properly implement the desired Minkowski boundary conditions, Eq. (\ref{eq:Mbdry-2}).  If $1<q <\bar q \rightarrow \infty$, choose $a=0$ for $g(q_<)$ and $b=0$ for $g(\bar q_>)$. If $1<\bar q<q\rightarrow \infty$ make the reverse choice. This leads to 
\begin{align}
&G^{(M)}_{(\Delta,\ell)} (u,v)
=  \widetilde k_{2(1-\lambda_+)} (q_<) \widetilde k_{2\lambda_-} (q_>) \nn
&=  \frac{\Gamma(3/2-\lambda_+)\Gamma(\lambda_-+1/2)}{2^{\ell-1}  \Gamma(1-\lambda_+)\Gamma(\lambda_-)}  \nonumber \\
&\qquad \qquad \times   Q_{-\lambda_+}(q_<) Q_{\lambda_--1}(q_>)  \, ,  \label{eq:d2MCBq}
\end{align}
in agreement with Eq. (\ref{eq:Mbdry-2}). 
\item{}
{\bf $d=4$:} It is possible to re-cast this case into one where $q$ and $\bar q$ equations again decouple as done in  \cite{Dolan:2011dv} working with $(x,\bar x)$.  The corresponding solution is
\begin{align}
G^{(M)}_{(\Delta,\ell)} (u,v) &={\rm sgn}(\bar q-q ) \Big(\frac{1}{q_>-q_<} \Big) \,  \widetilde k_{2(1-\lambda_+)} ( q_<) \nonumber \\
&\qquad \times\,\widetilde k_{2(\lambda_--1)} ( q_>)  \nn
&= 2^{2-\ell} \frac{\Gamma(3/2-\lambda_+)\Gamma(\lambda_- -1/2)}{\Gamma(1-\lambda_+)\Gamma(\lambda_--1)}\nonumber \\
&\quad \times {\rm sgn}(\bar q-q ) \Big(\frac{1}{q_>-q_<} \Big) \nn
&\quad \times   Q_{-\lambda_+}(q_<)\, Q_{\lambda_--2}(q_>)  \,. 
\label{eq:d4MCBq}
\end{align}
One can  verify that Eq. (\ref{eq:Mbdry-2}) is again satisfied.

\item{}{\bf General $d$: } The $d=1$ case is of special interest for SYK-like models, and will be discussed separately in Sec. \ref{sec:symmetric-d1}. For general $d$,  it is not possible to reduce the solution to a combination of simple products of hypergeometric functions.  However, an iterative procedure will be discussed in Sec. \ref{sec:symmetric} by treating $(q,\bar q)$ symmetrically. One finds that the general structure for the leading behavior in $u\rightarrow 0$ is shared  for all $d$.

We stress here that the physical region of s-channel scattering corresponds to $1<q,\bar q<\infty$, where $\widetilde k_{2\lambda}(q)$, Eq. (\ref{eq:q-function}), is real.   It follows that  $G^{(M)}_{(\Delta,\ell)} (u,v)$ is also real. However, since $\widetilde k_{2\lambda}(q)$ is defined as a real analytic function, it can be continued into the complex plane and acquire a phase for general $\ell$.  We will turn  to this question in Sec. \ref{sec:application-I},  as well as in Appendix \ref{sec:MCFB-append}.
\end{itemize}

The boundary conditions for the Euclidean conformal blocks, Eq. (\ref{eq:CB}) (a), can similarly be expressed as $G^{(E)}_{(\Delta,\ell)} (q,\bar q)  \sim q_<^{\lambda_+} q_>^{\lambda_-}$. The same approach as above then leads to \footnote{The connection to the canonically defined functions in terms of $\{x,\bar{x}\}$ is shown in Appendix \ref{sec:MCFB-append}}
\begin{equation}\label{eq:d2ECB}
G^{(E)}_{(\Delta,\ell)} (u,v) = \widetilde k_{2\lambda_+}(q) \widetilde  k_{2\lambda_-}(\bar q)+\widetilde  k_{2\lambda_+}(\bar q) \widetilde  k_{2\lambda_-}(q)\, ,
\end{equation}
for the $d=2$ case, and to
\begin{align}\label{eq:d4ECB}
&G^{(E)}_{(\Delta,\ell)} (u,v)  = \frac{1}{q-\bar q} \Big( \widetilde  k_{2\lambda_+}(q) \widetilde  k_{2(\lambda_--1)}(\bar q) \nn
&\qquad\qquad\qquad - \widetilde  k_{2\lambda_+}(\bar q) \widetilde  k_{2(\lambda_-- 1)}(q) \Big),
\end{align}
for the $d=4$ case.

In an Euclidean treatment, $q$ and $\bar q$ are complex conjugates, $q^*=\bar q$, thus not independent. 
It has been suggested that Minkowski conformal blocks are simply the analytic continuation of Euclidean conformal blocks, which changes the boundary conditions from Eq. (\ref{eq:CB}) (a) to (b). Appendix \ref{sec:MCFB-append} gives the detailed relation between $G^{(M)}_{(\Delta,\ell)} (u,v) $ and $G^{(E)}_{(\Delta,\ell)} (u,v)$ in terms of their analytic structure.  We will demonstrate explicitly that Eqs. (\ref{eq:d2MCBq}) and (\ref{eq:d4MCBq}) are not given by a direct analytic continuation of the corresponding ECB, Eqs. (\ref{eq:d2ECB}) and (\ref{eq:d4ECB}). 

To state it more succinctly, in performing analytic continuation from Euclidean to Minkowski limit, the Euclidean boundary condition  does not transform precisely into that for Minkowski limit. For instance, for $d=2$, starting with $G^{(M)}_{(\Delta,\ell)} (u,v)$ and following normal path of continuation,  one arrives in the Euclidean region where
\begin{align}
&G^{(M,continued)}_{(\Delta,\ell)} (u,v) \nn
&= c\, \widetilde  k_{2\lambda_+} (q_{<}) \widetilde  k_{2\lambda_-} (q_>) + d\, \widetilde  k_{2(1-\lambda_+)} (q_{<}) \widetilde  k_{2\lambda_-} (q_>)
\end{align}
  where $c= i \sqrt \pi \frac{\Gamma(1/2-\lambda_+)}{\Gamma(1-\lambda_+)^2}$ and $d=(-1)^{1-\lambda_+}/\cos\pi\lambda_+$. It does not lead to $G^{(E)}(u,v)$.
Conversely, an additional   prescription  is required in relating ECB to   the desired MCB.  This will be touched upon further in Sec. \ref{sec:SW}. 
    
Let us end  by examining  the constraints on $F^{(M)}(u.v)$ due to crossing symmetry. As mentioned earlier, for a t-channel OPE, we are interesting  in s-u crossing, which corresponds to interchanging either $1\leftrightarrow 2$ or  $4\leftrightarrow 3$. This leads to, as already discussed in Sec. \ref{sec:Crossing}, $u\rightarrow u/v$ and $v\rightarrow 1/v$, or,
\begin{align}
(q,\bar q) \quad &\Leftrightarrow \quad (-q,-\bar q), \quad {\rm and} \nn
 (w,\sigma) \quad &\Leftrightarrow \quad (-w,\sigma)
\end{align}
Consider the OPE expansion (\ref{eq:MOPE}), with $\ell$ integer. From Eq. (\ref{eq:q-function}), under s-u crossing,
\be
G^{(M)}_{(\Delta,\ell)} (u,v) = (-1)^{1-\ell} G^{(M)}_{(\Delta,\ell)} (u/v,1/v) \, .\label{eq:su-crossing2}
\ee
As we show more explicitly in the next section, this pattern holds for all $d$.

\subsection{Symmetric Treatment}\label{sec:symmetric}

We next turn to a  symmetric construction of MCB for general $d$.  This was first advocated for ECB in \cite{Dolan:2011dv}, starting with  $(x,\bar x)$, by shifting to new variables $u=x\, \bar x$ and $\sigma_0= (\sqrt {x/\bar x}+\sqrt {\bar x/x})/2$. This approach was pursued further in \cite{Hogervorst:2013sma}. As explained in \cite{Dolan:2011dv}, this approach has the advantage of being able to extend the boundary condition, for ECB, Eq. (\ref{eq:CB}) (a), to the region where $x=O(\bar x)$, and the desired boundary condition translates into that in the limit $\sigma_0\rightarrow \infty$.  

We will begin  with $(q,\bar q)$ and focus on the conformal blocks in the Minkowski limit.
Recall that, for the DLC limit, we are interested in first taking the $w=\sqrt {q\bar q} \simeq  \sqrt u^{-1} \rightarrow \infty $ limit, with the resulting  boundary condition specified at $\sigma = (\sqrt{q/\bar q}  + \sqrt{\bar q/q})/2 \simeq \sigma_0 \rightarrow \infty$.  In what follows, we shall  adopt $(w,\sigma)$ as two independent variables with the physical region specified by $1<w<\infty$ and $1<\sigma<\infty$.  \footnote{As we shall demonstrate below, for holographic CFTs, this allows a simpler representation connecting in the leading order to the Euclidean $AdS_{d-1}$ bulk-to-bulk propagator.} We will mainly work this asymptotic limiy, but show a higher order expansion can be obtained formally on an equal-footing for all d.

The differential operator ${\cal D}$, Eq. (\ref{eq:D}), when expressed in terms of $(w,\sigma)$, becomes a sum of three terms, ${\cal D}=({\cal L}_{0,w}+{\cal L}_{0,\sigma} +w^{-2}{\cal L}_2)/2$ where ${\cal L}_{0,w}(w\partial_w)$,  ${\cal L}_{0,\sigma}(\partial_\sigma,\sigma)$ and ${\cal L}_2(w\partial_w,\partial_\sigma, \sigma)$ are homogeneous in $w$~\footnote{See \cite{Hogervorst:2013sma} for a related treatment in ``radial quantization".}, 
\begin{align}
{\cal L}_{0,w}(w\partial_w)&= ( w\partial_w)^2+d \,w\partial_w,  \label{eq:leadingL0}\\
{\cal L}_{0,\sigma}(\partial_\sigma,\sigma)&=  (\sigma^2-1)\partial_\sigma^2 + (d-1)\sigma \partial_\sigma , \label{eq:Legendre}
\end{align}
and
\begin{align}  \label{eq:L2}
{\cal L}_2(w&\partial_w, \partial_\sigma,\sigma)=-(2\sigma^2-1)(w \partial_{w})^2 \nn
& + (4(\sigma^2-1)+d)w \partial_{w} +4\sigma(\sigma^2-1)w \partial_{w}\partial_{\sigma} \nn
& - (2\sigma^2-1)(\sigma^2-1)\partial^2_{\sigma} -(6(\sigma^2-1) \nn
&+ (d-1)) \sigma\partial_{\sigma}\, .
\end{align}
It thus suggests the following expansion for conformal blocks, 
\begin{align}
G^{(M)}_{(\Delta,\ell)}(u,v) =& w^{s}\Big( g_0(\sigma) + w^{-2} g_1(\sigma) \nn
&\qquad + w^{-4} g_2(\sigma) +\cdots\Big)\nn
=&\sum_{n=0}^\infty  w^{s-2n} g_n(\sigma).   \label{eq:symmetric}
\end{align}
Here,  each $g_n(\sigma)$ also depends on $\ell$, $\Delta$ and $d$, which will be exhibited explicitly when necessary for clarity. They can be found recursively, with the leading order term satisfying a relatively simple D.E.,
  \be
  {\cal L}_{0,\sigma } \, g_0(\sigma) =(2 C(\ell,\Delta)- s(s+d) )\, g_0(\sigma)\, . \label{eq:symmetric-leadingorder}
  \ee
This differential operator, $ {\cal L}_{0,\sigma }$, is of the same form as $D_0$, (\ref{eq:LegendreDO}), and it will appear repeatedly under several contexts, with solutions expressible in terms of hypergeometric functions. (However, each case may impose a different boundary condition.  See Sec. \ref{sec:symmetric-higherorder} and Appendix \ref{sec:hypergeometric} for further discussion.)   We will treat solutions to (\ref{eq:symmetric-leadingorder}) with appropriate  boundary conditions in Sec. \ref{sec:symmetric-leadingorder}.

  Before proceeding to solving for $g_0(\sigma)$, we mention first  the ease of exhibiting  crossing under this symmetric approach.  As we shall demonstrate below, for Minkowski conformal blocks, this corresponds to  the choice $s=\ell-1$. Under  s-u crossing, $w\rightarrow -w$ and $\sigma\rightarrow \sigma$,  (\ref{eq:symmetric}) leads to
\be
G^{(M)}_{\Delta, \ell}(-w,\sigma)  = (-1)^{\ell-1} G^{(M)}_{\Delta, \ell}(w,\sigma) \, . \label{eq:su-crossing3}
\ee
As expected, it is odd for $\ell$ even and  even for $\ell$ odd.

\subsubsection{Leading Order:}\label{sec:symmetric-leadingorder}

In implementing seperation of variables, Eq. (\ref{eq:indicial1}) suggests separating  $\Delta$ and $\ell$ dependences. A direct indicial analysis at $w\rightarrow \infty$ leads to several degenerate possibilities, with $s$ taking on (a) $s_a=-\Delta$, (b) $s_b=\ell-1$, (c) $s_c=\Delta-d$, and (d) $s_d=1-d-\ell$. The corresponding solutions to Eq. (\ref{eq:symmetric-leadingorder}),  labelled by $g_{0a}$, $g_{0b}$, $g_{0c}$ and $g_{0d}$ respectively, are 
\begin{align}
\begin{split} \label{eq:EMf}
(a)\,\,& \Big({\cal L}_{0,\sigma } -\ell(\ell + d-2)\Big) \,  g_{0a}(\sigma)=0\, \\
(b)\,\,&\Big({\cal L}_{0,\sigma } \, - (\Delta-1) (\Delta - d + 1)  \Big)\,  g_{0b}(\sigma)= 0\, , \\
(c)\,\,&\Big({\cal L}_{0,\sigma }\, -\ell(\ell + d-2) \Big)\, g_{0c}(\sigma)= 0\, , \\
(d)\,\,&\Big({\cal L}_{0,\sigma } \, - (\Delta-1) (\Delta - d + 1)\,\Big)\, g_{0d}(\sigma)=0 \, .
\end{split}
\end{align}
Note that indicial condition $(c)$ is conjugate of  $(a)$, under $\Delta \leftrightarrow d - \Delta$,  ($\widetilde \Delta\leftrightarrow -\widetilde \Delta$), leading to identical DE in $\sigma$, thus corresponding to the respective shadow blocks. Similarly, $(d)$ is conjugate of $(b)$  under $\ell \leftrightarrow -\ell -(d-2)$, ($\widetilde \ell \leftrightarrow -\widetilde \ell$). 
We will therefore concentrate on case (a) and (b). Of these two solutions, by switching back to $u$ and $v$ and comparing to Eq. (\ref{eq:CB}), we find that case (a) is appropriate for Euclidean conformal blocks, and case (b) is  appropriate for the Minkowski limit.

Observe  that these differential equations  are even under 
$\sigma \leftrightarrow - \sigma$. General solutions to these equations can be expressed in terms of hypergeometric functions and we need to impose the respective boundary conditions, Eq. (\ref{eq:CB})~\footnote{For example, for   $g_{0a}(\sigma)$, 
$
g_{0a}(\sigma)  \simeq  a \sigma^{\ell} (1+ O(\sigma^{-2}) )+ b \sigma^{1-\ell} (1+ O(\sigma^{-2}))
$,
as $\sigma\rightarrow \infty$.  For $g_{0b}(\sigma)$, one replaces  $\ell$ with $\Delta$.  
 For case (a), $s_a=-\Delta$,  the solution is that appropriate for Euclidean conformal blocks, with coefficient $b=0$. This leads to polynomial solutions for integral $\ell$, e.g., for $d=4$, Gegenbauer polynomials. As $\sigma\rightarrow \infty$, 
$
G^{(E)}_{(\Delta,\ell)}(u.v)\simeq w^{-\Delta} g_{0a}(\sigma) \sim w^{-\Delta} \,\sigma^\ell
$.
 }.   Let us focus here on Eq. (\ref{eq:EMf}) (b). 
With $s_b=\ell-1$,  the solution   is
\begin{align}
g&_{0b}(\sigma; \Delta,d)\nn
&= \sigma^{ 1-\Delta}\, _2F_1(\frac{\Delta-1}{2}, \frac{\Delta}{2}; \Delta-\frac{d}{2} +1; \sigma^{-2}) \, ,
\label{eq:symmetric-1}
\end{align}
with $g_{0b}$ real for $\sigma>1$.  For all $d$, at $\sigma$ large, $g_{0b}(\sigma; \Delta,d)\sim \sigma^{ 1-\Delta}$, corresponding to 
\be
G^{(M)}_{(\Delta,\ell)}(u.v)\simeq w^{\ell-1} g_{0b}(\sigma)\, \sim w^{\ell-1} \, \sigma^{1-\Delta},
\ee
in the limit $w\rightarrow \infty$ and $\sigma\rightarrow \infty$, as promised. Solutions for  $d=4$ and $d=2$ can be expressed simply as 
\begin{align}
g_{0b}(\sigma; \Delta,4) &= \frac{e^{-(\Delta-2)\xi}}{\sinh\xi}\, ,\quad {\rm and} \nn
g_{0b}(\sigma; \Delta, 2)& =  e^{-(\Delta-1)  \xi}\, .   \label{eq:adseven}
\end{align}
Similarly, for $d=3$ and $d=1$, the solutions also simplify to
 \begin{align}
g_{0b}(\sigma; \Delta, 3) &= Q_{\Delta-2} ( \sigma)\, ,  \quad {\rm and} \nn
g_{0b} (\sigma; \Delta, 1)&= \sinh \sigma  \,Q^{(-1)}_{\Delta-1} ( \sigma) \nn
&= \frac{d Q_{\Delta-1}(\sigma)}{d\xi}. \label{eq:adsodd}
\end{align}
Here $Q_{\nu} ( \sigma)$ is the Legendre function of the second kind. The case of $d=1$ is discussed in more detail in Sec. \ref{sec:symmetric-d1}, as it relates to SYK-like models.  

Since $d$ enters into the differential equation as a simple parameter, the following relation holds:
 \be
\frac{d}{d\sigma} g_{0b} (\sigma; \Delta-1,  d-2)=  -(\Delta-2)\,  g_{0b}(\sigma; \Delta, d) \, , \label{eq:raising}
\ee
or, equivalently,
\be
 g_{0b} (\sigma; \Delta-1,  d-2)=  (\Delta-2)\,  \int\limits_{\sigma}^\infty d \sigma \, g_{0b}(\sigma; \Delta, d) \,. \label{eq:lowering}
\ee
These relations hold for general $d$,  thus allowing one to find $g_{0b}(\sigma; \Delta, d)$ for even and odd integral $d$ iteratively given $g_{0b}(\sigma; \Delta, 2)$ and $g_{0b}(\sigma; \Delta, 3)$.  

We also point out that  Eq. (\ref{eq:symmetric-1}) corresponds precisely to an Euclidean bulk-to-bulk  scalar propagator in $AdS_{d-1}$, or more precisely $H_{d-1}$, with conformal dimension $\Delta-1$.  Here Eq. (\ref{eq:symmetric-1})  is derived purely from a CFT perspective, with $\sigma -1 $ plays the role of a chordal distance or, in an alternative  mathematical  usage, $\xi=\cosh^{-1} \sigma$ is the geodesic in $H_{d-1}$. (See also Appendix \ref{sec:holography}.) This  connection is meaningful physically only for $d\geq 2$ but can be extended formally to all $d$.  As such,  $g_{0}(\sigma; \Delta,d)$ is singular at $\sigma=1$, consistent with with our procedure of fixing a vanishing boundary condition at $\sigma=\infty$. 

Lastly, with  $\sigma  = \cosh \xi$,   Eq. (\ref{eq:EMf}) (b) can also be expressed as
\begin{align}
 \Big(&\partial^2_\xi  + (d-2) \coth \xi \partial_\xi \nn
 &- (\Delta-1) (\Delta - d + 1) \Big ) g_{0b}(\cosh \xi)   =0.
\end{align}
Introducing a reduced function $f(\xi)$, with 
$  g_{0b}(\cosh \xi) = (\sinh\xi)^{-\frac{d-2}{2}} f(\xi) $,
one has
\be
\Big( \partial_\xi ^2 - (\Delta-d/2)^2 + \frac{ (d-2)(d-4)}{4 \sinh^2\xi} \Big ) f(\xi) = 0.\label{eq:keyDE}
\ee
This simplifies for $d=2$ and $d=4$, leading to 
\be
f(\xi) = e^{-|\Delta-d/2| \, \xi}.
\ee 
This agrees with the desired result given above, and also serves as the leading term in an higher order expansion, (\ref{eq:higherorder2}) and (\ref{eq:higherorder4}) respectively.  For general $d$, the solution can be expressed in terms of associated Legendre functions.

\subsubsection{Higher Order Expansion:}\label{sec:symmetric-higherorder} 
As mentioned in the introductory remarks for this section, higher order expansion in (\ref{eq:symmetric}) can be found iteratively. Once the leading index, $s_b=\ell-1$, is identified and the corresponding solution $g_{0b}(\sigma)$ is found, it is possible to solve each expansion function $g_n(\sigma)$ for $1\leq n$.  With  (\ref{eq:leadingL0}) and (\ref{eq:L2}), one finds
\be
[{\cal L}_{0,\sigma} - m^2(\ell,\Delta)]  g_{n}(\sigma) = J_n(\sigma)\label{eq:recursion}
\ee
with $m^2(\ell,\Delta) =(\Delta-1) (\Delta - d + 1)  - 2 n (2n -d-2\ell +2).$ The source for $g_n$ is $J_n(\sigma)= -{\cal L}_2(\ell+1-2n,\partial_\sigma,\sigma) g_{n-1}(\sigma)$. I.e., given in terms of $g_{n-1}(\sigma)$. Therefore, one can proceed iteratively.   

Focusing on $1<\sigma<\infty$, Eq. (\ref{eq:recursion}) can be solved formally via a standard Green's function procedure with
\be
g_{n}(\sigma) = \int\limits_1^\infty d\sigma' G_0(\sigma, \sigma') J_n(\sigma')\,
\ee
where
\be
[{\cal L}_{0,\sigma} - m^2(\ell, \Delta)] G_0(\sigma, \sigma')=\delta(\sigma-\sigma')\,.
\ee
For $d=2$ and $d=4$, since explicit solutions are already known. We will instead demonstrate that they can be re-expressed in the symmetric form, Eq. (\ref{eq:symmetric}), in Appendix \ref{sec:MCFB-append}.  For general d,  care must be exercised in defining appropriate boundary conditions.  We will leave the case of general d to a future study, and will focus next on the case of $d=1$.

\subsubsection{The Case  of \texorpdfstring{$d=1$}{d=1}:}\label{sec:symmetric-d1} 

Due to the existence of a kinematic constraint, the leading order solution for $d=1$ requires additional examination. From the perspective of Euclidean $SO(d+1,1)$ invariance, the physical region is bounded by 
$(1-v+u) \leq 2\sqrt u$, with the equality holding at the kinematical boundary~\cite{Dolan:2011dv}. This boundary also defines the $d=1$ limit, which can be expressed more usefully as $\sqrt v =1- \sqrt u$. For Lorentzian vectors, a similar analysis yields $(1-v+u) \geq 2\sqrt u$, again leading to a kinematical relation $\sqrt v = 1- \sqrt u $ as  a  constraint for $d=1$. (See Appendix \ref{sec:d1}.)  In terms of $q$ and $\bar q$, both cases lead to  $q=\bar q$ and $ \sigma = \cosh \xi=1$. It follows that there is only one independent variable, instead of two. Therefore for $d=1$, conformal blocks, as functions of $w$ and $\sigma$, must satisfy a constraint: $\partial_\sigma G(w,\sigma)=0$. This means that care must be taken when re-interpreting the above results for the case of $d=1$.  

There are two possibilities.
\begin{itemize}
 \item{(a)} The Euclidean option corresponds to keeping only $\ell =0,1$, and going to higher order expansion in $w$, 
\be
G^{(E)}_{\Delta,\ell=0}(w,\sigma=1) = w^{-\Delta} \sum^\infty_{n=0} g_{a,n}(\sigma=1) w^{-2n}.\label{eq:Esum}
\ee
It is easily shown that, with $d=1$,  $g_{a/c,0}(\sigma)$ being a constant is a consistent solution to (\ref{eq:EMf}) for  $\ell=0,1$.
\item{(b)} The Minkowski option corresponds to having  $\Delta =0,1$, and
\be
G^{(M)}_{\Delta=0,\ell}(w,\sigma=1) =w^{\ell-1}  \sum^\infty_{n=0} g_{b,n}(\sigma=1) w^{-2n}. \label{eq:Msum}
\ee
It is also easily shown that $g_{b/d,0}(\sigma)$ being a constant is a consistent solution to (\ref{eq:EMf}), with $\Delta=0,1$. In Sec. \ref{sec:CFT-1} we show that this limit is  more appropriate in treating scattering for SYK-like 1-d models. 
\end{itemize}

For both cases it is necessary to go beyond the leading order to obtain the proper sums, Eqs. (\ref{eq:Esum}) and (\ref{eq:Msum}).  In terms of cross ratio, $w$, the residual symmetry for both cases is $O(1,1)$, with dilation for case (a) and Lorentz boost for case (b).
  
Let us treat case (b) first. We denote $G^{(M)}_{\Delta=0,\ell}(w,\sigma=1)$ simply as $G^{(M)}_\ell(w)$ and consider the limit $d=1$.  The series (\ref{eq:Msum}) can be obtained by working with   an ODE  
$
 {\cal D}_w G^{(M)}_\ell(w) =\ell(\ell-1) G^{(M)}_\ell(w)
$,
 where ${\cal D}_w$ is obtained from ${\cal D}(d=1)$ by acting on functions of $w$ only.   A bit of algebra then leads to 
\be
  {\cal D}_w =  (w^2-1) \frac{d^2}{dw^2} +2 w\frac{d}{dw},  \label{eq:Legendre2}
\ee
which is of identical form as Eq. (\ref{eq:Legendre}), with $d=3$, and also with Eq. (\ref{eq:LegendreDO}), i.e.,  it is again that for Legendre functions. It follows from Eq. (\ref{eq:q-solution}), for $w\rightarrow \infty$, there are in general two independent solutions, leading to
\be
G^{(M)}_\ell(w) = a\,  \widetilde k_{2\ell} (w)  + b\, \widetilde k_{2(1-\ell)} (w) \label{eq:1dSYK}
\ee
where $k_{2\ell} (w)$ is again given by Legendre function of second kind, $Q_{\ell}(w)$, Eq. (\ref{eq:q-function}).

 In Secs. \ref{sec:SYK1},  we consider Minkowski scattering, with  the physical region arranged   to lie in the region $1\leq w\leq \infty$. Therefore, our choice corresponds to $a=1$ and $b=0$. The normalization corresponds to having  conformal blocks
\begin{align}
G^{(M)}_\ell(w)&=2^{1-\ell} \frac{\Gamma(3/2-\ell)}{\pi^{1/2}\Gamma(1-\ell) } \, Q_{-\ell}(w) \nn
&\equiv c_\ell \,  Q_{-\ell}(w) \equiv \bar  Q_{-\ell}(w)  \, ,\label{eq:1dMCB}
\end{align}
so that $G^{(M)}_\ell(w)\simeq w^{\ell-1}$, with unit coefficient as $w\rightarrow \infty$. We shall also restrict  ${\rm Re} \,\widetilde \ell>0$, (thus ${\rm Re} \,\ell>1/2$).  From the identity
\be
\frac{\pi\, P_{\ell}(z)}{ \tan{\ell \pi}}= Q_{\ell}(z) -Q_{-\ell-1}(z),  \label{eq:PQidentity}
\ee
one has   $P_{\ell}(z)=P_{-\ell-1}(z)$, and   $Q_{-\ell}$ has poles at non-negative integers, $\ell=0,1,\cdots$.  However, since $c_\ell\sim 1/\Gamma(1-\ell)$,  it follows $ \bar  Q_{-\ell}(w)$ is analytic  for ${\rm Re} \,\ell>1/2$.
For  positive integral values, $\ell=n$, $n=1,2,\cdots$,
\be
G^{(M)}_n(w)= d_n\, P_{n-1}(w)   \label{eq:P-function} 
\ee
where  $d_n=  \pi^{1/2}  2^{1-n} \frac{\Gamma(n) }{\Gamma(n-1/2)}$.   

Let us briefly return to case (a), appropriate for an Euclidean treatment. As pointed out in~\cite{Dolan:2011dv}, the solution to the $\ell=0$ problem has a direct solution. Following the same approach as above, one finds
$
 {\cal D}_w G^{(E)}_\Delta(w) =(\widetilde \Delta^2 - 1/4) G^{(E)}_\Delta(w)=\Delta (\Delta-1) G^{(E)}_\Delta(w)
$,
leading to the same result as given in \cite{Dolan:2011dv},  with  $w$ replaced by $x=2/(w+1)$. The solution can also be obtained from that for the case (b), with $\Delta$ replacing $\ell$.

\section{Minkowski OPE and Scattering}\label{sec:application-I}

Let us now return to discuss how OPE in a Minkowski setting, Eqs. (\ref{eq:newgroupexpansion}) and (\ref{eq:newgroupexpansionDisc}), can be applied to high energy scattering. In this Section, we  focus on certain formal steps necessary before Minkowski OPE can be applied. The emphasis will be on first developing a Mellin-like representation for the OPE sum so that it applies to the physical scattering region~\footnote{We emphasize that the discussion here focuses on Mellin amplitudes that are distinct from the Mellin representation discussed in \cite{Mack:2009mi,Paulos:2011ie,Nandan:2011wc,Rastelli:2016nze,Fitzpatrick:2011ia}.}. This formulation stresses the importance of a spectral curve, $\Delta(\ell)$, and its relation to effective spin, $\ell^*$.  An equally important and related issue discussed is the relation between the t-channel OPE in a Minkowski setting to the principal series for an unitary irreducible representation~\footnote{The case of $d=1$ provides an  explicit illustration. This will be carried out in Appendix \ref{sec:green}. A more detailed discussion for $d\geq 2$ will be reported separately.}  of non-compact $O(4,2)$, leading to Eq. (\ref{eq:trueinvariance}). Applications of  dimensional reductions are discussed in Appendix \ref{sec:application-II} for $d=2$ and the case of DIS, and in Sec. \ref{sec:CFT-1} for $d=1$ scattering for SYK-like models.

\subsection{Kinematics}

Before applying (\ref{eq:MOPE}) to  high energy near-forward scattering, it is important to address the issue of the phase of $F^{(M)}(u,v)$.  As pointed out in Sec. \ref{sec:intro}, MCB, $G^{(M)}$ are  real valued functions over the physical region for s-channel scattering where $1<w<\infty$ and $1<\sigma< \infty$.  It follows that  the contribution to (\ref{eq:MOPE}) from each conformal primary is also real.  As a scattering amplitude, however, $F^{(M)}(u,v)$ is in general complex.  A complex phase emerges as a consequence of  re-summation. This can be carried via complex-$\ell$ utilizing the technique of Sommerfeld-Watson transform, which we turn to next.  Through this procedure, one also allows an natural continuation in $w$ from the s-channel where $1<w<\infty$ to the u-channel phyiscal region where $-\infty<w<-1$. it is therefore useful to have a closer examination of the symmetry involved under s-u crossing.

In the DLC limit,   Feynman amplitudes and CFT correlators have opposite parity under crossing. A crossing even Feynman amplitude, $T$, corresponds to a CFT invariant function $F^{(M)}$ which odd under s-u exchange. This is due to the presence of the extra factor of $w$ in T, for example, Eq. (\ref{eq:T-chi}). A general amplitude can have both components. A 4-point function for identical conformal primaries is s-u crossing symmetric. This corresponds to an even $T$, and thus $F^{(M)}(u,v)$ will be crossing odd: $F^{(M)}(u,v) = - F^{(M)}(u/v,1/v)$. It follows that if $F^{(M)}(u.v)$ is anti-symmetric under $s-u$ exchange, then, from Eq. (\ref{eq:su-crossing2}), only even $\ell$  contribute. Conversely, for $F^{(M)}(u,v)$ symmetric under $s-u$ exchange, only odd $\ell$  contribute.

In this study, we are mostly dealing with scattering amplitudes $T$ which are crossing even. If we are more explicit in identifying s-channel and u-channel amplitudes separately by $F^{(M)}_s$ and $F^{(M)}_u$, we have
$
 F_s^{(M)}(u,v) = - F_s^{(M)}(u/v,1/v)= F_u^{(M)}(u',v')
$,
with $(u',v')$ identified with $(u/v,1/v)$. In what follows, we will always work with s-channel amplitudes,  $F_s^{(M)}$, while dropping the subscript. When expressed in terms of variables $(w,\sigma)$ and analytically continued, one has
\be
F^{(M)}(w,\sigma) = - F^{(M)}(-w,\sigma).  \label{eq:crossingodd-2}
\ee

\subsection{Sommerfeld-Watson Transform}\label{sec:SW}
We begin by first re-grouping the OPE in Eq. (\ref{eq:MOPE}) as
\be
F(w,\sigma) = \sum_{\alpha}  \sum_\ell \,   a^{(12),(34)}_{\ell, \alpha} \, G(w,\sigma; \ell, \Delta_{\ell, \alpha})\, .
\label{eq:ConformalBlock2}
\ee
We  have re-expressed the partial-wave amplitude as $a_{\ell, \alpha}^{(12),(34)}$, switched the dependence on cross ratios from $(u,v)$ to $(w,\sigma)$ and have also re-grouped the sum in a form which allows a re-summation, leading to a representation for $F(w,\sigma)$ valid  for $w\rightarrow \infty$.  

Consider the case where scattering amplitude is even under  crossing~\footnote{In general, the sum can  be separated into a sum of even spins or another over odd spins. The case of crossing odd has been treated in~\cite{Brower:2014wha}.}, thus $F(w,\sigma)$ odd in $w$.  From (\ref{eq:su-crossing3}), the sum is over even $\ell$  only. Sommerfeld-Watson transform corresponds to turning this discrete sum over $\ell$ into an integral over complex-$\ell$ plane and then opening up the contour into an integral along a vertical line. This vertical line is chosen initially for convergence, leading to  a Mellin-like integral, i.e., 
\be
 \sum_{\ell=2n} \rightarrow \sum_{\ell =2n< L_0} - \int\limits_{L_0 -i\infty}^{L_0 + i\infty}  \frac {d\ell}{2 i} \frac {1- e^{i \pi(1- \ell )} }{\sin\pi \ell }   \, .\label{eq:SW+}
\ee
We assume that  $F(w,\sigma)$ is polynomially bounded at $w=\infty$, i.e., $ | F(w,\sigma)|<O(w^{N})$, thus  $N+1<L_0$.  This allows one to represent $F^{(M)}(w,\sigma)$ by
\begin{align}
&F^{(M)}(w,\sigma)=  F_0^{(M)}(u,v)+\int\limits_{L_0-i\infty}^{L_0+i\infty}  \frac{d \ell}{2 i}  \nn    
&\times \frac{-(1 +  e^{-i \pi \ell}) }{\sin\pi \ell }     \sum_\alpha    a(\ell, \Delta_\alpha(\ell))\,\,  {G} (\ell,\Delta_\alpha(\ell) ; w,\sigma), \label{eq:newgroupexpansion}
\end{align}
The assumption of polynomial boundedness can also translate into having $a_{\ell, \alpha}^{(12),(34)}$ analytic in $\ell$ for $N+1<\ell$. 
The residue   $F^{(M)}_0(w,\sigma)$ represents the original finite sum with $\ell<L_0$.

In Eq. (\ref{eq:ConformalBlock2}), we have also separated the sum into a sum over families conformal primaries. This was described in the introduction in discussing anomalous dimensions for the leading twist conformal primaries. Conformal dimensions for each family can be interpolated by their spins continuously by $\Delta(\ell)$, i.e., leading to a spectral curve. There will be many families and each family is labeled by an index $\alpha$~\footnote{In the absence of interactions, the label $\alpha$ is simply the twist, $\tau_0\equiv \Delta-\ell$, and with additional index for other  families of conformal primaries of the same twist.  For simplicity, we shall keep in what follows only one family for each twist. }. The separation in Eq. (\ref{eq:newgroupexpansion}) into two terms is at first necessary  due to possible existence of singularities for ${\rm Re} \ell<L_0$.  It is nevertheless interesting to note, since $F_0(w,\sigma)$ is real, one always has
 \begin{align}
{\rm Im}\, F(w,\sigma) =&   \sum_\alpha \int\limits_{L_0 -i\infty}^{L_0 + i\infty}  \frac {d\ell}{2 i} a^{(12),(34)}(\ell, \Delta_\alpha(\ell)) \nn&
\times G(w,\sigma; \ell,\Delta_\alpha(\ell))  \, .\label{eq:ImConformalBlock.c}
\end{align}
with $ L_0$ sufficiently large.

A brief discussion on various formal  assumptions necessary in carrying the above analysis is in order.  Here we summarize a few key points:\\
(a) There exists a unique analytic continuation away from integral values for $\ell$ while satisfying the constraint of ``Carlson's theorem"~\cite{Chew,Squires}, with  $a_\alpha^{(12),(34)}(\ell)$  polynomially bounded as ${\rm Re} \ell \rightarrow \infty$.  This, in general, requires the separation of the sum over $\ell$ into even and odd parts. Since we are dealing odd amplitude, this step is not necessary.\\
(b) Eq. (\ref{eq:ImConformalBlock.c}) corresponds to a sum over    unitary irreducible representations of the non-compact group $O(d+1,1)$,
\be
F(w,\sigma) = \sum_{\ell} \int\limits^{\infty}_{-\infty} \,\frac{d \nu}{2\pi } \, a  (\ell, \nu) \, {\cal G} (\ell,\nu ; w,\sigma)\,.
\label{eq:groupexpansion}
\ee 
This principal series combines a discrete sum in the spin $\ell$ and a  Mellin transform in  a complex $\Delta$-plane,
with $\widetilde \Delta \equiv  i\nu = \Delta - d/2$.  As stressed by Mack~\cite{Mack:2009mi}, this representation should serves as a more general starting point for CFT, with standard OPE a consequence of this representation, (See (d) and (e) below and also \cite{Mack:2009mi} for more discussions.)\\
(c) Conformal Regge theory assumes  a meromorphic representation in the $\nu-\ell$ plane, with
poles specified by the collection of allowed spectral curves,
$\Delta_\alpha(\ell)$, e.g., 
\begin{align}
a&(\ell,\nu) =\sum_\alpha \frac{r_\alpha(\ell)}{\nu^2 + \widetilde \Delta_\alpha(\ell)^2 }\nn
&= \sum_\alpha \frac{r_\alpha(\ell)}{2\nu }\Big(  \frac{1}{\nu +i \widetilde \Delta_\alpha(\ell) }+ \frac{1}{\nu - i\widetilde \Delta_\alpha(\ell) }\  \Big) \, .\label{eq:trueCFT}
\end{align}
The  spectral curve associated with the energy-momentum tensor plays the dominant role, with $\Delta_P(2)=4$, and on which the Pomeron singularity lies, as in Fig.\ref{fig:BFKLDGLAP} ~\footnote{For more discussion see~\cite{Brower:2006ea,Brower:2010wf,Costa:2013uia,Brower:2014wha}.}.\\
(d)  To recover the standard conformal block expansion, it  is conventional to close   the contour in the lower-half $\nu$-plane~\footnote{Due to conformal invariance,  the integrand  is even in $\nu$, or, equivalently, symmetric in $\Delta\leftrightarrow 4-\Delta$. The contour can be closed either in the upper or the lower  half $\nu$-plane.  The poles in the upper half $\nu$-plane corresponds to ``shadow" operators.} (equivalently, closing the contour in the $\Delta$-plane to the right) picking up only
dynamical poles in $a(\ell, \nu)$, at $\nu(\ell)=-i( \Delta(\ell)-2)$.  It has also been explained in \cite{Mack:2009mi} that this closing of contour is allowed after separating the unitary representation function ${\cal G} (\ell,\nu ; w,\sigma)={\cal G}^{(+)} (\ell,\nu ; w,\sigma)$+${\cal G}^{(-)}  (\ell,\nu ; w,\sigma)$, where ${\cal G}^{(+)} (\ell,\nu ; w,\sigma)={\cal G}^{(-)}  (\ell,-\nu ; w,\sigma)$, with ${\cal G}^{(+)}$ leading to convergence in the lower $\nu$-plane and ${\cal G}^{(-)}$ in the upper $\nu$-plane~\footnote{Such procedure was carried out in \cite{Brower:2006ea,Brower:2007qh,Brower:2007xg}, based on a treatment equivalent to  keeping Minkowski conformal block under the leading order approximation, Sec. \ref{sec:symmetric-leadingorder}. A simpler example for this separation is Eq. (\ref{eq:PQidentity}), appropriate in treating for $d=1$ and also used in traditional Regge analysis. A related discussion has also been carried out recently \cite{Murugan:2017eto} by identifying ${\cal G}^{(-)}$ as corresponding to the shadow blocks.}. Closing in either lower or the upper $\nu$-plane leads to Eq. (\ref{eq:MOPE}), with $a^{(12),(34)}(\ell, \Delta_\alpha(\ell))=  r_\alpha(\ell)$  and $ G(w,\sigma; \ell,\Delta_\alpha(\ell)) = i \nu^{-1}{\cal G}^{(+)} (\ell,\nu ; w,\sigma)$. Summing over all $\alpha$  leads to a sum over allowed conformal primaries, with dimension $\Delta_\alpha(\ell)$ and spin $\ell$.\\
(e) By combining Eq. (\ref{eq:groupexpansion}) with a Sommerfeld-Watson transform for the angular momentum, it is possible to formally represent the conformal invariant amplitude $F^{(M)}$ in a double-Mellin form~\footnote{ This procedure was first discussed by M. Toller in the context of Lorentz symmetry, $SO(3,1)$, \cite{Toller1,Toller3,Toller2,Herman}. See also a related discussion for $O(4,2)$ in \cite{Mack:2009mi}. We have also simplified the analysis by keeping only crossing even contribution, $\ell$ even in Eq. (\ref{eq:MOPE}). The contribution with $\ell$ odd leads to the so-called Odderon contribution, which has been discussed in~\cite{Djuric:2014wzs,Brower:2008cy,Brower:2014wha}.},
 \begin{align}
F&(w,\sigma)
= F_{Regge}(w,\sigma)- \int\limits_{- i\infty}^ {i\infty}  \frac {d \tilde  \ell}{2 i}\frac {1+ e^{-i \pi \ell } }{\sin\pi \ell } \nn
&\times \int\limits^{i\infty}_{-i\infty} \,\frac{d \widetilde \Delta }{2\pi i} \, a  (\ell, \nu) \, {\cal G} (\ell,\nu ; w,\sigma)\,.\label{eq:trueinvariance}
\end{align}
This representation can formally be regarded as the principal series for a unitary irreducible representation for the non-compact group, $SO(4,2)$, with $a(\ell, \Delta_i(\ell))$ an analytic  function of $\ell$. The contour is along the imaginary axis, distorted to include all participating families of conformal blocks, appropriate for describing amplitudes which are bounded  at $u=0$ by a power $(1/\sqrt u)^{N}$, or $w^N$ at $w=\infty$. In pushing the $\ell$-contour to ${\rm Re}\, \tilde \ell=0$, or ${\rm Re}\,\ell= - (d-2)/2$, the term $F_{Regge}(w,\sigma)$ corresponds to contribution coming from all singularities in $d/2< {\rm Re}\,\ell< L_0$. The contour distortion allows one to include all participating families of conformal blocks, appropriate for describing divergent amplitudes at $w=\infty$.\\
(f) Finally, the above analysis also suggests a procedure to resolve the issue of the connection between MCB and ECB. A more desirable, top down approach is to first construct the principle series for a unitary irreducible representation of $SO(d,2)$, i.e. finding ${\cal G} (\ell,\nu ; w,\sigma)$ in Eq. (\ref{eq:trueinvariance}). Then by reverse engineering, one arrives at MCB and ECB respectively. (See Sec. 2 of \cite{Brower:2014wha} for a related discussion.)

\subsection{Spectral Curve}\label{sec:SpectralCurve}

In Eq. (\ref{eq:newgroupexpansion}), the integration contour can be pushed  further to the left, with pole contributions from the contour passing $j=2n$ cancelling that in $F_0(w,\sigma)$. This can be done until $L_0<0$ thus removing  $F_0$ entirely. However, one has to   pick up  contributions from possible $\ell$-plane singularities for $0<{\rm Re} \ell<L_0$ which  might enter  through $a^{(12),(34)}(\ell, \Delta_\alpha(\ell))$ and also that from the conformal block through $\Delta_P(\ell)$. These contributions will be collectively denote as $A_{\rm Regge}(w,\sigma) $. In terms of $\widetilde \ell = \ell+(d-2)/2$, we can shift  the contour until   the integration path is over ${\rm Re}\, \widetilde \ell=0$, or ${\rm Re}\,\ell=-(d-2)/2$, arriving at 
Eq.   (\ref{eq:trueinvariance}). 
For $d=4$, the continuum corresponds to an integral along  the line ${\rm Re}\, \ell= -1$. (For $d=3$, the path is along  ${\rm Re}\, \ell=-1/2$, the traditional ``background integral" in a Regge representation.)

Let us turn next to ${\rm Im}\, F(w,\sigma) $. As one pushes the contour to the line along ${\rm Re} \,\widetilde \ell=0$, one finds that
\begin{align}
{\rm Im}&\, F(w,\sigma) = {\rm Im}\, F_{\rm Regge}(w,\sigma)  \nn
&+ \sum_\alpha\int\limits_{-(d-2)/2- i\infty}^{-(d-2)/2+ i\infty} \frac {d  \ell}{2 i}\nn
&\times  a^{(12),(34)}(\ell, \Delta_\alpha(\ell)) G(w,\sigma; \ell,\Delta_\alpha(\ell))  \, .\label{eq:ConformalBlock.e}
\end{align}

Here let us examine the limit $w\rightarrow \infty$.   With ${\rm Re}\, \ell = -(d-2)/2$, the background integral is of the order $O(w^{1+(d-2)/2})$, and is bounded for $1\leq d$.
 It follows that a divergent contribution to $F(w,\sigma)$ occurs only if there exists a singularity in $\ell$ in the regions $ -(d-2)/2< {\rm Re}\, \ell <2$.  This is entirely analogous to a conventional Regge theory where  the sub-dominant contribution from the continuum is  referred to as that from the background integral.  

Without dynamical inputs, it is not possible  to specify what singularities might exist in the region to the right of ${\rm Re}\, \ell = -(d-2)/2$ for a general CFT.  Let us consider first ${\cal N}=4$ SYM at $d=4$ and  focus on the leading twist-two contribution, which interpolates the  stress-energy tensor, with the associated  spectral curve denoted $\Delta_P(\ell)$. Based on weak-coupling perturbation analysis,  there exists at least one singularity $\ell_{eff}$, to the right of  $\ell=1$ \cite{Fadin1975,Balitsky:1978ic,Kuraev:1977fs}. This enters  through a branch point of $\Delta_P(\ell)$. At strong coupling, the location of the corresponding singularity is bounded from above, $\ell_{eff} <2$.  (See Fig. \ref{fig:BFKLDGLAP} for a schematic representation.)  The spectral curve in weak coupling can be found by solving the BFKL equation, Eq. (\ref{eq:BFKL-DGLAP1}) and, in strong coupling, via AdS/CFT~\cite{Brower:2006ea,Kotikov:2004er}. This singularity is historically referred to as the Pomeron. 
  
It can be shown that this singularity, in the immediate neighborhood of $\ell=2$ and at large $\lambda$,  is of the square-root type~\cite{Brower:2006ea,Brower:2014wha,Basso:2011rs,Gromov:2014bva}. Using the effective AdS mass introduced in (\ref{eq:adsmass-2}), i.e., $m^2_{eff}(\ell)= - 4 +(\ell-\ell_{eff})B^2(\ell,\lambda)$ the leading $\lambda$ spectral curve becomes
\be
\Delta_P(\ell) = 2 + {B(\ell,\lambda)} \sqrt {\ell-\ell_{eff}},   \label{eq:Delta-j-analyticity}
\ee
with  $B(\ell,\lambda)$ analytic at $\ell=2$. It follows that the leading behavior as $w\rightarrow \infty$ for the correlation functions $F$ is
\be
{\rm Im}\, F(w,\sigma) \sim \, \frac{w^{\ell_{eff}-1}}{|\ln w|^{3/2}}\,.
\label{eq:reducedF}
\ee

 As stressed in \cite{Brower:2006ea}, for any d-dim CFT with a gravity dual, this singularity can be associated with string modes interpolating graviton in AdS.  With $m^2(2)=0$, it follows that its location  is bounded from above,
  \be
\ell_{eff}<2.
  \ee
The deviation from $\ell_{eff}=2$, $\delta\equiv 2-\ell_{eff}>0$,   can be attributed to   stringy corrections. 

To demonstrate the generality of (\ref{eq:Delta-j-analyticity}), let us recall that for flat-space string theory, massless graviton lies on a linear trajectory,  
\be
\ell= 2 +(\alpha'/2) t\, , 
\ee
with $\sqrt {\alpha'}=\ell_{string}$ 
 providing a length scale.  This can be understood by the on-shell condition for graviton, $1=L_0=\bar L_0= \ell/2 -(\alpha'/4) t$, where $L_0$ and $\bar L_0$ are generators of dilation in a world-sheet treatment. 
Consider next  string living on a curved background, e.g., $AdS_5$.  In the weak-curvature limit,  $\ell_{st}^2 /R^2_{ads}=1/ \sqrt \lambda <<1$, one has $1=L_0=\bar L_0\simeq \ell/2 -(\alpha'/4R_{ads}^2) \widetilde \nabla^2(\ell)$, with $ \widetilde \nabla^2(\ell)$ dimensionless and $ \widetilde \nabla^2(2)= \Delta (4-\Delta)$, with stress-energy tensor having conformal dimension $\Delta=4$.  This leads to a leading $\lambda$ spectral curve condition
\be\label{eq:speccurve}
\Delta (4-\Delta)\simeq  2\sqrt{\lambda}(\ell-2)\, , 
\ee
which is parabolic, for $(\ell-2)^2<<1$ and symmetric about $\Delta=2$, as dictated by conformal invariance.  This relation is an expansion in \emph{both} $\lambda^{-1/2}$ and $(\ell-2)$. It can next be cast in the form (\ref{eq:Delta-j-analyticity}), with
\be
B(\ell, \lambda) \simeq \sqrt 2 \, \lambda^{1/4} \,,
\ee
and  the branch point at $\ell_{eff} = 2- 2/\sqrt \lambda$.  

As already stressed, Eq. (\ref{eq:Delta-j-analyticity}) can be interpreted as having an $\ell$-dependent effective $AdS$ mass. For deformed or thermal theories mass/momentum modes will be shifted. For general $d$ and/or deformed AdS-background, it can be shown that 
\be
\ell_{eff} = 2- (d/2)^2 /2\sqrt {\lambda_{eff}}
\ee
where $\lambda_{eff}$ may depend on temperature. We return to this point in Secs. \ref{sec:CFT-1} \& \ref{sec:discuss}.

\section{Scattering for CFT at \texorpdfstring{$d=1$}{d=1}}\label{sec:CFT-1}

The importance of the CFT 4-point function has appeared recently in the study of CFTs dual to AdS spaces with a black hole. This set up has become tantalizing for two main reasons: first, as a model of holography it is relatively simpler than the canonical $\mathcal{N}=4$ SYM duality and one might be able to demonstrate more general aspects of holography. Secondly, using the conformal symmetry on the boundary one might learn about the information loss paradox.  In this section we apply Minkowski conformal blocks to $d=1$ CFTs.  The primary example of interest is the SYK theory \cite{kitaev1,kitaev2,Sachdev:1992fk}, however, as most of our results apply to CFTs more generally, we refer  to these as SYK-like models \footnote{For a brief description of the SYK theory relevant to this work see Appendix \ref{sec:green}}. Here we address the issue of $d=1$ CFT directly as a scattering problem where one explores the $SO(1,1)$ Lorentz boost symmetry. The related CFT invariant four point correlation function, expressed in terms of appropriate cross ratio, $\Gamma(w)$, has a power behavior, as $w\rightarrow \infty$, 
\be
\Gamma (w)\simeq w^{\ell_{eff}-1}\, , \label{eq:chaos}
\ee
with $\ell_{eff}\leq 2$. The precise definition for the variable $w$ is given  by Eq. (\ref{eq:w-variable}), in analogy to that defined for $d\geq 2$ in Eq. (\ref{eq:qqbar}).  When $\ell_{eff} \rightarrow 2$, this can be interpreted as corresponding to dual graviton contribution which saturates the chaos bound~\cite{Shenker:2014cwa,Shenker:2013pqa}. There are two challenging aspects involved. The first  is to find what kind of CFT  leads to such power-behavior.  An equally challenging task is, given a specific 1-d CFT, how best to identify the effective spin.  In this section, we will focus on the latter aspect. 

The relevant kinematics for a 1-d scattering is discussed in Sec. \ref{sec:SYK1} where we provide a Mellin-like representation for $\Gamma(w)$, as well as  ${\rm Im} \, \Gamma(w)$, analogous to Eq.  (\ref{eq:RedConformalBlock-Disc}). In Sec. \ref{sec:SYK} we examing the Schwinger-Dyson equation for SYK-like models. This corresponds to treating a set of Wightman functions for which the integral equation simplifies since it involves only amplitudes over physical scattering regions. This integral equation can be diagonalized in $\ell$, and the associated partial-wave amplitude, $A(\ell)$, can be then found algebraically.  For these models, one can show that  $\Gamma(w)$ is power-behaved, as in Eq. (\ref{eq:chaos}),  and  the leading effective spin can be identified simply. 

Our treatment  parallels to that carried out in \cite{Polchinski:2016xgd,Maldacena:2016hyu} but differs significantly  in how  the relevant spectral analysis is carried out.  In working with $ {\rm Im} \, \Gamma(w)$,  we can in principle deal only with the Hilbert space of square-integral functions defined over half-line, $1<w<\infty$, instead of  over the whole line, $-\infty<w<\infty$, as done in \cite{Polchinski:2016xgd,Maldacena:2016hyu}.  This leads to a significant simplification.  In Appendix \ref{sec:green}, 1-d Green function for $D_w$, Eq. (\ref{eq:Legendre2}),   is discussed from a conventional  spectral analysis, before generalizing  to the case of functions which are polynomially bounded.  Here, we follow more directly  the procedure of re-summation performed in the last section for $d\geq 2$, starting with the result of Sec. \ref{sec:symmetric-d1}, which agrees with the result followed from a Hilbert space treatment, Eq. (\ref{eq:spectralrep3}).

\subsection{Kinematics}\label{sec:SYK1}

In analogy to the case of $d\geq 2$, we consider a reduced  invariant 4-point function, $ \Gamma$, as a function of a cross ratio for the process $1+3\rightarrow 2+4$.  There are several options for cross ratios~\footnote{We have switched to using a timelike coordinates, $t$ and $\tau$, to emphasize that we are interested in the Lorentz boost symmetry and to conform with the more standard notation found in the literature. See also Appendix  \ref{sec:d1}.}, e.g.,
  \be
\tau=\frac{t_{21}t_{43}}{t_{23}t_{41}} \, \quad {\rm or} \quad \tau_c=\frac{t_{13}t_{42}}{t_{23}t_{41}} \, .\label{eq:tau1}
\ee
In 1-d, however, only one is independent due to the constraint $|\tau_c|+|\tau|=1$.  Let us first adopt $\tau$ as the independent variable. For s-channel scattering, $1+3\rightarrow 2+4$, we require $t_4,t_2>t_3,t_1$.  Without loss of generality, we consider the causal limit of $t_4>t_2>t_1>t_3$.  Reparametrization  invariance  allows one to keep three points fixed. We choose the ansatz $t_4=\infty$, $t_3=0$, $t_2=1$, with $t_1\equiv t$ as the independent variable, thus $0\leq  t\leq 1$.    In terms of invariant cross ratios,  $\tau=1-t$ and $\tau_c=t$, and the s-channel physical region corresponds to $0\leq \tau\leq  1$, and $\tau+\tau_c=1$.

The u-channel physical region, with $3\leftrightarrow 4$ interchanged, corresponds to $\tau\rightarrow \tau'\equiv \frac{t_{12} t_{34}} {t_{13} t_{24}} = \tau/(\tau-1)$. The u-channel physical region, $0\leq \tau' \leq 1$, then leads to $-\infty<\tau<0$. In order to exhibit crossing more symmetrically, it is again convenient  to adopt a new variable (similar to Eq. (\ref{eq:qqbar}))
\be
w\equiv (2-\tau)/\tau. \label{eq:w-variable}
\ee
The s-channel physical region corresponds to 
\be
1<w<\infty\, ,
\ee
with u-channel region given symmetrically by $-\infty<w<-1$.  

Let us consider both  the invariant amplitude, $\Gamma(w)$,  and its imaginary part in the s-channel physical region,  ${\rm Im}\,\Gamma(w)$. 
We assume that $\Gamma(w)$ is polynomial bounded, $| \Gamma(w)|< w^{N}$, at $w=\infty$.  As before, this invariant function has a factor of $w$ removed from the 4-point correlator, as in Eq. (\ref{eq:T-chi}). For correlators which are  crossing even, $\Gamma(w)$ will be  crossing odd, $\Gamma(-w)=-\Gamma(w)$. As shown in Sec. \ref{sec:symmetric-higherorder},  eigen-functions for the quadratic Casimir for 1-d scattering processes can be chosen to be  Legendre functions of the second kind, $G_{1d}(w)=\bar Q_{-\ell}(w)=c_\ell Q_{-\ell}(w)$, with normalization $c_\ell$ specified in Eq. (\ref{eq:1dMCB})~\footnote{The utility of using Legendre functions of the second kind to describe $\Gamma(w)$ has been noted by other authors studying the SYK model \cite{antal}. Also see \cite{Das:2017wae,Das:2017hrt,Das:2017pif}.}.  In terms of the Mellin representation, described in Sec. \ref{sec:SW}, reparametrization invariance leads to
\begin{align}
\Gamma(w)&= \sum_{0<\ell<L_0, \,{\rm even}} a(\ell)\, \bar  Q_{-\ell}(w)  - \int\limits_{L_0 -i\infty}^{L_0 + i\infty}  \frac {d\ell}{2 i} \nn
&\times\frac {1+ e^{-i \pi \ell} }{\sin\pi \ell } a(\ell)\, \bar Q_{-\ell}(w)  \, .\label{eq:1D-ConformalBlock}
\end{align}
where $N+1<L_0$. For $w$ large, from Eq. (\ref{eq:1dMCB}), the Legendre function simplifies to
$\bar Q_{-\ell}(w)\simeq w^{\ell-1}(1+O(w^{-2}))$.

For its imaginary part, defined for $1<w<\infty$,   the sum over $\ell<L_0$ does not contribute, and 
\begin{align}
 {\rm Im}& \, \Gamma(w)= - \int\limits_{L_0 -i\infty}^{L_0 + i\infty}  \frac {d\ell}{2 i}  a(\ell)\, \bar Q_{-\ell}(w)\nn
 &=  - \int\limits_{L_0 -i\infty}^{L_0 + i\infty}  \frac {d\ell}{2\pi i} (2\ell-1)\,A(\ell)\, \, \frac{\tan \ell\pi}{\pi} Q_{-\ell}(w) \, .\label{eq:1D-ImConformalBlock}
\end{align}
where we have also brought the representation into a more conventional form~\footnote{Our notation here conforms to  that adopted in  Appendix \ref{sec:spectral2}, in particular,  factoring out the term $(2\ell -1)$. A more natural choice would be shifting $\ell$ by one unit, leading to a more familiar factor of $2\ell+1$.  We have not done so due to  our choice of  maintaining symmetry about ${\rm Re} \tilde \ell=0$. for $d=1$, this corresponds to ${\rm Re}\, \ell=(2-d)/2$.}, with  $\frac{\tan \ell\pi}{\pi} Q_{-\ell}(w)$ regular in $\ell$ at positive integers. By taking advantage of the identity in Eq. (\ref{eq:PQidentity}), it is useful to simplify  (\ref{eq:1D-ImConformalBlock}) further  in terms of Legendre function, $P_\ell(w)$, with $\ell$ complex and $1<w<\infty$,
 \be
 {\rm Im} \, \Gamma(w)=   \int\limits_{L_0 -i\infty}^{L_0 + i\infty}  \frac {d\ell}{2 \pi i} (2\ell-1) A(\ell)\, P_{\ell-1}(w)  \, .\label{eq:1D-ImConformalBlock-2}
\ee
 This follows from the fact that one can add terms under the integral in (\ref{eq:1D-ImConformalBlock}) which will then vanish by closing contour to the right. In applying to 1-d scattering, it is this Mellin representation which will be particularly useful.  As shown in Appendix \ref{sec:green}, this representation can also be arrived at through a Hilbert space treatment over $1<w<\infty$.
 
 Making use of a standard orthogonality condition, 
\be
\int\limits_1^\infty P_\nu(x) Q_\sigma(x) dx = \frac{1}{(\sigma-\nu)(\sigma+\nu+1)}
\ee
 one has the following inversion formula
\be
A(\ell) = \int\limits_1^\infty d w\,  {\rm Im} \, \Gamma(w) \, Q_{\ell-1}(w). \label{eq:inversion-2}
\ee
  Conversely, we can examine the full amplitude, 
\begin{align}
\Gamma(w)&= \sum_{2\leq \ell<L_0, \, {\rm even}} A(\ell)\, P_{\ell-1}(w)   \nn
&- \int\limits_{L_0 -i\infty}^{L_0 + i\infty}  \frac {d\ell}{2 i} \frac {1+ e^{-i \pi \ell} }{\sin\pi \ell } (2\ell-1) A(\ell)\, P_{\ell-1}(w) \, ,\label{eq:1D-ConformalBlock-2}
\end{align}
and examine the continuation to  the region $-1<w<1$, away from the s,u-channel physical regions. From Eq. (\ref{eq:inversion-2}), $A(\ell)$ is polynomial bounded as ${\rm Re} \, \ell \rightarrow  \infty$.  Similarly, $P_{\ell-1}(w)$ vanishes exponentially for ${\rm Re}\, \ell$ large when  $-1<w<1$. Closing the contour in Eq. (\ref{eq:1D-ConformalBlock-2}) to the right leads directly to a real amplitude
\be
 \Gamma(w)= \sum_{2\leq \ell=2n}  (2\ell-1)\, A(\ell)\, P_{\ell-1}(w) \label{eq:EuclideanPartialWave}
\ee
with ${\rm Im}\, \Gamma(w)=0$.  

Finally we note that a 4-point correlator symmetric under s-u crossing, $T(w)=T(-w)$, can be defined by $T(w)=w\Gamma(w)$.  In continuing to the Euclidean region, from Eq. (\ref{eq:1D-ConformalBlock-2}),  it takes on a more conventional form
\be\label{eq:tw}
T(w) = w \Gamma(w) =  \sum_{ 0\leq \ell,\,  {\rm even}}(2\ell+1)\, B(\ell)\, P_{\ell}(w)\, 
\ee
where
\be
B(\ell) = \frac{2\ell-1}{2\ell+1} A(\ell) + \frac{(\ell+1)^2}{(2\ell+1)^2} A(\ell+2) 
\ee
with $A(0)=0$ \footnote{The final relation comes from the two-term recursion relation
$
P_{n+1}(w)=  w P_n (w) - \frac{n^2}{4n^2-1} P_{n-1}(w)\, .
$ The contributions from $\ell = 0$ and $\ell=2$ require a more careful treatment. We defer to the comments after Eq. (\ref{eq:EuclideanPartialWave-2})}.

\subsection{Scattering for SYK-Like Models}\label{sec:SYK}

In the low temperature limit of the SYK model there is a \emph{near} conformal symmetry that is both explicitly and spontaneously broken. The theory is close to a 0+1 dimensional conformal theory.  Within this approximation, the invariant function $\Gamma$ can be obtained by solving a ladder-type Bethe-Salpeter equation~\footnote{This type of integral equation has a long history of use in particle physics. For a different example see Appendix \ref{sec:BFKL-DGLAP} where it is discussed in the context of deep inelastic scattering.},  which can be represented schematically as
\be
 \Gamma=   \Gamma_1+    K_0\, \otimes \, \Gamma     \label{eq:SD-W-1}
\ee
where $\Gamma$ is a function of 4 coordinates: $\Gamma(t_1,t_2, t_3, t_4)$. We shall refer to models that admit such a representation as {\it SYK-like models.}  Reparametrization invariance allows  diagonalization of this integral equation.  This has indeed been carried out in \cite{Polchinski:2016xgd,Jevicki:2016bwu,Maldacena:2016hyu}, primarily in an Euclidean setting.  To make contact with the chaos bound, an appropriate analytic continuation to the Minkowski setting has to be performed.

As a scattering problem, Eq. (\ref{eq:SD-W-1}) has the structure of a ``box-diagram", and it can be expressed equivalently as an analogous integral equation for its absorptive part, the imaginary part of $\Gamma$ in the physical region. Schematically, this can be expressed as 
\be
{\rm Im}\, \Gamma=  {\rm Im}\, \Gamma_1+   \widetilde K_0\, \otimes' \, {\rm Im}\, \Gamma     \label{eq:SD-ImW-1}
\ee
with the convolution, $\otimes'$, is over the physical region only.  

We will discuss the integral equation, Eq. (\ref{eq:SD-W-1}), as a scattering problem in real time, and study its solution in the high energy limit. As such, it is analogous to the BFKL type integral equation when applied to DIS structure functions, Appendix \ref{sec:application-II}.  We will spell out more precisely the nature of this integral equation after first discussing the simplification due to reparatrization invariance, allowing one to express $\Gamma$ in terms of invariant cross ratios, Eq. (\ref{eq:S-D-full}). Reparametrization invariance leads to a simpler diagonalization procedure in arriving at the desired solution in identifying the leading effective spin, $\ell^*$ which directly controls the large-time behavior for $\Gamma$.

It is also instructive to express the solution to Eq. (\ref{eq:SD-ImW-1}) formally as an iterative sum, 
\bea
{\rm Im}\, \Gamma &=& \sum^\infty_{n=1} ({\rm Im}\, \Gamma )_n, \label{eq:unitaritysum}\\
({\rm Im}\, \Gamma )_{n} &= & \widetilde K_0\, \otimes' \, ({\rm Im}\, \Gamma)_{n -1}. \label{eq:MPM}
\eea
The process can be viewed as producing n-lumps of particles, with $\widetilde K_0$ providing  the relative probability of producing an additional ``lump". Each lump is irreducible, and the allowed intermediate states consists of n  such  lumps, with $n=1,2,3,\cdots$.  This interpretation is analogous to the case of $d=2$ forward scattering discussed in Appendix \ref{sec:application-II}.

More precisely, Eq. (\ref{eq:MPM}) can be identified with  the early work of Amati, Fubini and Stangelini~\cite{MPM,Bali:1967zz} in constructing multiperipheral models (MPM) of particle production. Their construction is equivalent to a Bethe-Salpeter equation for the absorptive part of the forward scattering amplitudes. Here $({\rm Im}\, \Gamma )_n\propto |T_{2\rightarrow n}|^2$, where a factorizable 2-to-n production amplitude, $T_{2\rightarrow n}$, leads to a recursive relation, Eq. (\ref{eq:MPM}).
Furthermore, consistency requires that $ \widetilde K_0\geq 0$, over the allowed phase region.  Given $ \widetilde K_0$ for a specific model, the challenge is to carry out the sum, Eq. (\ref{eq:unitaritysum}). In exact analogy to the BFKL integral equation, Eq. (\ref{eq:BFKL-DGLAP1}), this integral equation can be solved by diagonalization due to $SO(1,1)$ symmetry.

\subsubsection{Kinematics of Integral Equation  for \texorpdfstring{${\rm Im \,\Gamma}$}{ImP}:}\label{sec:SYK2}

In order to understand the kinematics of the integral equation for ${\rm Im}\, \Gamma$, it is useful to first examine the Schwinger-Dyson equation in $d=1$ for the full amplitude $\Gamma$. Consider a theory defined in such a way that it is given by a sum of ladder graphs, i.e. schematically it is analogous to that for BFKL integral equation, Eq. (\ref{eq:BFKL-full}). For SYK-like models, the integral equation can be written for  an amputated 4-point function, with labels corresponding to ``scattering" in the s-channel of $1+ 3 \rightarrow  2+ 4$. Explicitly, Eq. (\ref{eq:SD-W-1}) becomes
\begin{align}
\Gamma& (t_2,t_1;t_4,t_3)= \Gamma _1(t_2,t_1;t_4,t_3) \nn
&+ \int d t_5 dt_6  K_0(t_2,t_1;t_6,t_5) \Gamma (t_6,t_5;t_4,t_3) \, . \label{eq:S-D-full}
\end{align}
Similarly to the BFKL case, the ladder sum is in the $t$-channel.  The first term on the right corresponds to an inhomogeneous contribution. The second term  corresponds to a convolution of  the ``2-particle irreducible" kernel,, $ K_0(t_2,t_1;t_6,t_5)$,    with   the connected full amplitude, $\Gamma (t_6,t_5;t_4,t_3)$.  Although the solution can also be expressed formally in an iterative sum, Eq. (\ref{eq:unitaritysum}), it is important to emphasize that the integration is over all values of $t_5$ and $t_6$. Some regions do not correspond to physical scattering, with $K_0$ taking on negative values. Reparametrization invariance dictates that $K_0(t_2,t_1;t_6,t_5)$ is a function of a cross ratio. For SYK like models, define
\begin{align}
K_0&= (1/\alpha_0(q) ) \Big(\frac{t_{21}t_{65}}{t_{25}t_{61}}  /\frac{t_{15}t_{62}}{t_{25}t_{61}}\Big)^{2/q} \nn
&=(1/\alpha_0(q ))\Big(\frac{\tau_k}{1-\tau_k}\Big)^{2/q},  \label{eq:realkernal}
\end{align}
where  $\alpha_0(q) = \frac{2\pi q }{(q-1)(q-2 ) \tan {(\pi/q)}}$, with $4\leq q<\infty$.   We have expressed $K_0$, which is derived in an Euclidean treatment,  such that $K_0>0$ over the physical region of $0<\tau_k<1$. As mentioned above, it is necessary to treat different kinematic regions differently, as done in \cite{Polchinski:2016xgd}. However, we are interested in treating this as a generic scattering problem for 1-d CFT, but will not address here the  question of how this type of model can arise from  a more fundamental perspective~\footnote{In moving to a Minkowski setting, the kernel should acquire an additional phase. We will return to this question below.}.

The structure of the integral equation for ${\rm Im}\, \Gamma$ is the same as for the full amplitude $\Gamma$, except the integration region in Eq. (\ref{eq:S-D-full}) has to be restricted to the physical region. For 1-d, it is possible to express both $\Gamma$ and  ${\rm Im}\, \Gamma$   in terms of a single cross ratio.  We will initially adopt $\tau$, Eq. (\ref{eq:tau1}), as the independent variable and use the ansatz described at the beginning of Sec. \ref{sec:SYK1}.  For the integrand in Eq. (\ref{eq:S-D-full}), we can construct invariants analogous to $\tau$:  $\tau_k=\frac{t_{21}t_{65}}{t_{25}t_{61}}$ and $\tau'=\frac{t_{65}t_{43}}{t_{63}t_{45}}$.  In terms of these invariants
\be
dt_5dt_6 \rightarrow \frac{d\tau_k d\tau' }{\sqrt {D}} \, ,
\ee
where the Jacobian is given by   $D(\tau,\tau', \tau_k)= 4(\tau^{-2}+\tau_k^{-2}+{\tau'}^{-2})+4  ( (\tau\tau')^{-2} +(\tau'\tau_k)^{-2} +(\tau\tau_k)^{-2}   )-6(\tau^{-1}+\tau_k^{-1}+{\tau'}^{-1})+3$.   The restriction to the physical region, for ${\rm Im}\, \Gamma$, corresponds to enforcing the constraint $D\geq 0$.

As discussed in Sec. \ref{sec:SYK1}, a more symmetric treatment for the physical scattering region can be carried out by working with variable $w= \frac{2-\tau}{\tau} $, with s-channel physical region $1<w<\infty$ and correspondingly $-\infty<w<-1$ for the u-channel. More directly, it can be shown that
 \be
w=\frac{2-\tau}{\tau} = \frac{t^2_{12} + t^2_{34}- (\bar t_{12} -\bar t_{34})^2}{2 t_{12} t_{34}} \, .
\ee
where $\bar t_{ij} = (t_i+t_j)$. Again, for the integrand in Eq. (\ref{eq:S-D-full}), we can construct
\begin{align}
w_k&=\frac{2-\tau_k}{\tau_k} = \frac{t^2_{12} + t^2_{56}- (\bar t_{12} -\bar t_{56})^2}{2 t_{12} t_{56}}\nn
w'&=\frac{2-\tau'}{\tau'} = \frac{t^2_{56} + t^2_{34}- (\bar t_{56} -\bar t_{34})^2}{2 t_{56} t_{34}}\, ,
\end{align}
with $1<w_k<\infty$ and $1<w'<\infty$  for s-channel physical regions and  $-\infty <w_k<-1$ and $-\infty <w'<-1$  for u-channel physical regions,  for $1+5\rightarrow 2+6$ and $5+3\rightarrow 6+4$ processes respectively.  The region will further be restricted by the requirement that the Jacobian of transformation from $t_5$ and $t_6$ to $w_k$ and $w'$ be real. It is not difficult to check that the Jacobian is given by 
$
dt_5dt_6 \rightarrow \frac{d w_k d w' }{\sqrt {  \bar D(w,w_k,w')}}
$,
where $\bar D(w,w_k,w')$  takes on a standard  form of a ``triangle function", 
\be
 \bar D(w,w_k,w') = w^2 +w_k^2+w'^2-1-2w\,w_k\, w'\, .
\ee  

The condition of integrating over physical region corresponds to  $\bar D(w,w_k,w')>0$, with $1<|w'|,\, |w_k|< w$.   The condition for physical scattering, designated by $ \theta^{+}(\bar D)$, further restricts $1\leq w_k,w'\leq w$, with contributions from  u-channel, $-\infty<w_k,w'<-1$,  folded into the $1<w_k,w'<w$ by symmetry.  Introducing Lorentz boost parameters, $ \cosh \beta=w$, $ \cosh \beta_k=w_k$ and $ \cosh \beta'=w'$,  the constraint due to the Jacobian can also  simply be expressed as a triangle inequality, $0<\beta_k+ \beta'\leq \beta$,  which respects the proper time-ordering. 
It follows that the integral equation for ${\rm Im}\, \Gamma (w)$,  Eq. (\ref{eq:unitaritysum}),  in the s-channel physical region can be   expressed as
 \begin{align}
{\rm Im}&\, \Gamma (w) =  {\rm Im}\, \Gamma_1 (w) \nn
&+ \int\limits_1^w \int\limits_1^w d\, w_k d\, w'\frac{\theta^{+}( D)}{\sqrt { D}}\, \widetilde K_0 (w_k)\, {\rm Im} 
\Gamma (w') ,   \label{eq:SD-ImW-2}
\end{align}
which is one of our key results. 
 
 Let us turn next to the integral kernel $\widetilde K_0(w_k)$ and its relation to the corresponding kernel $K_0 (w_k)$ for the full amplitude. In higher dimensions, $\widetilde K_0$ would be appropriate discontinuities of $K_0$ across various physical regions. For CFT, these discontinuities are all equal, up to a constant numerical factor.  We will therefore replace $\widetilde K_0$ in Eq. (\ref{eq:SD-ImW-2}) by
$
\widetilde K_0 (w_k)= C\,  K_0 (w_k) . 
$
with $K_0$ real and positive, given by Eq. (\ref{eq:realkernal}),  in the s-channel  scattering region of $1<w_k<\infty$. For 1-d, to keep track of the contributing phases, we need to take account of time-ordering across two edges of the ladder. Denote the signs for $t_{53}$ and $t_{64}$ by $\pm$. There are 4 distinct time orderings: $(++), (--), (+-),$ and $(-+)$, which contribute to ${\rm Im}\, \Gamma (w) $ in the s-channel~\footnote{In contrast, for the integral equation for the full amplitudes, depending on the counting schemes, there are many more configurations to consider  in an Euclidean treatment~\cite{Polchinski:2016xgd,Maldacena:2016hyu}.}. For SKY-like models, the combined contribution becomes
$
  C = (e^{-i\pi /q }+e^{i\pi /q })  (e^{-i\pi /q }+e^{i\pi/q})^*= 4 \cos^2 ( \pi /q)
$, with $C$ real and positive. This amounts to adopting an approach using a retarded propagator, as done in~\cite{Maldacena:2016hyu}.  It follows that
 \begin{align}
\widetilde K_0(w_k)= &  \frac{2^{1+1/q}(q-1)(q-2)}{q^2\Gamma(1+2/q)\Gamma(1-2/q)} \nn
& \quad\times (w_k-1)^{-2/q} \theta(w_k-1)\,. \label{eq:SYK-kernel}
\end{align}
This  shall serve as our ansatz for SYK-like models.

\subsubsection{Diagonalization:}
We proceed first to a general solution to this integral  equation for ${\rm Im} \Gamma(w)$, Eq. (\ref{eq:SD-ImW-2}),  taking advantage of our analysis in Sec. \ref{sec:SYK1}. Just like the situation for $d\geq 2$, we will consider the class of problems where $\Gamma (-w) =-\Gamma (w) $ and $\Gamma (w)$ is polynomially bounded as $w\rightarrow \infty$. It follows that ${\rm Im}\, \Gamma (w) $ can be represented by an inverse Mellin transform, Eq. (\ref{eq:1D-ImConformalBlock-2}). A similar representation can be written for the kernel,
\be
\widetilde K_0(w_k)= \int\limits^{L_0+ i\infty} _{L_0-i\infty} \frac{d\ell}{2\pi i } (2\ell-1)k(\ell) P_{\ell-1}(w_k).
\ee
and also for the inhomogeneous term, $\Gamma_1$.  The inversion formula is given by Eq. (\ref{eq:inversion-2}), and, for the kernel and the inhomogeneous term, 
\bea
k(\ell) &=&    \int\limits_1^\infty dw\, Q_{\ell-1}(w)  \,  \widetilde K_0(w)      \, ,  \label{eq:kernel}\\
A_1(\ell) &=& \int\limits_1^\infty dw Q_{\ell-1}(w)  {\rm Im} \, \Gamma_1 (w)
\eea
We emphasize again, for both $A_1(\ell)$ and $k(\ell)$, the integrals are over the physical regions only. With  $k(\ell)$  given by a single integral, Eq. (\ref{eq:kernel}), this is a significant simplification when compared to the comparable treatment involving the full amplitude, $\Gamma$~\cite{Polchinski:2016xgd,Maldacena:2016hyu}.

Let us apply the integral transform $\int_1^\infty dw Q_{\ell-1}(w)$ to Eq. (\ref{eq:SD-ImW-2}). The left-hand side leads to $A(\ell)$.  With the help of the identity~\cite{Abarbanel:1970ef,Abarbanel:1971nj,Saunders:1971tz}
\be
 \int\limits_1^\infty \frac{d\, w}{ \sqrt {D(w,w_k,w')}}  \theta^{(+)} (D)  Q_{\ell}(w)  = Q_{\ell}(w_k)Q_{\ell}(w')
 \ee
valid for ${\rm Re}\,\ell>1$, the non-trivial term on the right can be diagonalized as $k(\ell)\, A(\ell)$, thus leading to 
\be
A(\ell)=A_1(\ell) + k(\ell,q)\, A(\ell).
\ee
The ability to reduce the integral equation, Eq. (\ref{eq:SD-ImW-2}), to an algebraic equation by a rather straight-forward analysis is another one of our key results. This process is analogous to that carried out in~\cite{Abarbanel:1970ef,Abarbanel:1971nj,Saunders:1971tz}. The treatment here corresponds to an harmonic analysis by non-unitary representation of $SO(1,1)$.

From this diagonalized equation, the partial-wave amplitude $A(\ell)$ can be  found simply,
\be
A(\ell) = \frac{A_1(\ell) }{1- k(\ell,q) }
\ee
with $k(\ell)$ given by Eq. (\ref{eq:kernel}).  This in turn allows one to recover ${\rm Im} \Gamma$ via Eq. (\ref{eq:1D-ImConformalBlock-2}) and/or the full amplitude ${\rm \Gamma}$  more directly from  via Eq. (\ref{eq:1D-ConformalBlock}).  

\subsubsection{Identifying the Leading Intercept \texorpdfstring{$\ell^*$}{l*}}

The leading effective spin, $\ell_{eff}$, corresponding to the rightmost singularity of $A(\ell)$, can be found by the condition
\be
k(\ell_{eff},q)=1,
\ee
where $k(\ell)$ is given by Eq. (\ref{eq:kernel}).   
Let us  focus on SYK-like models where  the kernel  $\widetilde K_0(w) $ is given by Eq. (\ref{eq:SYK-kernel}). This leads to
\be
k(\ell,q) =\frac{2^{1+2/q}(q-1)(q-2)}{q^2\Gamma(1+2/q)\Gamma(1-2/q)}  I(\ell,q)\, ,
\ee
where we have exhibited its dependence on the parameter $q$, with $I(\ell, q)$ given by a single integral
\be
  I(\ell, q)=\int\limits_1^\infty \, dw\, ({w-1})^{-2/q}\, Q_{\ell-1}(w).
  \ee
 This integral converges for $0<1-2/q<{\rm Re} \,\ell$  at $w=1$ and $w=\infty$.  Therefore, $k(\ell,q)$ is analytic to the right of ${\rm Re} \,\ell=1-2/q$.

It is easy to check that  $k(2,q)=1$, therefore $A(\ell)$ has a pole   at $\ell=2$.  This can also be verified  by evaluating $ I(2, q)=\int_1^\infty \, dw\, ({w-1})^{-2\delta}\, Q_{1}(w)$ via a contour integral. 
For general $\ell$, $0<1-2/q <{\rm Re} \,\ell$, the integral  $  I(\ell, q)$ can also be evaluated explicitly  by expanding $Q_{\ell-1}(w)$  and summed up to 
$
  I(\ell, q)
 =  2^{-2/q}( \Gamma(1-2/q))^2  \frac{ \Gamma(\ell-1+2/q)        }{\Gamma (  \ell +1-2/q)      }
$. 
This leads to 
\bea
k(\ell,q)
  &=&   \frac{\Gamma(3-2/q)}{\Gamma(1+2/q)}  \frac{ \Gamma(\ell-1+2/q)        }{\Gamma (  \ell +1-2/q)      }\, ,\label{eq:kernel-ell}
\eea
which agrees with that found earlier~\cite{Polchinski:2016xgd,Jevicki:2016bwu,Maldacena:2016hyu}. The amplitude $A(\ell)$ is thus
\begin{align}
&A(\ell)=\nn
&\frac{\Gamma(1+2/q) \Gamma(\ell+1-2/q)       } {\Gamma(1+2/q) \Gamma (  \ell +1-2/q)- \Gamma(3-2/q)\Gamma(\ell_1+2/q)}\nn
&\quad \times  A_1(\ell)  \, . \label{eq:A-ell}
\end{align}
As a further check, for $\ell=2(n+1)$, $n=1, 2,\cdots$, with $q=4$,  and $A_1(\ell)= \, k(\ell)$, 
\be
A(\ell) = 6\, \frac{(n+3/4)^2}{n}\, , \label{eq:A-ell-2}
\ee
which agrees with Eq. (4.23) of \cite{Polchinski:2016xgd}, up to a normalization constant.  Note that $A(\ell)$ is singular at $n=0$, corresponding to $\ell_{eff}=2$.

Since $k(\ell,q)$ is monotonic in ${\rm Re} \,\ell$ and vanishing at ${\rm Re} \,\ell=\infty$,  $\ell_{eff}=2$ is the leading singularity to the right of ${\rm Re}\, \ell=1$. From Eq. (\ref{eq:1D-ImConformalBlock-2}), one has, as $w\rightarrow \infty$,
\be
{\rm Im} \, \Gamma (w) \rightarrow \gamma \,w + O(w^{1-2/q}).  \label{eq:graviton}
\ee
with $\gamma$ given by residue of $ A(\ell) $ at $\ell=2$. 
Corrections to the leading order comes from a singularity at $\ell=1-2/q$, driven by the Born term, ${\rm Im}\, \Gamma_1$.  For the full amplitude, Eq. (\ref{eq:graviton}) corresponds to having a leading behavior
\begin{align}
\Gamma (w )\simeq& - \pi^{-1} \gamma w [\log (1-w) + \log (w-1)] \nn
&+ \gamma{'} w +  O(w^{1-2/q}), \label{eq:full-graviton}
\end{align}
with $\gamma'$ given by residue of $(\ell-2) A(\ell) $ at $\ell=2$. The emergence of log is perhaps puzzling. It is mathematically necessary due to the fact that ${\rm Im} \, \Gamma (w) $ grows linearly with $w$.  This  is also related to the fact that Eq. (\ref{eq:A-ell-2}) is singular at $n=0$.  To clarify this issue further, we turn next to a brief discussion on the continuation into the region $|w|<1$, conformal symmetry breaking, and stringy corrections.

\subsection{Analyticity and Corrections}

Let us end this section with several additional comments.

\paragraph{Hilbert Space Treatment:}

 Our treatment  for ${\rm Im} \, \Gamma (w)$ can be framed in the context of an harmonic analysis over the non-compact group, $SO(1,1)$, as carried out in Appendix  \ref{sec:green}. We begin first with a spectral analysis  over the Hilbert space of square-integrable functions over the interval $ (1,\infty)$,~ Sec. \ref{sec:spectral1}.  The framework is next  extended  to allow functions which are polynomially bounded at $w=\infty$, in Sec. \ref{sec:spectral2}, leading to the representation (\ref{eq:1D-ImConformalBlock}-\ref{eq:1D-ImConformalBlock-2}).  As explained in~\cite{Abarbanel:1970ef,Abarbanel:1971nj,Saunders:1971tz}, this can be regarded as an harmonic analysis by non-unitary representation of $SO(1,1)$. 
 
 Let us contrast our scattering treatment with other related  Euclidean treatments, for example that of \cite{Polchinski:2016xgd,Maldacena:2016hyu}, which can also be framed in a Hilbert space treatment. In \cite{Maldacena:2016hyu}, one first considers the space of functions defined over $0<\tau<2$, which corresponds to $0<w<\infty$. It is then extended to the whole range in $\tau$ by symmetry. With $w=(2-\tau)/\tau$, this corresponds to reflecting $w\leftrightarrow -w$, leading to the whole range $-\infty<w<\infty$.   As a consequence, eigenfunctions contain log-singularities at $\tau=1$ ($w=1$). In contrast, our eigenfunctions, (\ref{eq:ON-w})-(\ref{eq:Completeness-w}), are defined  over the interval $(1,\infty)$.  In order to extend their treatment to include functions which are more singular at $\tau=1$, e.g., $(\tau-1)^{-N}$, $N>0$, ($\sim |w|^N$ as $|w|\rightarrow \infty)$, it is implicit that an extension of the standard  spectral analysis  also has to be  made.  Therefore, our treatment here is in some sense no less general than that carried out in \cite{Polchinski:2016xgd,Maldacena:2016hyu}.
 
\paragraph{Continuation to Euclidean Region:}
 
Given ${\rm Im} \, \Gamma (w)$, for the region $1<|w|<\infty$, let us examine  the specification of  the full amplitude $\Gamma(w)$.  From Eq. (\ref{eq:1D-ConformalBlock-2}), it is possible to extend $\Gamma(w)$ to the complex $w$-plane, as a real analytic function with branch cuts for $1<|w|$. For the case of Eq. (\ref{eq:graviton}), one can initially choose $2<L_0<4$.  In continuing to the region $-1<w<1$,  the contour can be closed to the right, Eq. (\ref{eq:EuclideanPartialWave}), arriving at
\begin{align}
\Gamma(w)&=   \bar A_2\, w + \sum_{ 2<\ell,\,  {\rm even}}  (2\ell-1)\, A(\ell)\, P_{\ell-1}(w)  \nn
&=\bar A_2\, w + \sum_{ 3\leq \bar \ell,\,  {\rm odd}}  (2\bar \ell+1)\, A_{\bar\ell+1}\, P_{\bar \ell}(w) \label{eq:EuclideanPartialWave-2}
\end{align}
In closing the contour, one finds ${\rm Im} \, \Gamma (w)=0$ in the region $|w|<1$, as indicated earlier. In the first line on the right,  $\bar A_2$ a constant, is {\it a priori} unspecified.  The second line is a re-write of the first, with $\bar \ell=\ell-1$ and $A_{\bar \ell+1}$ summed only over $\bar \ell$ odd integers.  This is precisely the ordinary Legendre expansion for square-integrable functions over $(-1,1)$ where $\Gamma(w)=-\Gamma(-w)$.

Under normal circumstance,  the sum over $\bar \ell$ can be extended to $\bar \ell=1$,  with $\bar A_2 = 3A(2)$  given  by the analytic continuation of $A(\ell)$ to $\ell=2$. However, this is not a necessity~\footnote{In a traditional Regge treatment, if this were to happen, it would correspond to the theory not being uniquely defined by the analytic S-Matrix~\cite{Chew},  $A(\ell)$ would contain a Kronecker-delta term, and  the theory would require ``Castillejo-Dalitz-Dyson" (CDD) poles. See \cite{Frautschi} for a historical discussion.}.   For SYK-like models, the situation is more robust since $A(\ell)$ has a pole at $\ell=2$. From Eq. (\ref{eq:1D-ConformalBlock}), there is a double-pole at $\ell=2$, leading to logrithmic behavior. Lastly, we note that Eq. (\ref{eq:EuclideanPartialWave-2}) can again be converted to a symmetric function, $T(w)$ , via Eq. (\ref{eq:tw}).

\paragraph{Conformal Symmetry Breaking:}
The existence of a pole at $\ell=2$, as explained in \cite{Polchinski:2016xgd,Maldacena:2016hyu}, is due conformal invariance, and it corresponds to the existence of a goldstone mode. For SYK models, this mode is unphysical and should be removed. There are several scenarios to consider. 
 One possibility is simply to define the theory with this mode removed. For example, in Eq. (\ref{eq:EuclideanPartialWave-2}), setting $\bar A_2=0$, as is done in \cite{Polchinski:2016xgd}.  In this case, it is possible to reverse the procedure in re-summing Eq. (\ref{eq:EuclideanPartialWave-2}) for the limit $w\rightarrow \infty$, leading to Eqs. (\ref{eq:graviton}) and (\ref{eq:full-graviton}) with $\gamma=0$. Analytically, this can be accomplished by introducing an extra zero to $A(\ell)$, e.g., $A(\ell) \sim \ell-2$, leading to $\gamma=0$ and $\gamma'\neq 0$. This will not alter the eigenvalue condition for $k(2,\delta)=1$. 
As a consequence, this ``weak breaking" scenario does note alter  the feature that $\ell_{eff}=2$; it leads to a situation where  the log term in Eq. (\ref{eq:full-graviton}) is removed.     A stronger modification to the kernel is required in order to change $\ell_{eff}$ from 2.

\paragraph{Stringy Corrections:} On the other hand, the model can be embellished   by considering stringy corrections as discussed in~\cite{Maldacena:2016hyu}, and, more generally, in ~\cite{Shenker:2014cwa}.   In this case, one has $\ell_{eff}<2$, which leads to
\begin{align}
\Gamma (w) \rightarrow \gamma'\, [&-(1-w)^{\ell_{ell}-1}+  (w-1)^{\ell_{ell}-1}] \nn
&+ O(w^{1-2/q}),
\end{align}
as $w\rightarrow \infty$. It follows that  the log-term in Eq. (\ref{eq:full-graviton})  is again removed. In analogy with ${\cal N}=4$ YM, it is tempting to refer to this singularity as due to stringy corrections, as the 1-d Pomeron~\cite{PomeronStory}. However, as discussed in the next section, thermal effects must be taken into account.

\section{Discussion and Summary}\label{sec:discuss}
In this paper we have focused on scattering in CFTs, for example off-shell photon-photon scattering, through an OPE with Minkowski conformal blocks. We review the major results here and discuss future connections and applications.

This paper consists of three main components. The first part, directly defining and calculating MCB, is shown in Secs. \ref{sec:MCB}.  MCB are solutions to the quadratic Casimir for the product of scalar conformal primaries.  We show how  ``scattering along light-cones", also realized by taking a double light-cone limit as discussed in Sec. \ref{sec:DLC},  selects a natural basis for MCB satisfying a different set of boundary conditions from that for Euclidean conformal blocks.  Due to the difference in boundary conditions, MCB are not given by a direct analytic continuation but only related to the ECB.
 
In an Euclidean setting, the DLC limit involves a single scale, dilatation, leading to a  single scaling limit.  In a Minkowski setting, there are two scaling limits: dilatation and boost.  The  dilatation limit is  characterized by a parameter  $\sigma\rightarrow \infty$,  and the second scaling limit,  $w\rightarrow \infty$, is analogous to a Lorentz boost.  Dilatation leads to scaling dependent on the conformal dimension $\Delta$, Eq. (\ref{eq:Mbdry-1}), while the Lorentz boost leads to a dependence on an effective spin $\ell_{eff}$,  Eq. (\ref{eq:PomeronIntercept}).  When a 4-point correlator  is expressed in terms of invariant cross ratios, for example $F(u,v)$ in Eq. (\ref{eq:A(X)}), these two scaling limits allow one to explore the consequence of the residual symmetry, $O(1,1)\times O(1,1)$. 
  
The second part deals with the application of Minkowski OPE for formal scattering processes. We explain how a t-channel OPE leads to a description for s-channel scattering amplitudes in terms of contributions from t-channel conformal primaries. In the DLC limit, the leading contribution comes from the stress-energy tensor.   For a realistic phenomenological treatment, stringy, and sometimes thermal, corrections must be included. This requires summing contributions from the t-channel OPE via the Sommerfeld-Watson transformation. This is discussed in Sec. \ref{sec:application-I}, leading to a complex $\ell$-plane representation, (\ref{eq:newgroupexpansionDisc}), as well as  a double-Mellin representation in complex-$\Delta$ and complex-$\ell$, Eq. (\ref{eq:trueinvariance}).    

The third part turns to specific applications.  The simplest phenomenological application, which directly explores the residual $SO(1,1)\times SO(1,1)$ symmetry, is Deep Inelastic Scattering (DIS). Both formal and phenomenological aspects, like the BFKL program, are summarized in Appendix \ref{sec:application-II}. In Sec. \ref{sec:CFT-1} we focus on scattering for 1-d SYK-like models.  We show how MCB can be used to simplify the dynamics for these models and how the leading effective spin, associated with the chaos bound, can be identified.  

We conclude here with a more detailed discussion on how stringy correction to $\ell_{eff}$ for SYK-like models can be framed through AdS/CFT along the formalism introduced in~\cite{Brower:2006ea,Brower:2007qh,Brower:2007xg}.  This issue has been previously addressed in \cite{Shenker:2014cwa} and \cite{Maldacena:2016hyu,Murugan:2017eto}. We add further discussion by exploring the consequences of  $SO(1,1)\times SO(1,1)$ symmetry via the spectral curve, $\Delta (\ell)$, for the leading twist conformal primaries interpolating the stress-energy tensor.  This was explored in Sec. \ref{sec:application-I}.

The importance of spectral curve for scattering in CFT can best be illustrated via DIS.  It is well known that anomalous dimensions of the leading twist-2 conformal primaries of conformal dimension $\Delta$ and spin $\ell$,  ${\cal O}_{\Delta,\ell}$, controls  the large $q^2$ dependence for the moments of structure functions. For example, for $F_2(x,q^2)$,  for $q$ large,
$M_\ell=\int_0^1 x^{\ell-2}F_2(x,q^2) dx \sim q^{-\gamma_\ell}$, as in Eq. (\ref{eq:f2DGLAP}), where $\gamma_\ell =\Delta(\ell) - \ell-d/2$. Due to crossing symmetry, only even $\ell$ enters. (See Appendix  \ref{sec:application-II}  for kinematic details.)  The positivity constraint, Eq. (\ref{eq:positivity}), for general $d$,  leads to $\Delta(\ell))\geq d/2$. The second scaling is related to the limit $x\rightarrow 0$ ($E_\gamma \rightarrow \infty$). Again, for $F_2$, this leads to Eq. (\ref{eq:BFKL-Regge2}). The effective spin can be found by solving an eigenvalue condition Eq. (\ref{eq:PomeronIntercept})~\cite{Brower:2006ea,Brower:2007xg}, which can be expressed more explicitly as $\Delta (\ell_{eff})=2$,  which saturates the positivity bound.

It is useful to provide  additional  discussion on the importance of the spectral curve,  $\Delta (\ell)$, for the leading twist conformal primaries interpolating the stress-energy tensor. This is most illuminating in the context of the AdS/CFT. In Sec. \ref{sec:SpectralCurve},  we have shown how, for string theories, world-sheet conformal invariance can be enforced by $L_0=\bar L_0=1 $.  This constraint  can be enforced by performing a spectral analysis for the propagator $G=\frac{\delta_{L_0,\bar L_0}}{L_0+\bar L_0-2}$, from which $\Delta(\ell)$ can be extracted. Due to conformal invariance, it is  symmetric under 
\be
\Delta(\ell) \leftrightarrow d-\Delta(\ell)\, ,  
\ee
for general d. (See Fig. \ref{fig:BFKLDGLAP}, for $d=4$.) In terms of the Poincare patch, this follows simply from $z\leftrightarrow z^{-1}$  symmetry.  
 
In flat space, this propagator, in a momentum space representation, leads to the Regge trajectory interpolating the graviton, $G(t, \ell) =  \frac{1}{\ell-2- (\alpha'/2)\, t}\,$, i.e. $\ell(t) = 2+\alpha'  t/2$, with mass-shell condition corresponding to poles for even $\ell=2n$, $n=1,2, \cdots$. Consider next strings propagating over $AdS_5$. In the weak curvature limit, this turns $G(\ell)$ into, as a differential operator~\cite{Brower:2006ea},
$
G(\ell) = \frac{1}{ \ell - 2 +  (\alpha' /2R^2_{ads}) \nabla^2(\ell)} 
$,
 where $\alpha' /R^2_{ads}=1/\sqrt \lambda$ and $\nabla^2(\ell)$ is the tensor Laplacian. For $\ell\simeq 2
 $,  after a similarity transformation, $\nabla^2(\ell)$ reduces to the scalar Laplacian, $\nabla_0^2$, with eigenvalue  $(d/2)^2+\nu^2$, $-\infty<\nu<\infty$, leading to a spectral representation in $\ell$,  
    \be
G(\ell)=\int\limits^\infty_{-\infty} \frac{d\nu}{2\pi i} \frac{|\psi(\nu\rangle \langle \psi(\nu)| }{\ell-2 + (1/2\sqrt \lambda) (\nu^2+ d^2/4)}
\ee
This corresponds to a continuous spectrum, $-\infty<\ell< \ell_{eff}$, with $\ell_{eff}=2- \frac{d^2}{8\sqrt \lambda}$, i.e., a branch cut in the complex $\ell$-plane over $(-\infty,\ell_{eff})$. With $\Delta=d/2 +i\nu$, this corresponds to the desired spectral curve~\footnote{For ${\cal N}=4$ SYM, by taking advantage of integrabililty, Eq. (\ref{eq:Delta-j-analyticity}) represents a systematic expansion at strong coupling and low spin.  Using this approach it has been possible to calculate the Pomeron and Odderon, associated with the anti-symmetric tensor, $B_{\mu\nu}$, intercepts to several high orders  in $1/\sqrt \lambda$~\cite{Brower:2014wha}.}, i.e., for $d=4$, Eq. (\ref{eq:Delta-j-analyticity}), with $B(\lambda, \ell) \simeq \sqrt 2\lambda^{1/4}$. More explicit spectral analysis can be carried out in a momentum representation, with $t=-(p_1-p_2)^2$, given by (\ref{eq:ell-spectral}).   

An alternative spectral representation in $t$ can also be obtained in terms of regular Bessel function, (\ref{eq:t-spectral}). At integral $\ell$,  they correspond to momentum space representation of bulk-to-boundary propagators. In the DLC limit where $p_1-p_2\simeq q_\perp$ is asymptotically transverse, performing a 2-dim Fourier transform, one finds, up to a factor of $(z\bar z)^2$ in reducing to scalar propopagator,
\be
G(\cosh \xi, \ell) =  \frac{e^{- (\Delta(\ell) - 2)\xi}}{\sinh \xi}
\ee
with $\cosh \xi =( z^2+ \bar z^2 + b_\perp^2)/2 z \bar z$, i.e., a formal solution expressible in terms of geodesic $\xi$ on $H_3$.  At $\ell=2$ and $d=5$, other than the extra factor $(z\bar z)^2$, this is nothing  but the scalar $AdS_3$ bulk-to-bulk propagator. 

Let us turn next to thermo-CFT-correlators and treat it similarly via AdS/CFT by considering a black-hole background. For $d\geq 3$ we can write the metric as \cite{Maldacena:2001kr}, 
\begin{align}
ds^2 = \frac{R_{eff}^2}{z^2} [& (1-z^{d+1})d\tau^2  + \sum^d_{i=1}dx_i^2 \nn
&+\quad  (1-z^{d+1})^{-1}dz^2]
\end{align}
where  we have scaled the horizon to $z=1$.  A similar spectral analysis in  $t$ and in $\ell$ can be carried out, extending the treatment of \cite{tanglueball}. Interestingly, the spectral in $t$ is now discrete, with $t=m_n^2>0$, $n=0, 1,\cdots$, and is analytic for $t<0$,
  \be
 G(t, \beta,z,\bar z,\ell) = \sum \frac{\Psi_n(z,\ell) \Psi_n^*(\bar z, \ell)}{m_n^2(\ell)-t}.\label{eq:thermo-glueball}
 \ee
For $\ell=2$, these correspond to tensor glueballs calculated in \cite{tanglueball}. There is  a finite mass gap, with $m_0^2>0$. 

For $t>0$, the spectrum in $\ell$ is also discrete for $\ell>0$, which, when combined with Eq. (\ref{eq:thermo-glueball}), leads to Regge trajectories, $\ell_n(t)$, $n=0,1,\cdots$.   
  However, for $t<0$, the t-dependent term in the Laplacian turns repulsive, and  the spectrum in $\ell$ is continuous. This leads to branch cut at $(-\infty,\ell_{eff})$,
\be
\ell_{eff} = 2- (d/2)^2 /2\sqrt {\lambda_{eff}}
\ee
where we have introduced an effective 't Hooft coupling, $\lambda_{eff} \equiv (R_{eff}/\ell_{string})^4$.   (This branch cut also persists for $t>0$.) 
A similar expression for $\ell_{eff}$ has also been arrived in \cite{Shenker:2014cwa} and \cite{Maldacena:2016hyu}.   In particular, \cite{Maldacena:2016hyu} finds that $\lambda_{eff}$ should  be temperature dependent.  We defer to a future study on how our analysis can be framed accordingly \cite{future}.

We end by pointing out that we have focused in Sec. \ref{sec:CFT-1}, on the large $w$ behavior for ${\rm Im}\, \Gamma$, with the real part given by dispersion relation. However, in ``lifting" the model to higher dimensions, the relation between the real part and  the imaginary part becomes more complex. In \cite{Brower:2007xg}, in the Regge limit, it was pointed out that the limit of large $s$ and large impact parameter, $b$, do not commute. Our study here takes the limit  $s$ large before $b$ is allowed to be large, corresponding to the limit $u\rightarrow 0$ with $(1-v)/\sqrt u$ initially fixed. It is nevertheless interesting to examine the large $s$ limit, but fixed, and taking $b$ large. In that case, one regains the single gravity exchange, with a cutoff in impact space controlled by the lowest tensor glueball mass, $m_0$, 
\be
 G(\vec b,z,\bar z,\ell=2)\sim e^{-m_0 |\vec b|}
 \ee
 with $m_0$ scaled by the inverse temperature $\beta$~\cite{Brower:2007xg}.
 It is equally important to recognize that this mass does not directly determine~\footnote{For a possibly different perspective, see \cite{Shenker:2014cwa}.} the stringy correction  to $\ell_{eff}$. 

\paragraph{Acknowledgments:} We would like to thank Antal Jevicki for useful discussions and for comments on an early draft of this paper.  We would also like to thank the Brown High Energy Theory group for comments on an early version of this work presented in seminar. C-I. T. would like to thank  J. C. Lee and Y. Yang of National Chiao-Tung University (NCTU), Taiwan,  and its Yau Center, for hosting and discussion where this work began in earnest during recent visits.  He is particularly indebted to J. Polchinski, as well as R. Brower and M. Strassler for collaborating on \cite{Brower:2006ea}, which shaped his interest in scattering in AdS/CFT.  He owes special thanks to R. Brower for collaboration and encouragement on issues relating to CFT in general and AdS/CFT in particular. T.R. is supported by the University of Kansas Foundation Distinguished Professor starting grant of Christophe Royon. Research for C-I. T. is supported in part  by the Department of Energy under
contact DE-Sc0010010-Task-A and Brown University research funds. 

\vskip30pt

\appendix
\section{More on DLC Kinematics}\label{sec:kinematics}
Here we provide more detail on many of the specific kinematic details, notations, and conventions, that are used throughout the main text.

\subsection{Channels}\label{sec:channels}

\begin{figure}[!ht]
\begin{center}
     \includegraphics[scale=.25]{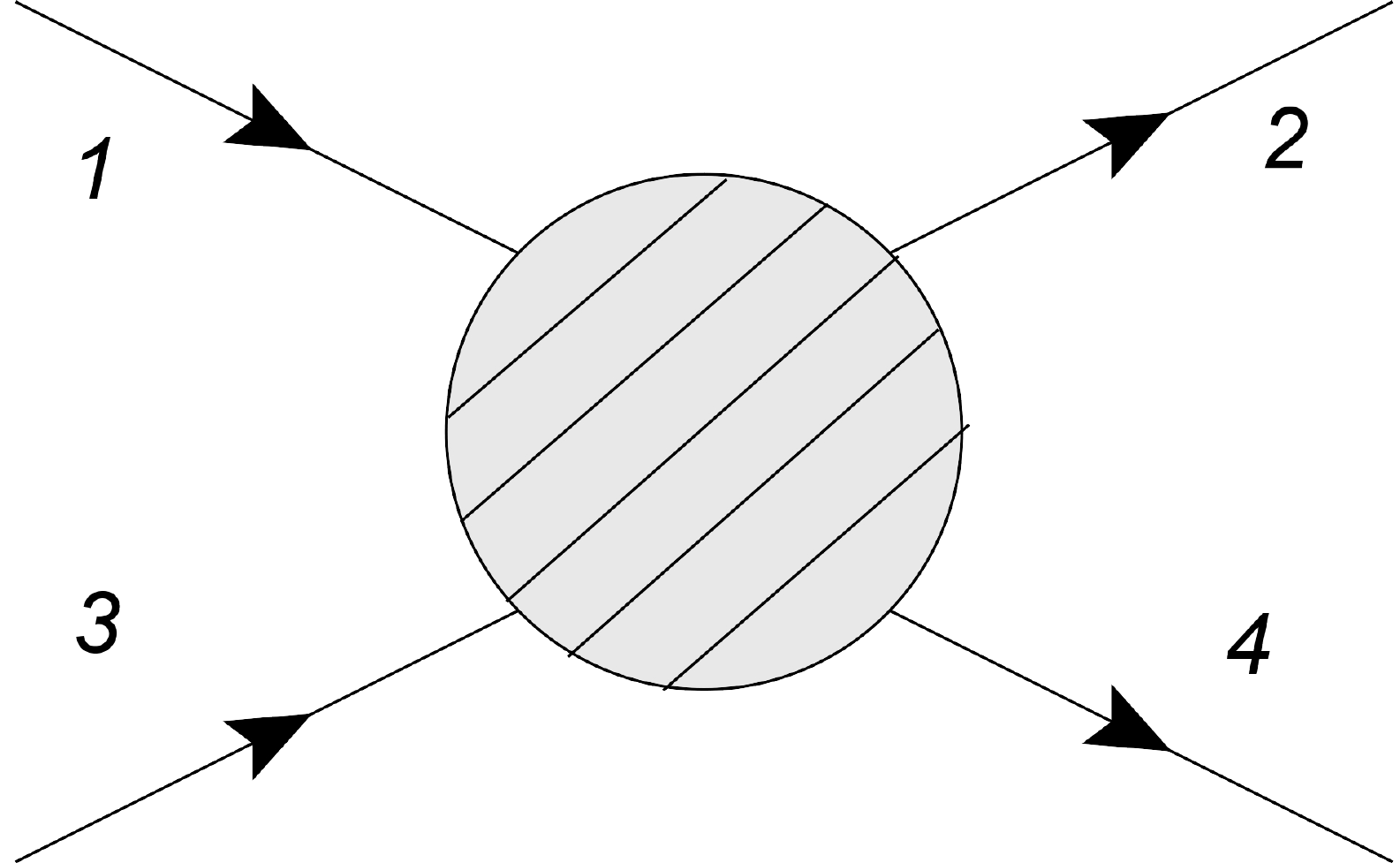}
  \hspace{15pt}
     \includegraphics[scale=.25]{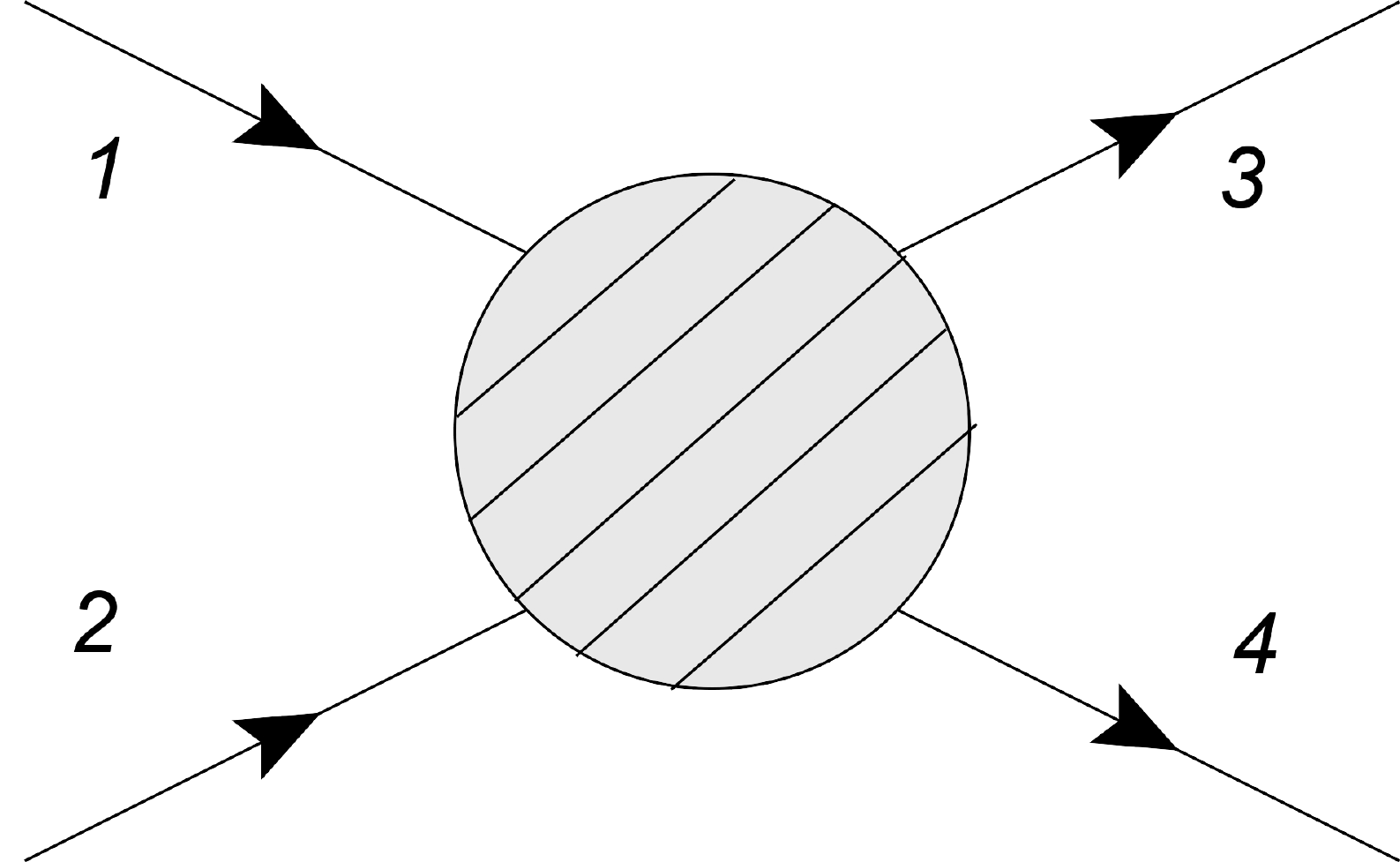}
       \hspace{15pt}
     \includegraphics[scale=.25]{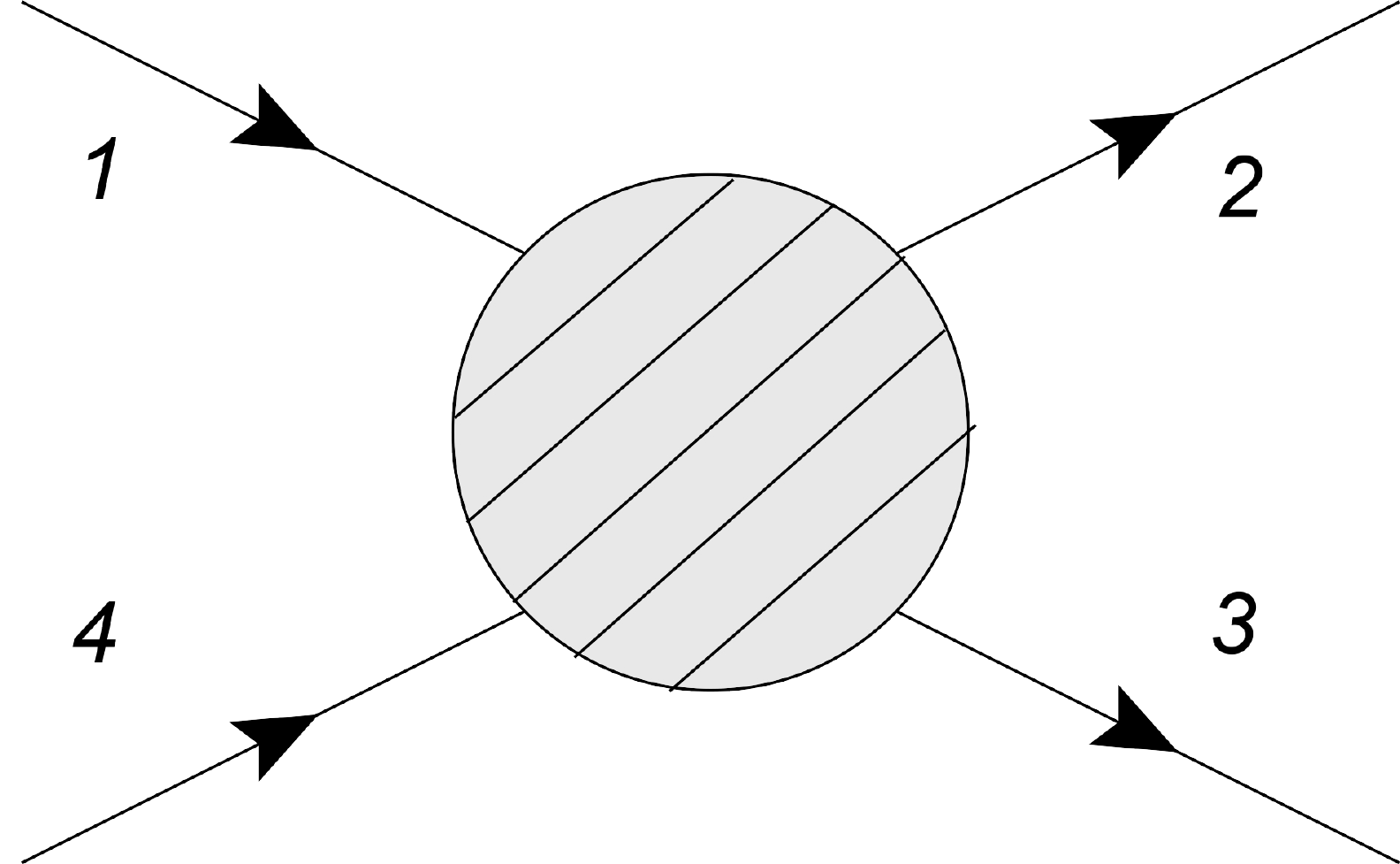}
\end{center}
\caption{\label{fig:chan-proc}(top) s-channel scattering, (middle) t-channel scattering, and (bottom) u-channel scattering.}
\end{figure}

\begin{figure}[!ht]
\begin{center}
     \includegraphics[scale=.25]{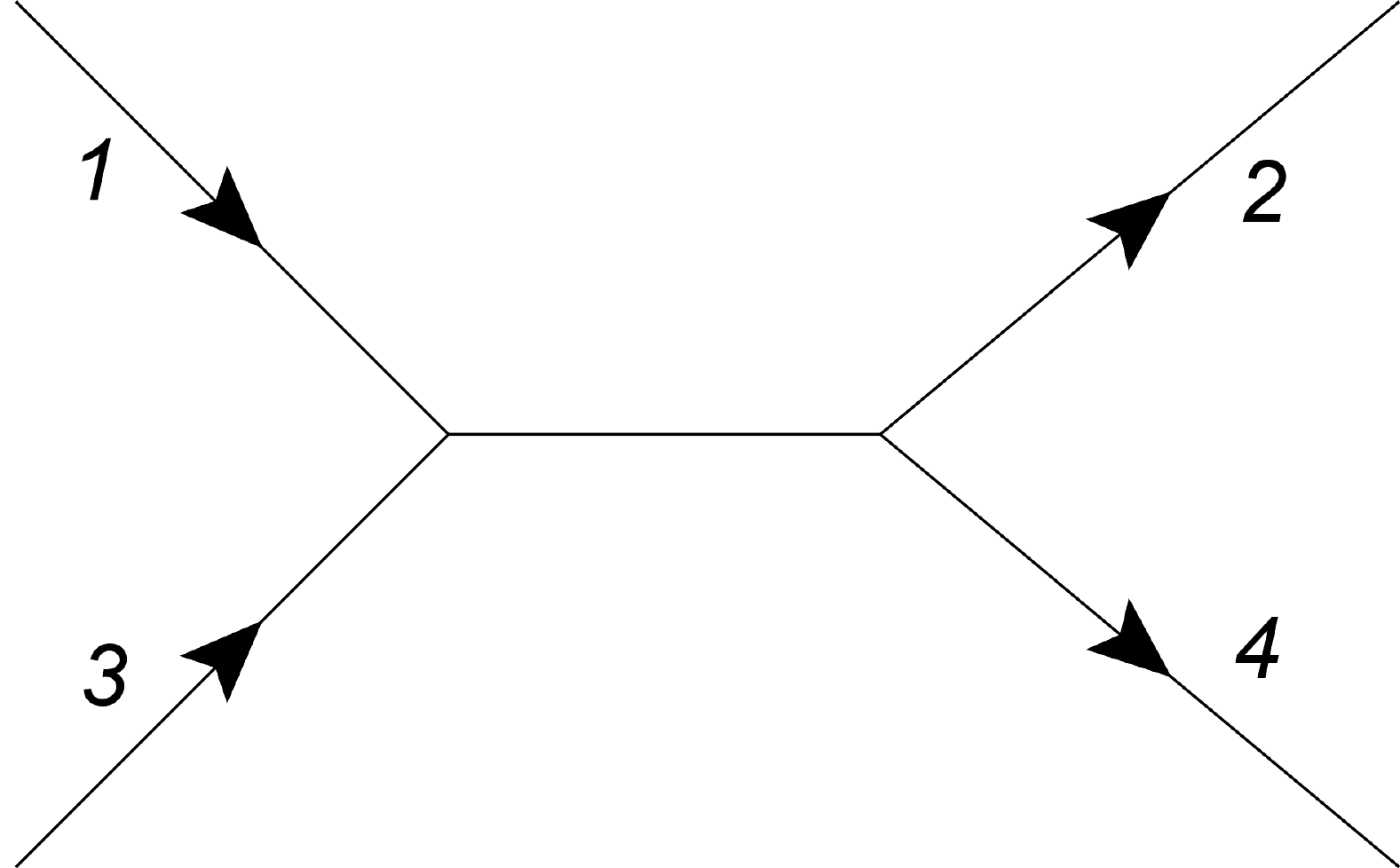}
  \hspace{15pt}
     \includegraphics[scale=.25]{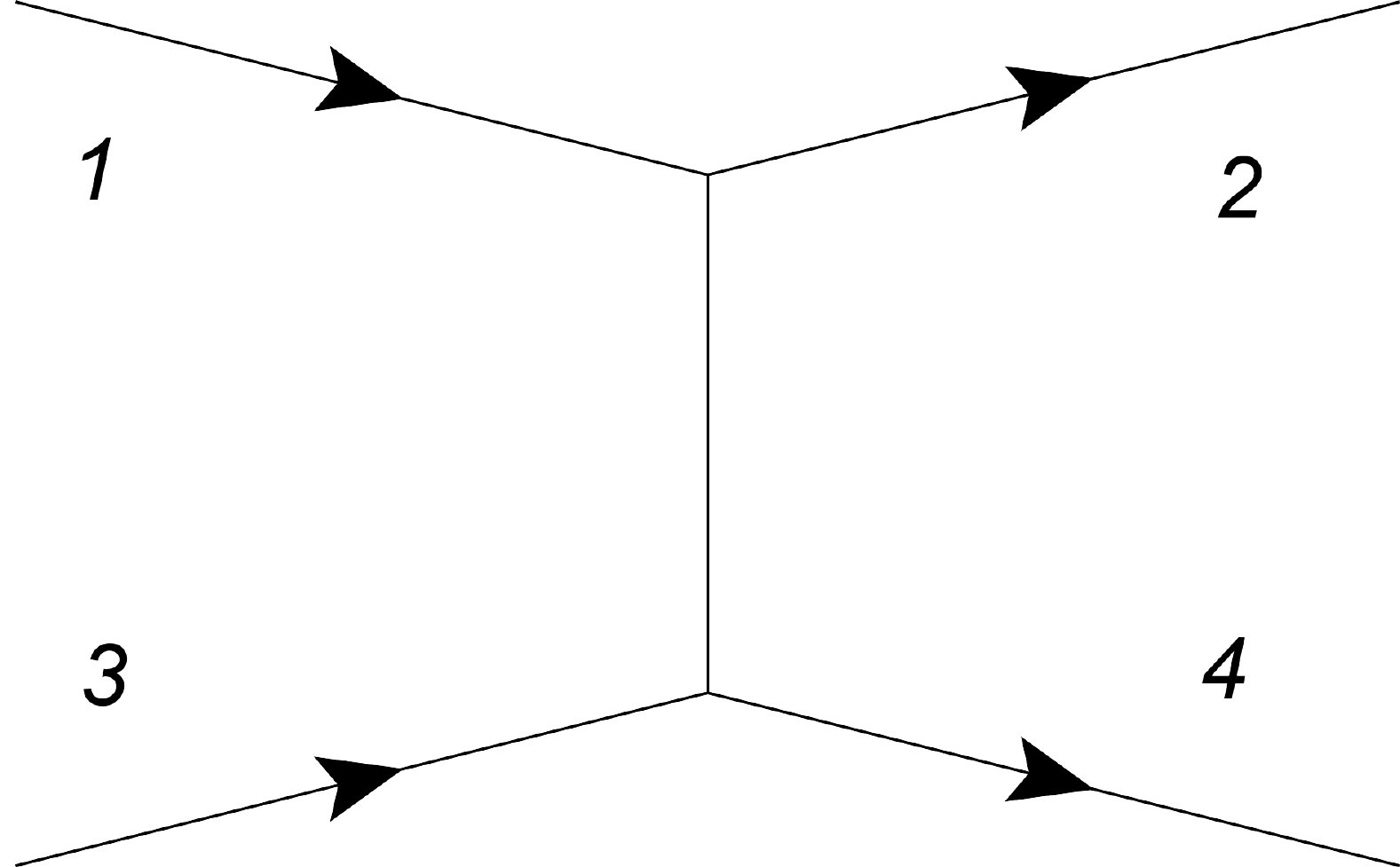}
       \hspace{15pt}
     \includegraphics[scale=.25]{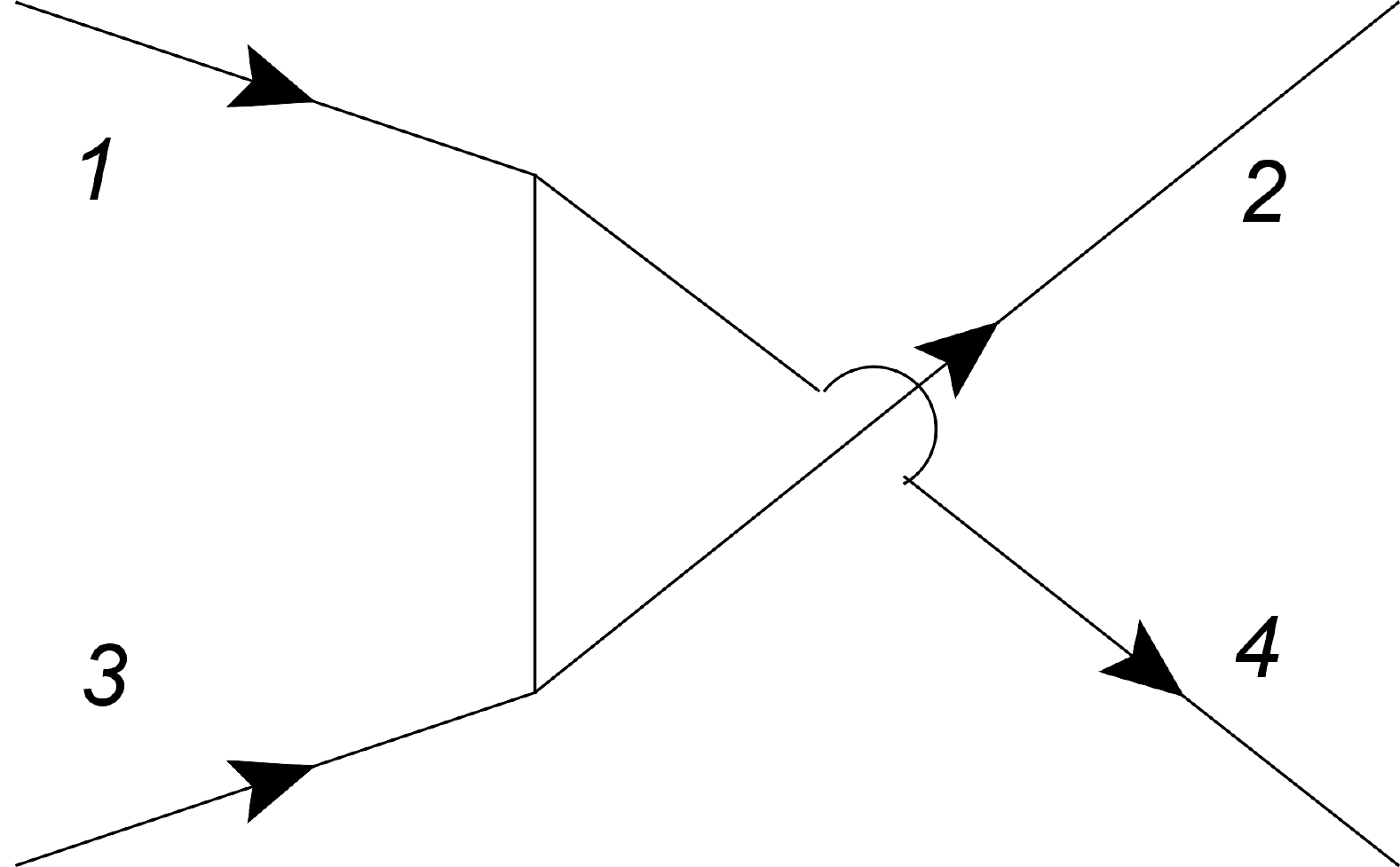}
\end{center}
\caption{\label{fig:sOPEs} s-channel scattering involving (top) an s-channel OPE, (middle) a t-channel OPE, and (bottom) a u-channel OPE.}
\end{figure}

\begin{figure}[!ht]
\begin{center}
     \includegraphics[scale=.25]{s-proc-t-OPE.pdf}
  \hspace{15pt}
     \includegraphics[scale=.25]{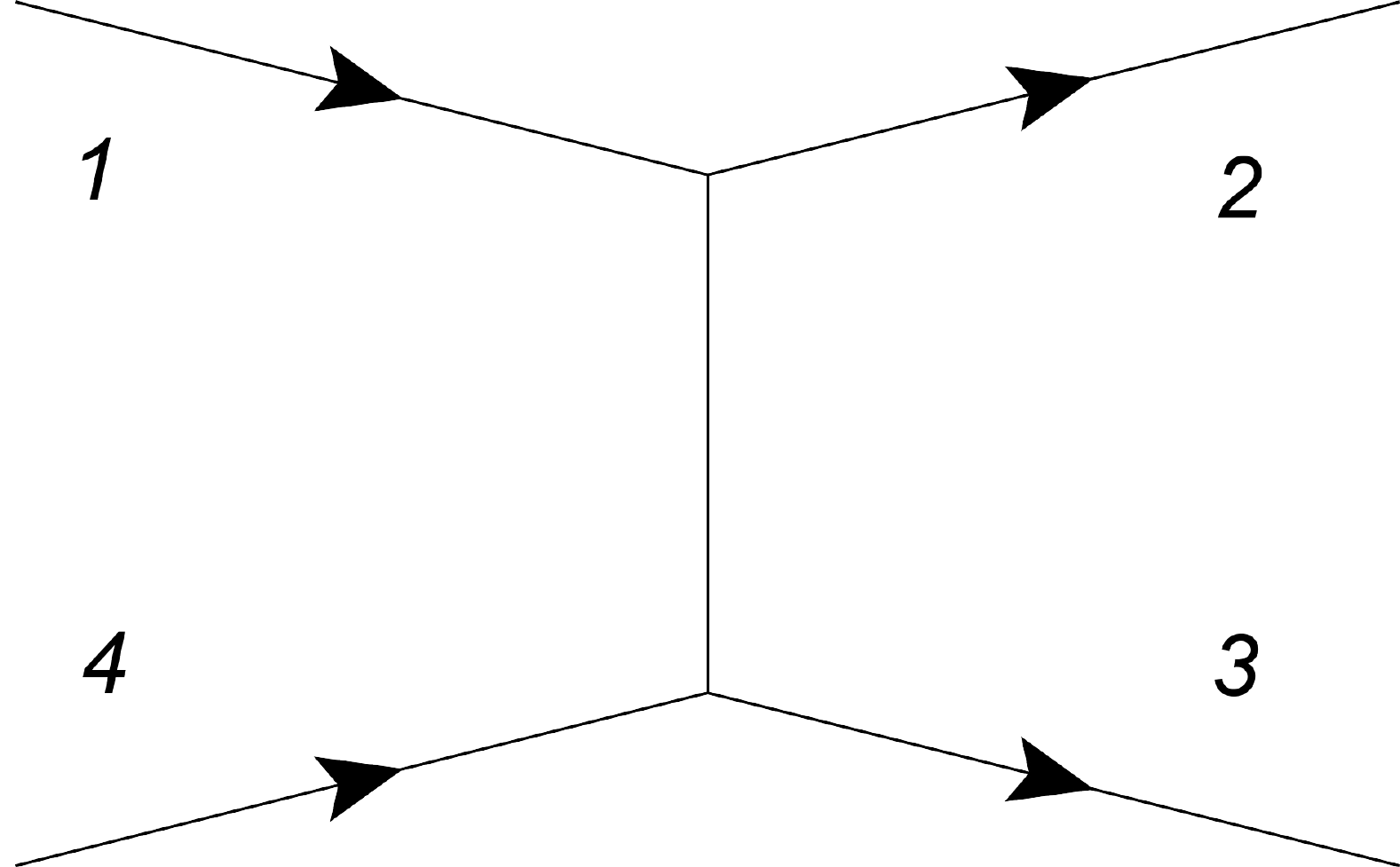}
\end{center}
\caption{\label{fig:su-tOPE}The t-channel OPE for the (top) s-channel scattering process and (bottom) u-channel scattering process are related.}
\end{figure}

An unfortunate side effect of a long history of literature is that similar syntax confusingly gets used to refer to different things. In an effort to clear up confusion, in this brief appendix we would like to highlight two different uses of the \emph{channel} of a scattering process that are used in the text.  We refer to the channel of a scattering process, for example s-channel scattering, as a reference to incoming and outgoing particles. Once a scattering channel  is defined,  the rest follow.  This can be seen in Fig.\ref{fig:chan-proc}. A single scattering process can be written in terms of different OPE combinations as seen in Fig.\ref{fig:sOPEs}.  Note that in a CFT, these contributions are \emph{not} summed; a single OPE prescription describes the entire correlation function.  Finally we note that the s-channel process and u-channel process can involve a similar t-channel OPE structure as in Fig.\ref{fig:su-tOPE}.

\subsection{Rindler Coordinates for the DLC Limit}\label{sec:DLC-append}

In this Appendix, we review the Rindler-like hyperbolic coordinate parametrization which is used extensively in Sec. \ref{sec:DLC}. We consider here s-channel scattering approaching the DLC limit.

Consider  LC coordinates, $ x=(x^+,x^-; x_\perp)$, $x^\pm = x_0\pm x_L$. Our convention for different time signatures is

\be
x\cdot y = 
	\begin{cases} &-\frac{1}{2}(x^+y^- + x^-y^+) + x_{\perp}^2 \quad \text{Minkowski}\\
&\,\,\,\, \frac{1}{2}(x^+y^- + x^-y^+) + x_{\perp}^2 \quad \text{Euclidean.}
	\end{cases}
\ee

For the Minkowski case, $x$ space-like implies  $x^2=-x^+x^-+ x_\perp^2>0$ and $x^+x^-<0$. 
For a 4-point Minkowski correlator, LC components can be parameterized by a conformal virtuality, Eq. (\ref{eq:virtuality}), and conformal rapidity, Eq. (\ref{eq:boostparam})  
 \begin{align}
 x_1=&( -r_1 e^{y_1}, r_1 e^{-y_1}; x_{1,\perp})\, ,\nn
 x_2=&( r_2 e^{y_2}, -r_2 e^{-y_2}; x_{2,\perp} )\, ,\nn
  x_3=&( r_3 e^{-y_3}, -r_3 e^{y_3}; x_{3,\perp})\, ,\nn
   x_4=&( -r_ 4 e^{-y_4}, r_4 e^{y_4}; x_{4,\perp} )\, , \label{eq:Melner}
  \end{align}
where the time and longitudinal coordinates are explicitly, 
\begin{align}
x_1^0 &= - r_1 \sinh y_1, \quad &x_1^L = - r_1 \cosh y_1, \nn
x_2^0 &=  r_2 \sinh y_2, \quad &x_2^L = r_2 \cosh y_2, \nn
 x_3^0 &= - r_3 \sinh y_3, \quad &x_3^L = r_3 \cosh y_3, \nn x_4^0 &=  r_4 \sinh y_4, \quad &x_4^L = - r_4 \cosh y_4. \label{eq:Melner2}
\end{align}
Sending all rapidities $y_i\rightarrow \infty$, with $x_{i\perp}$ fixed,  leads to (\ref{eq:DLC2}).

\subsubsection{Relating to Invariant Cross Ratios for General \texorpdfstring{$d$}{d}:}
For invariants, we have, for general $d$, with transverse coordinates $x_{\perp,i}$ kept for $d> 2$,
\begin{align}
u&=\frac{x_{12}^2x_{34}^2}{x_{13}^2x_{24}^2} \nn
&= \frac{[\cosh y_{12} + R(1,2)] [\cosh y_{34} + R(3,4)]}{ [ \cosh \bar y_{13}  + R(1,3) ][  \cosh \bar y_{24}  + R(2,4)  ]  }\\
v&= \frac{x_{23}^2x_{14}^2}{x_{13}^2x_{24}^2} \nn
&=\frac {[ \cosh \bar y_{23}  -R(2,3) ][  \cosh \bar y_{14}  -R(1,4)  ]  }{[\cosh \bar y_{13} + R(1,3)] [\cosh \bar y_{24} + R(2,4)]}\label{eq:u-v-local}
\end{align}
where 
\begin{align}
y_{ij}&=y_i-y_j\, ,\quad \bar y_{ij}=y_i + y_j\, ,\quad b_{ij}=x_{\perp i}-x_{\perp j}\, ,\nn
R&(i,j)=\frac{r_i^2+r_j^2+ b_{ij}^2}{2r_ir_j}\, . 
\end{align}
To simplify the discussion, we will  adopt a frame where $x_{i,\perp}=x_{2,\perp}$ and $x_{3,\perp}=x_{4,\perp}$, i.e., $b_{12}=b_{34}=0$,  with $b_{\perp}= x_{1,\perp}-x_{3\perp}$ as the relative separation between $(1,2)$ and $(3,4)$ in the transverse impact space.

In terms of the global rapidity $y$ and virtuality $r_i$ for each coordinate, cross ratios $u$ and $v$ take on relatively simple forms, 
For all $y_i$ taking on a global value $y$,   thus $y_{ij}=0$ and $\bar y_{ij}=2y$, 
\begin{align}
u&=\frac{x_{12}^2x_{34}^2}{x_{13}^2x_{24}^2} \nn
&= \frac{[( r_1+r_2)^2/(r_1r_2)]}{   (e^{2y} +2R(1,3) + e^{-2y})]  },\nn
&\qquad \times\frac{[(r_3+r_4)^2 /(r_3r_4)]}{[  (e^{2y} +2R(2,4) + e^{-2y}) ]} \label{eq:u-global}\\
&\simeq  ( \sqrt{r_1}+\sqrt{r_2})^2 (\sqrt{r_3}+\sqrt{r_4})^2 \,  e^{-2y} +  O(e^{-4y}),\\
v&= \frac{x_{23}^2x_{41}^2}{x_{13}^2x_{42}^2} \nn
&= \frac{[   e^{2y } -2R(1,3) + e^{-2y}] }{[ e^{2y} +2R(2,3) + e^{-2y}]}, \nn
&\qquad \times \frac{[   (e^{2y} -2R(2,4) + e^{-2y}) ]}{[e^{2y} +2R(1,4) + e^{-2y})]} \label{eq:v-global}\\
&\simeq 1 +  O(e^{-2y}),
\end{align}
where the transverse separation enters through $R(i,j)$. 
  In the case of two pairs of equal virtuality, $r_1=r_2$ and $r_3=r_4$, these can further be simplified, leading to (\ref{eq:uvrange}).
The limit $u\rightarrow 0$ can therefore be achieved either by $y\rightarrow \infty$ or $r_i\rightarrow 0
$, or by $b_\perp^2 \rightarrow \infty $ first. For near-forward scattering, or the DLC limit, we consider the first scenario of $y\rightarrow \infty$ with $r_i$ and $b_\perp^2$ fixed.  The limit $u\rightarrow 0$ therefore exploits the scaling limit of Lorentz boost.

It is straight forward to calculate the combination $\frac{1-v+u}{2\sqrt u}$. The physical region is constraint to satisfy
\be
1-v+u\geq 2 \sqrt u  \label{eq:bound}
\ee
As an illustration, we consider the limit where $r_1=r_2$ and $r_3=r_4$, where one finds
\bea
\frac{1-v+u}{2\sqrt u}&=&\frac{ 2 \cosh (2y) R(1,3) +4}{2\cosh (2y) + R(1,3)},
\eea
leading to (\ref{eq:sigma}) in the limit $y\rightarrow \infty$.

\subsubsection{\texorpdfstring{$d=1$}{d=1} }\label{sec:d1}

By keeping only time components $x^{(0)}_i$, from (\ref{eq:Melner2}), one has
for cross ratios, 
\begin{align}
u&=\frac{x_{12}^2x_{34}^2}{x_{13}^2x_{24}^2}\nn
&= \frac{(r_1\sinh y_1+r_2 \sinh y_2)^2 }{r_1\sinh y_1- r_3 \sinh y_3)^2 }\, ,\nn
&\qquad\qquad \times \frac{(r_3\sinh y_3+r_4 \sinh y_4)^2 }{(r_2\sinh y_2- r_4 \sinh y_4)^2}\nn
v&=\frac{x_{23}^2x_{41}^2}{x_{13}^2x_{42}^2} \nn
&= \frac{(r_2\sinh y_2+r_3\sinh y_3)^2 }{(r_1\sinh y_1- r_3 \sinh y_3)^2 }\, \nn
&\qquad\qquad \times \frac{(r_1\sinh y_1+r_4 \sinh y_4)^2 }{(r_2\sinh y_2- r_4 \sinh y_4)^2 }\, .
\end{align}
It is easy to check that equality for (\ref{eq:bound}) holds, leading to $(1-\sqrt u)^2=v$, or $1=\sqrt u\pm \sqrt v$. Therefore, only one cross ratio is independent.

Since each coordinate now has only a single component, we consider cross ratios defined in (\ref{eq:tau1}). From (\ref{eq:Melner2}), keeping parametrization for   $t_i = x^{(0)}_i$,  one has
\begin{align}
\tau&=\frac{t_{21}t_{43}}{t_{23}t_{41}} \nn
&=  \frac{(r_1\sinh y_1+r_2 \sinh y_2) }{(r_2\sinh y_2+r_3\sinh y_3) }\, , \nn
&\qquad\qquad \times \frac{(r_3\sinh y_3+r_4 \sinh y_4)}{(r_1\sinh y_1+r_4 \sinh y_4)} \, , \nn
\tau_c&=\frac{t_{13}t_{42}}{t_{23}t_{41}}\nn
&=  \frac{(r_1\sinh y_1- r_3 \sinh y_3)}{(r_2\sinh y_2+r_3\sinh y_3) }\nn
&\qquad\qquad \times \frac{(r_2\sinh y_2- r_4 \sinh y_4) }{(r_1\sinh y_1+r_4 \sinh y_4)} \, .
\end{align}
One easily checks that
\be
\tau + \tau_c =1\, . \label{eq:1dConstraint}
\ee
As expected,  there is only one independent cross ratio.  Lastly, for Eq. (\ref{eq:w-variable}), one has
\begin{align}
&w= 1 +\nn
&+ 2\frac{(r_3\sinh y_3-r_1 \sinh y_1) (r_4\sinh y_4-r_2 \sinh y_2) }{(r_2\sinh y_2+r_1 \sinh y_1) (r_4\sinh y_4+r_3 \sinh y_3)}
\end{align}
For either $t_3<t_1<t_2<t_4$ or $t_1<t_3<t_4<t_3$, one has $1<w<\infty$.

\section{Scattering in CFT and Holography}\label{sec:holography} 

In order to infer the desired boundary condition for Minkowski conformal blocks,  (\ref{eq:Mbdry}), we need to consider scattering amplitudes at high energy and large but fixed impact separation. The amplitude,  in a ``shock-wave"~\footnote{The shockwave set up was originally formulated by 't Hooft and Dray \cite{DRAY1985173,Dray1985}.} treatment, can be characterized by an eikonal phase,   $\chi(s,\vec b,z_{12}, z_{34})$.    At a large  impact separation, $\chi(s,\vec b,z_{12}, z_{34})$ is small and can be treated perturbatively, i.e.,  Eq. (\ref{eq:T-chi}).
This representation can also be interpreted holographically as scattering in the AdS bulk~\cite{Brower:2006ea,Brower:2007qh,Brower:2007xg,Cornalba:2006xm}. Here we provide a short summary, following that done in \cite{Brower:2014wha}, with some notational changes adopted in the current paper.

\subsection{Impact Parameter Representation and Holography}
 
Consider the Fourier transform of the connected correlation function   defined in (\ref{eq:A(X)}),
\begin{align}
&\left\langle  {\cal O}_1(p_1) {\cal O}_2(p_2) {\cal O}_3(p_3) {\cal O}_4(p_4) \right\rangle_c\, \nn
&=\, (2\pi)^4\, \delta^{(4)}\!\left( \sum p_j \right) i\, T(p_1, p_2, p_3, p_4).
\end{align}
The  amplitude $T(p_j)$  can be expressed as a function of Mandelstam
invariants $s$, $t$, and $p_j^2$. The  Regge limit corresponds to $s$ large, which defines a light-cone  direction, with $t<0$ and $p_j^2$ fixed. In this limit,  the momentum transfer is asymptotically transverse, with $t=(p_1+p_2)^2\approx - q_\perp^2$.  In a coordinate representation, this corresponds to the DLC limit, discussed in Sec. \ref{sec:DLC} and explained further in Appendix \ref{sec:DLC-append}.
Using conformal symmetry,  it is possible to express the amplitude $T(p_j)$ as
\begin{align}
T&(s,t,p^2_i)  \approx \int \frac{dz}{z^5}\, \frac{d{z'}}{{z'}^5} \, \nn
&\Phi_1(z,p_1^2)\Phi_{2}(z,p^2_2) \,  { \cal K}(s,t,z,\bar z)   \, \Phi_{3}(\bar z,p_3^2)\Phi_{4}(\bar z,p_4^2)\, ,
\end{align}
where ${ \cal K}(s,t,z,\bar z) $ corresponds to a Pomeron-Regge kernel  which in the 
Regge limit  admits an  impact parameter representation,~\cite{Cornalba:2007fs,Cornalba:2008qf,Cornalba:2009ax,Costa:2012cb}
\be
 { \cal K}(s, t,z,\bar z) =  (z\bar z)^2s\,  \int \frac{d^2b_\perp}{4 \pi^2}\, e^{\textstyle i q_\perp \cdot b_\perp}  {\cal F} (S,\sigma_0) \; , \label{eq:IPrep}
\ee
with $b_\perp$ the two-dimensional impact parameter.  
The amplitude ${\cal F}(S, \sigma_0)$  encodes all dynamical information and, due to conformal symmetry, depends only on 
the variables
$
S=z\bar z s $ and $\sigma_0\simeq\frac{z^2+\bar z^2+b_\perp^2}{2z\bar z}$, (\ref{eq:sigma0}).
It is important to note that the conformal  representation (\ref{eq:IPrep}) of the amplitude is valid for any value of the coupling constant, since it relies only on conformal invariance. 
The same representation was  obtained  through  direct  AdS/CFT considerations, \cite{Brower:2006ea,Brower:2007qh,Brower:2007xg}, 
leading to an identical Regge kernel, ${\cal K}( s, b_\perp, z,z')$. 
Up to irrelevant constants, this kernel is related to  ${\cal T}(S,\sigma_0)$ by  
\be
{\cal K}(s,b_\perp,z,\bar z) \sim  \,{(z\bar z)^2\, s}\,\,{\cal F}(S,\sigma_0) \,.
\ee
The Regge limit is now $S\to \infty$ with fixed $\sigma_0$.

We have therefore two representations of the correlation function in the Regge limit. One
derived from the CFT analysis in position space $F^{(M)}(u,v)$, given by Eqs. \ref{eq:A(X)}) and
(\ref{eq:newgroupexpansion}), and another from a computation in
momentum space with a clear geometrical interpretation as a scattering
process in $AdS$, given by Eq. (\ref{eq:T-chi}). This establishes a
dictionary, where, in the Regge limit, 
\begin{align}
F^{M)}(u,v)&\leftrightarrow{\cal F}(S,\sigma_0)\nn
&=N^{-2}\, (z\bar z )^{-2} s^{-1} {\cal K}( s,b^2_\perp, z, \bar z).
\end{align}
We will also identified, as done in Sec. \ref{sec:DLC},   $w\approx 2\sqrt u^{-1}  \leftrightarrow S= z \bar z s$ and $ \frac{1-v}{2\sqrt u} \leftrightarrow  \sigma =\cosh \xi \sigma_0\simeq \frac{b_\perp^2 + z^2 + {\bar z}^2}{2 z \bar z}\,. $
More details can be found in~\cite{Brower:2014wha}. It is also possible to carry out a more formal analysis in
establishing this equivalence~\cite{Cornalba:2009ax} and a useful more recent review can also be found in~\cite{Shenker:2014cwa}.  It
suffices to emphasize the exact equivalence of the two approaches  to identify
the spectral curve, $\Delta(\ell)$ in Fig. \ref{fig:BFKLDGLAP},  which serves as the common link between them.

\subsection{\texorpdfstring{$AdS_{d-1}$}{AdSd-1} Bulk-to-Bulk Propagator and the Pomeron Intercept}\label{sec:propagator}

A more precise relation  discussed above can best be illustrated by the well-known example of one graviton-exchange contribution through AdS/CFT. 
The contribution  is proportional to the traceless-transverse bulk-to-bulk graviton propagator in $AdS_5$, which can be identified with   the leading  contribution to the eikonal, (\ref{eq:T-chi}). In the near forward limit, the momentum transfer is small and transverse to the LC, and one finds, the net contributions reduces to~\footnote{The transition fro $AdS_5$ to $AdS_3$ propagator can be understood best in a momentum treatment. See \cite{Brower:2007qh,Brower:2007xg}.},
\begin{align}
\chi &\simeq (z \bar z s)^{\ell-1} G_{ads_3}(\sigma, \ell,\Delta)\, \quad {\rm and}\nn 
G_{ads_3}(\sigma,\ell,\Delta) &= \frac{e^{-(\Delta(\ell) -2) \,  \xi}}{\sinh \xi}      \label{eq:AdS3}
\end{align}
with $\ell=2$ and $\Delta(2)=4$ for graviton in $d=4$.  $G_{ads_3}$ is  the Euclidean $AdS_3$ scalar propagator for conformal dimension $\Delta-1$, with  $\sigma=\cosh \xi$ related to the chordal distance.

 As explained in Sec. \ref{sec:discuss}, one can begin by restricting   the string propagator $G=\frac{\delta_{L_0,\bar L_0}}{L_0+\bar L_0-2}$ on $AdS_5$ to the graviton sector. In strong coupling, it can be reduced to that involving scalar AdS Laplacian 
$
 \frac{1}{ \ell - 2 +  (1/2\sqrt \lambda) \nabla_0^2} .
$
In a momentum-space representation, one has  $\nabla_0^2=\nabla^2_{0,radial}  -z^2 t$ and $\nabla^2_{0,radial}=z^5\partial_z z^{-3}\partial_z$.  A self-adjoint spectral analysis can be carried out in $\ell$ with $t<0$ and also in $t$ with $\ell>2$.

For $AdS_5$, one finds~\cite{Brower:2006ea,Brower:2007qh,Brower:2007xg}, with $t<0$,
\begin{align}
G&(t, z,\bar z,j) \nn
&=\frac{(z\bar z)^2}{R_{ads}^4} \int \frac{d\nu}{2\pi i} \frac{ K_{i\nu}(qz) K_{i\nu}(q\bar z)}{\ell-2 + (1/2\sqrt \lambda) (\nu^2+ d^2/4)}, \label{eq:ell-spectral}
\end{align}
with $d=4$. Here $K_{i\nu}(qz) $ is the  modified Bessel function, where $q=\sqrt {-t}$, and it corresponds to the  scalar bulk-to-boundary  propagator 
in the momentum representation.  Note that this representation can be expressed in  form given earlier, (\ref{eq:trueCFT}). 
 With  $\Delta=(d/2) +i\nu$ and $d=4$, this corresponds to the desired spectral curve, 
$
\Delta(\ell) \simeq 2 +  \sqrt{2} \lambda^{1/4}   \sqrt {\ell-\ell_{eff}}. 
$
Alternatively, one can carry out a spectral analysis in $t$, while holding $j>2$, leading to~\cite{Brower:2006ea,Brower:2007qh,Brower:2007xg},  
\be
G(\ell, z,\bar z,t)=\frac{(z\bar z)^2}{R_{ads}^4}\int_0^\infty d k^2 \frac{J_{\widetilde \Delta(\ell)}(z k)J_{\widetilde \Delta(\ell)}(\bar z k) }{k^2-t-i\epsilon}\, ,
\ee  \label{eq:t-spectral}
where  $\widetilde \Delta (\ell)=\Delta(\ell)-2$. 
One  finds that $G(\ell, z,\bar z,  t)$ has a continuous spectrum for $0<t$.     

Finally in the DLC limit where $p_1-p_2=q_\perp$ is asymptotically transverse, performing a (d-2)-dim Fourier transform, one finds
\be
G(z,\bar z, \ell) = \frac{(z\bar z)^2}{R_{ads}^4} G_{ads_{(d-1)}}(\sigma,\ell,\Delta)
\ee
with $G_{ads_{(d-1)}}(\sigma,\ell,\Delta)$ given earlier, i.e., a formal solution expressible in terms of geodesic $\xi$ on $H_{d-1}$.  At $\ell=2$ and $d=5$, other than the factor $(z\bar z)^2R_{ads}^{-4}$, this is nothing  but the scalar $AdS_3$ bulk-to-bulk propagator.  For general $\ell$, it corresponds to having  AdS mass, $m^2(\ell)$, (\ref{eq:adsmass-2}).

\section{Minkowski Conformal Blocks and Analytic Continuation} \label{sec:MCFB-append}

\subsection{Useful Mathematical Facts}\label{sec:hypergeometric}
We summarize several useful facts relating to  hypergeometric DE. First,  differential equation for associated Legendre functions, $P_\nu^\mu(q)$ or $Q_\nu^\mu(q)$,  is
\begin{align}
(1-q^2) \frac{d^2 P(q)}{dq^2} &- 2 q\frac{d P(q)}{dq} \nn
&+ [(\nu(\nu+1) - \frac{\mu^2}{1-q^2}] P(q)=0. \label{eq:Lengendre}
\end{align}
We will  work here mostly with $\mu=0$, leading to   either  $P_\nu(q)$ or $Q_\nu(q)$, depending on appropriate boundary conditions.  More generally, Eq. (\ref{eq:EMf}) (b), is of the form 
\be
(1-q^2) \frac{d^2 P(q)}{dq^2} - (d-1)\, q\frac{d P(q)}{dq} +  m^2  P(q)=0.  \label{eq:AdS}
\ee
It can be shown to correspond to  the DE  for  $AdS_{d-1}$ propagator, on $H_{d-1}$, with  geodesics 
$\xi=\cosh^{-1} q$ and AdS mass, $m$.  They can be related to hypergeometric DE by a change of variable.  For instance,  in terms of $y=q^2$, it leads to the standard  DE, $\epsilon(d) = (d-2)/2$,
\be
y(1-y) \frac{d^2 F(y)}{dy^2} +[c- (a+b+1) y] \frac{d F(y)}{dy} - ab F(y)=0\label{eq:Hyper}
\ee
with $ a =\epsilon(d)/4 +\sqrt{ \epsilon(d)^2/4+ m^2}/2$, $ b =\epsilon(d)/4 -\sqrt{ \epsilon(d)^2/4+ m^2}/2$ and $c=1/2$. 

For general values of $a,b,c$, the regular solution at $y=0$ is denoted by the standard notation of 
$
_2F_1(a,b;c;y) = \frac{\Gamma(c)}{\Gamma(a)\Gamma(b)} \sum_n \frac{\Gamma(n+a)\Gamma(n+b)}{\Gamma(n+c)}\frac{y^n}{\Gamma(n+1)}
$. 
However, we are interested in the region $1<y<\infty$.     Two independent solutions can be chosen as
\begin{align}
F_{1,\infty}(y) &= y^{-a} F(a,a-c+1, a-b+1; y^{-1}) \\ 
F_{2,\infty}(y) &= y^{-b} F(b,b-c+1, b-a+1; y^{-1})
\end{align}
Since $ab<0$, adopting the convention $a>0$, the solution where  $P(q)$ vanishing at $q\rightarrow \infty$ corresponds to  $F_{1,\infty}(q^2)$,
\be
q^{-2a} F\Big(a, a +\frac{1}{2} , 2a -\frac{d-4}{2} , q^{-2}\Big)\, .
\ee
  In particular, for $d=3$, this leads to $Q_\nu(q)$, with $\nu=-1/2 +\sqrt{ 1/4+ m^2}$,
\begin{align}
Q_\nu&(q) =  \pi^{1/2} \frac{\Gamma(\nu+1)}{\Gamma(\nu+3/2)} (2q)^{-(\nu+1)} \nn
&\times F\Big(\frac{\nu+1}{2}, \frac{\nu}{2}+1; \nu+\frac{3}{2}; \frac{1}{q^2}\Big).
\end{align}

Alternatively,  in terms of $z=(q+1)/2$, the resulting DE  also take on the same form, 
with $ a =\epsilon(d)/2 +\sqrt{ \epsilon(d)^2/4+ m^2}$, $ b =\epsilon(d)/2 -\sqrt{ \epsilon(d)^2/4+ m^2}$   and $c=(\epsilon(d)+1)/2$. For $d=3$, one has, for $1<z<\infty$, 
\begin{align}
Q_\nu&(z) =  \pi^{1/2} \frac{\Gamma(\nu+1)}{\Gamma(\nu+3/2)} (2z)^{-(\nu+1)}\nn
&\times  F\Big(\nu+1, \nu+1; 2\nu+2; \frac{1}{z}\Big).
\end{align}
This leads to a useful identity in changing variable from $z$ to $q=2z-1$, which, more generally, corresponds to the identity 
\begin{align}
F&(a,b,2b,w)\nn
&=(1-\frac{w}{2})^{-a} F\Big(\frac{a}{2}, \frac{a+1}{2}, b+\frac{1}{2}, \frac{w^2}{(2-w)^2}\Big).  \label{eq:z-2-q}
\end{align}

\subsection{Standard Differential Equation for Conformal Blocks}
The Casimir differential operator,
$
D^{\epsilon}(a,b) \,  G(u,v) =C_{\Delta,\ell} \, G(u,v)\, 
$,
 can be expressed, either in terms of $(x,\bar x)$ or $(q,\bar q)$, as  a sum of terms~\cite{Dolan:2011dv},
$
D^{\epsilon}(a,b)  = D_0(a,b) + \bar D_0(a,b) +   D^{\epsilon}_1.  \label{eq:Dxq}
$
The first term involves $x$ or $q$,
\begin{align}
D_0(a,b) &= x^2(1-x) \frac{d^2}{dx^2} -(a+b+1) x^2 \frac{d}{dx} - ab x \label{eq:D0x}\\
&=  (q^2-1) \frac{d^2}{dq^2} + 2 (q+a+b) \frac{d}{dq}  - \frac{2ab}{q+1}\label{eq:D0q}
\end{align}
and the same for $\bar D_0(a,b)$ with $\bar x$ and $\bar q$ replacing $x$ and $\bar x$n respectively. 
The mixed term, $D^{\epsilon}_1$, is 
\begin{align}
D^{\epsilon}_1&=2\epsilon \frac{x\bar x}{x-\bar x} \Big( (1-x) \partial_x - (1-\bar x) \partial_{\bar x}\Big) \label{eq:Dmixedx}\\
&=2 \epsilon \frac{1}{q-\bar q} \Big( (q^2-1) \partial_q - (\bar q^2-1) \partial_{\bar q}\Big)\label{eq:Dmixedq}
\end{align}
where $\epsilon = (d-2)/2$.  
Here $a$ and $b$ stand for more general four point function where $a= -\Delta_{12} $ and $b=\Delta_{34}/2$. We will restrict in what follows to the case where $a=b=0$.  

Solutions to $D_0(x)f(x) =  \lambda(\lambda-1) f$  can be expressed in terms of Hypergeometric functions, i.e., with $f(x)=x^\lambda \bar f(x)$, DE for $\bar f(x)$ becomes
\be
x(1-x) \bar f^{''}(x) + (2\lambda- (2\lambda+1)x) \bar f'(x) -\lambda^2 \bar f(x)=0\, ,
\ee
i.e., again in the hypergeometric form, Eq. (\ref{eq:Hyper}). 
A general solution can be expressed as
\be
f(x) = a\,   k_{2\lambda} (x)  + b\,  k_{2(1-\lambda)} (x)    \label{eq:x-solution}
\ee
where
\be
k_{2\lambda} (x) =x^\lambda\,\,  _2F_1(\lambda,\lambda; 2\lambda; x). 
\ee

The corresponding differential equation in terms of variable $q$ is 
\be
 (1-q^2) \frac{d^2 g(q)}{dq^2} - 2 q \frac{d g(q)}{dq} + \lambda (\lambda -1) )g(q)=0
\ee
with boundary condition specified at $q\rightarrow \infty$. This is precisely that  for Legendre function of the second kind, Eq. (\ref{eq:Hyper}). A general solution can also be expressed in terms hypergeometric funvtions, e.g., $\widetilde k_{2\lambda}(q)$, Eqs. (\ref{eq:q-solution}-\ref{eq:q-function}).
Eq. (\ref{eq:Hyper}), valid for $|q|>1$, is particularly useful in considering continuation from $1<q<\infty$ to $-\infty<q<-1$.

\subsection{Comparison with Analytically Continued Euclidean Conformal Blocks}

It has been suggested that Minkowski conformal blocks are simply an appropriate analytic continuation of Euclidean conformal blocks, which changes the boundary conditions in  Eq. (\ref{eq:CB}) from (a) to (b).  We demonstrate below that Eq. (\ref{eq:d2MCBq}) and Eq. (\ref{eq:d4MCBq}) are not given by a direct analytic continuation of the corresponding ECB, Eq. (\ref{eq:d2ECB}) and Eq. (\ref{eq:d4ECB}). 
 
The discussion for analytic continuation  is normally framed in terms of variables $x$ and $\bar x$.  One can easily transition from $(q,\bar q)$ to $(x,\bar x)$, with 
\be
 k_{2\lambda}(x)  =\widetilde k_{2\lambda}(q),
\ee
due to the identity, Eq. (\ref{eq:z-2-q}), where $z=x^{-1}$. We examine more closely here the relation of  between  $G^{(M)}_{(\Delta,\ell)} (x,\bar x) $ and $G^{(E)}_{(\Delta,\ell)} (x,\bar x) $ in terms of their analytic structure in $x$ and $\bar x$. 
 It should be emphasized  that, by treating $x$ and $\bar x$ as independent complex variables, one necessarily extends beyond the Euclidean limit where $\bar x = x^*$.  For Minkowski limit both $x$ and $\bar x$ are real but independent.  To examine their possible connection, both $x$ and $\bar x$ are to be treated as independent complex variables. In order to make this demonstration explicit, we will focus on the case of $d=2$. The case $d=4$ can also be dealt with explicitly. 
 
It is sufficient  to examine the analytic structure of $k_{2\lambda}(x)$ as one circles around its branch point at $x=1$, or equivalently, at $x=\infty$, while holding $\bar x$ fixed. It is useful to expose square-root singularities at $x=1$ and $x=\infty$  by a mapping
\be
\rho =\frac{1-\sqrt{1-x}}{1+\sqrt{1-x}}
=  \frac{x}{(1+\sqrt{1-x})^2}
\ee
\begin{figure}[ht]
\begin{center}
\includegraphics[width = .4\linewidth]{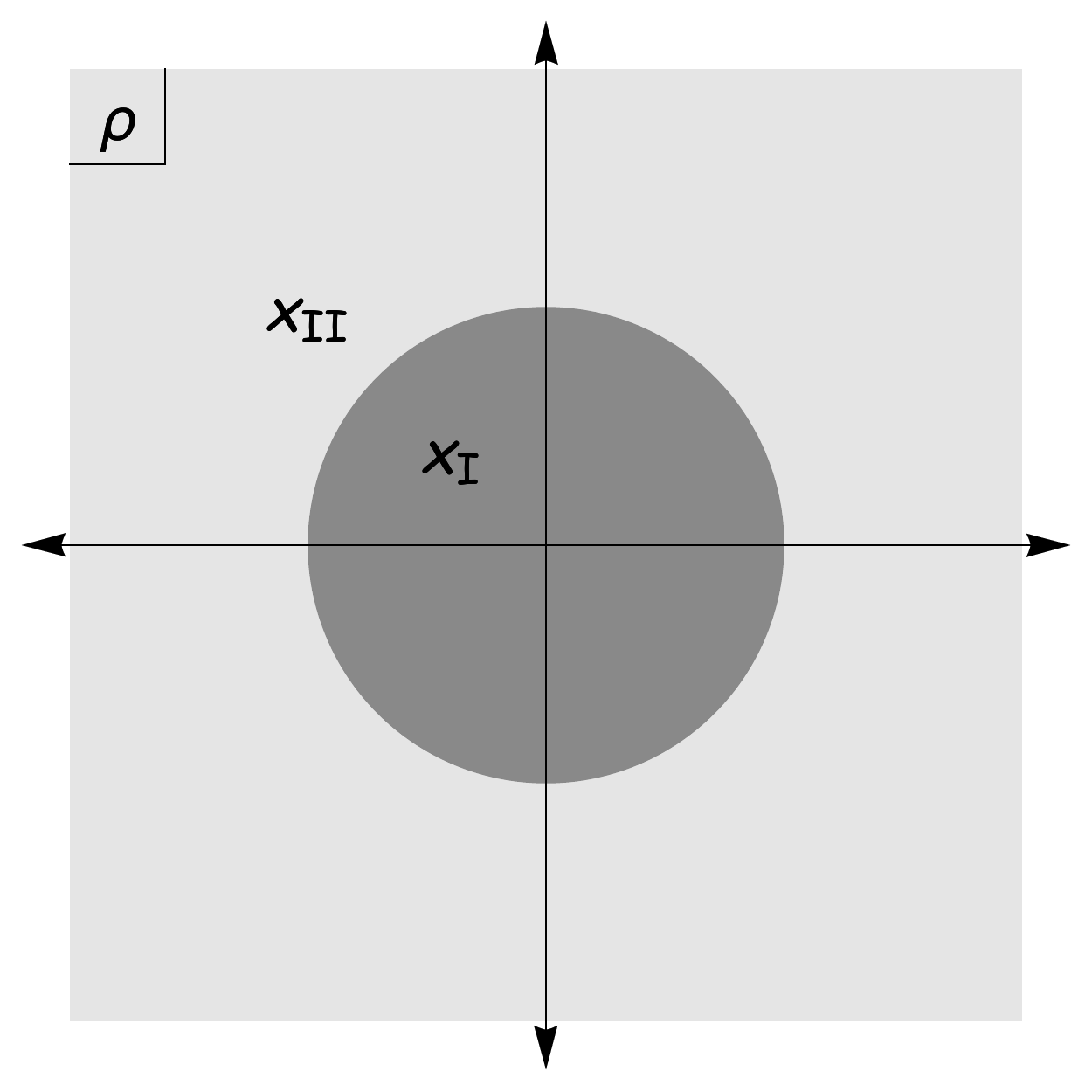}
\caption{$\rho$ plane indicating regions of $x_I$ and $x_{II}$.}
\label{fig:rhoplane}
\end{center}
\end{figure}
This map exposes a  double-sheet structure (See Fig.(\ref{fig:rhoplane})). In general $\rho$  maps the first sheet of $x$, denoted as the physical sheet,  to the region within the unit circle, $|\rho|<1$,  and the second sheet in $x$, analytically continued through the cut $(1,\infty)$, to the region $|\rho|>1$, outside of the unit circle.
For $x=0$, on the physical sheet sheet, $\rho\simeq 0$. On the other hand, when continued to the second sheet, the limit $x\rightarrow x_{II}\rightarrow 0$ maps $\rho\simeq 4/x_{II}\rightarrow \infty$. This map exposes a  double-sheet structure. In general $\rho$  maps the first sheet of $x$, denoted as the physical sheet,  to the region within the unit circle, $|\rho|<1$,  and the second sheet in $x$, analytically continued through the cut $(1,\infty)$, to the region $|\rho|>1$, outside of the unit circle.
For $x=0$, on the physical sheet sheet, $\rho\simeq 0$. On the other hand, when continued to the second sheet, the limit $x\rightarrow x_{II}\rightarrow 0$ maps $\rho\simeq 4/x_{II}\rightarrow \infty$. Therefore, to understand the analytic continuation, it is simpler working directly with  the variable $\rho$. 
 
It is useful to again relate both MCB and ECB directly to hypergeometric functions. The differential operator  $D_0$ simplifies in terms of $\rho$. For $d=2$ it becomes
\be
D_0 = \frac{\rho^2}{1-\rho^2} \frac{d}{d\rho}(1-\rho^2) \frac{d}{d\rho}\, , \label{eq:rhoD0}
\ee
and the relevant eigenvalue condition becomes
\be
\frac{\rho^2}{1-\rho^2} \frac{d}{d\rho}(1-\rho^2) \frac{d}{d\rho} = \lambda(\lambda-1)  G(\rho)\,.\label{eq:rhoD0ev}
\ee 
It is convenient to introduce new notation,
$
K_{\alpha}(\rho^2) = k_{2\alpha} (x)
$.
From  Eq. (\ref{eq:rhoD0ev}), one finds
$
K_{\alpha}(\rho^2) = \rho^{\alpha}\, _2F_1(1/2, \alpha; \alpha + 1/2; \rho^2)
$.
In terms of $K_{\alpha}(\rho^2)$, the Minkowski conformal block, for $d=2$, is given by
\begin{align}
G^{(M)}_{(\Delta,\ell)} (x,\bar x) &=  K_{(1-\lambda_+)} (\rho^2_>) K_{\lambda_-} (\rho^2_<)  
\label{eq:d-2MCB3}
\end{align}
where $\rho_<$ is the small of $(\rho, \bar \rho)$, and $\rho_>$ the other. 
 In contrast,  for  the Euclidean conformal block, 
\begin{align}
G&^{(E)}_{(\Delta,\ell)} (x,\bar x) \nn
&= K_{\lambda_+}(\rho^2) K_{\lambda_-}(\bar \rho^2)+K_{\lambda_+}(\bar \rho^2) K_{\lambda_-}(\rho^2) \label{eq:d-2ECB3}
\end{align}

 We stress, for both Eq. (\ref{eq:d-2ECB3}) and Eq. (\ref{eq:d-2MCB3}), we are, for now,  restricted to the region $|\rho|<1$ and $|\bar \rho|<1$ so that $x$ and $\bar x$ remain on the physical sheet.   The advantage of using $K_{\alpha}(\rho^2)$ over $k_{2\alpha}(x)$ lies in the fact that  analytic continuation to the second sheet in $x$ simply corresponds to moving outside of the unit circle, $\rho=1$.

Let us focus on the analytic continuation of Eq. (\ref{eq:d2ECB}).  For convenience,   we will analytically continue $x$ from the first sheet to the second sheet, $x\rightarrow x_{II} \rightarrow 0$,  with  $\bar x\rightarrow 0$ on the first sheet.   In terms of Eq. (\ref{eq:d-2ECB3}), this corresponds to taking $\rho\rightarrow \infty$ and  $\bar \rho  \rightarrow 0$.
 That is, in continuing  $x$ from first sheet to second sheet, $x\rightarrow x_{II} \rightarrow 0$, 
\begin{align}
\rho \rightarrow \rho_\infty &= \frac{1+\sqrt{1-x_{II}}}{1-\sqrt{1-x_{II}} } = \frac{1}{\rho} \nn
&=  \frac{x_{II}}{(1-\sqrt{1-x_{II}})^2}\rightarrow \infty.
\end{align}
The analytically continued ECB becomes
\begin{align}
G&^{(E,continued)}_{(\Delta,\ell)} (x_{II},\bar x)\nn
&=  K_{\lambda_+}(1/\rho^2) K_{\lambda_-}(\bar \rho^2)+K_{\lambda_+}(\bar \rho^2) K_{\lambda_-}(1/\rho^2)
\end{align}
This can be converted  via the inversion identity, Eq. (\ref{eq:inversion-2}),  to a simpler representation amenable to  expansions about $x_{II} =0$ and $\bar x=0$.
The  continued conformal block  consists of  four groups of terms, 
\be
G^{(continued)}_{(\Delta,\ell)} (x,\bar x) = \sum_{i,1,2,3,4} G^{(i)}_{(\Delta,\ell)}(x,\bar x), 
\ee
each with unique small $x$ and $\bar x$ behavior
\begin{align}
G^{(1)}_{(\Delta,\ell)} (x,\bar x) &\sim  (\sqrt {x\bar x})^{(1-\ell)}\, (x/\bar x)^{(1-\Delta)/2}, \label{eq:d-2ECB4}\\
G^{(2)}_{(\Delta,\ell)} (x,\bar x) &\sim  (\sqrt {x\bar x})^{(1+\ell)}\, ( x/ \bar x)^{(1+\Delta)/2}, \label{eq:d-2ECB5}\\
G^{(3)}_{(\Delta,\ell)} (x,\bar x) &\sim    (\sqrt {x\bar x})^{\Delta} \,(x/\bar x)^{\ell/2}  ,   \label{eq:d-2ECB6}\\
\text{and} \qquad G^{(4)}_{(\Delta,\ell)} (x,\bar x) &\sim  (\sqrt {x\bar x})^{\Delta} \,( x/\bar  x)^{-\ell/2} .   \label{eq:d-2ECB7}
\end{align}
Of these, only the first  term has the desired  dependence, Eq. (\ref{eq:Mbdry-2}). Therefore, an analytically  continued ECB  does not lead to the desired Minkowski conformal block.

As another check on the fact that $G^{(M)}(x,\bar x)$ and $G^{(E)}(x,\bar x)$ are related but not directly given via analytic continuation, it is instructive to carry out the following exercise by starting with our $G^{(M)}$ as defined on the ``second sheet" and analytically continued it back to the first-sheet and then compare with $G^{(E)}(x,\bar x)$. This can again be done by using $\rho$-representation  by first continuing from $\rho_{>II}=0$ to $\rho_\infty$. Next making use of inversion formula, Eq. (\ref{eq:inversion-2}), to bring it back to a representation amenable to  an expansion around $x_>=0$.  Consider, for $d=2$, 
$
G^{(M)}(x,\bar x)=k_{2(1-\lambda_+)} (x_{>II}) k_{2\lambda_-} (x_<)
$.
One finds,
\begin{align}
  G&^{(M,continued)}_{(\Delta,\ell)} (x,\bar x) \nn
  &= a\, k_{2\lambda_+} (x_{>}) k_{2\lambda_-} (x_<) \nn
  &\qquad\qquad+ b\, k_{2(1-\lambda_+)} (x_{>}) k_{2\lambda_-} (x_<)
  \end{align}
  where $a= i \sqrt \pi \frac{\Gamma(1/2-\lambda_+)}{\Gamma(1-\lambda_+)^2}$ and $b=(-1)^{1-\lambda_+}/\cos\pi\lambda_+$. It does not lead to $G^{(E)}(x,\bar x)$.

\subsection{Symmetric Treatment}  
Once the leading index $\gamma_b=1-\ell$ is identified,  it is possible to solve each expansion function $g_n(\sigma)$ iteratively.  We will not do it here in general except for the case of $d=1$.  For $d=2$ and $d=4$, since explicit solutions are already known, we will instead demonstrate that  they can be re-expressed in the symmetric form, Eq. (\ref{eq:symmetric}).

Let us begin with Eq. (\ref{eq:d2MCBq}) and  Eq. (\ref{eq:d4MCBq}). With $d=2$ and  consider the case $\bar q>q$,  expanding hypergeometric functions   leads to 
   an expansion
\begin{align}
G^{(M)}_{(\Delta,\ell)} (w,\sigma)&= w^{\ell-1} e^{(1-\Delta)\xi} \nn
& \quad \times \sum_{n=0}^\infty a_n q^{-2n}\sum_{m =0}^\infty b_m {\bar q}^{-2m}
\end{align}
with $a_n$ and $b_m$ given by the standard coefficient functions.  By re-grouping, this can be re-expressed as
\be
G^{(M)}_{(\Delta,\ell)} (w,\sigma)=  w^{\ell-1} e^{(1-\Delta)\xi} \sum_{n=0}^\infty (w e^{\xi})^{- 2n } \sum_{p=0}^{2n} c_p e^{p \xi}   \label{eq:higherorder2}
\ee
 with $c_p$ given by a finite sum of products $a_r b_s$, with $r+s=n$.
Turning next to the case of $d=4$ and consider again the case $\bar q>q$. One finds, 
\begin{align}
G&^{(M)}_{(\Delta,\ell)} (w,\sigma)   \nn
&=  w^{\ell-1}  \frac{e^{(2-\Delta)\xi}} {\sinh \sigma}  \sum_{n=0}^\infty (w e^{\xi})^{-2n}  \Big(\sum_{p=0}^{2n} c'_p e^{p \xi}
\Big). \label{eq:higherorder4}
\end{align}
with coefficients $c'_n$ again given by a finite sum.

\section{Deep Inelastic Scattering (DIS)}\label{sec:application-II}

In Sec. \ref{sec:boost} we saw that a Lorentz boost plus dilatation correspond to a $SO(1,1)\times SO(1,1)$ subgroup of the full conformal transformations, $SO(4,2)$. It has long been known that approximate $O(2,2)$ symmetry is an important feature of QCD near-forward scattering at  high energies~\cite{PomeronBook}. To exemplify this, let us turn first to deep inelastic scattering (DIS), which corresponds to a measurement of a total  cross section, $\sigma^{total}_{\gamma^* p}$, for a virtual photon with momentum $q$ scattering off of a proton of momentum $p$. The measure of photon  ``off-shellness", characterized by $1/q^2$,  is  referred to as its virtuality. This serves as the scale in probing short-distance behavior of the product of two local currents involved.  There also exists another scale in the problem, the photon energy, $E_\gamma$. The  limit of  $q^2$ and $E_\gamma$ both large, with the ratio  $x\sim q^2/E_\gamma \rightarrow 0$,  leads to another scaling behavior.  These scalings are related through the t-channel OPE for electromagnetic currents $J^\mu(x)J^\nu(0)$. To be more explicit, it is described in \cite{Polchinski:2002jw} that anomalous dimensions of the leading twist conformal primaries, ${\cal O}_{\Delta,\ell}$, control the large $q^2$ dependence for the moments of hadronic structure functions.
In this appendix we first review the direct amplitude calculation of DIS structure functions, revealing a Mellin representation. Next we apply the approach of Sec. \ref{sec:application-I} to extract  BFKL and DGLAP physics. Finally we examine the BFKL-DGLAP integral equation itself which is analogous in treatment to techniques used in evaluating SYK-like models.

\subsection{Direct Computation}\label{sec:DIS-append}

We provide below a brief review, following that of \cite{Polchinski:2002jw}, with some notational change adopting that used in this paper. We are focused on the limit $q^2\rightarrow \infty$ with $x=q^2/s$ fixed\footnote{Here $q^2>0$ for spacelike $q$.}.  The hadronic tensor is $U^{\mu\nu}(p,q)$, defined as the Fourier transform of the current  commutator, $  \langle p |[J^\mu(x) ,J^\nu (0)] |p\rangle$, and it can be written in terms of two scalar structure functions, ${\cal U}_\alpha$, $U^{\mu\nu} = {\cal U}_1(x,q^2) \Big(g_{\mu\nu}-\frac{q_\mu q_\nu}{q^2}\Big) + {\cal U}_2(x,{q^2})\Big(p_\mu+\frac{q_\mu}{2x}\Big)\Big(p_\nu+\frac{q_\nu}{2x}\Big).\label{eq:DISstructure}$ Through the optical theorem, ${\cal U}_\alpha$ can be identified as the imaginary part of the forward 
virtual Compton scattering $T^{\mu\nu} (p,q; p', q')$, i.e. in the limit  $p=p'$ and $q=q'$.  In this limit, $T^{\mu\nu}$ has a Lorentz covariant expansion similar to that of $U^{\mu\nu}$, with structure functions ${\cal T}_\alpha(x,q^2)$ replacing ${\cal U}_\alpha(x,q^2)$. Treating  $U_\alpha(x,q^2)$ as real-analytic functions  of $x$ with a branch cut over $[-1,1]$, the relation between the two structure functions is
  \be
{\cal U}_\alpha(x,q^2) = 2 \pi\,{\rm Im}\, {\cal T}_\alpha(x,q^2)\, .
\ee
  These discontinuities can also directly be related to $\sigma_T$ and $\sigma_L$ for transverse and  longitudinal off-shell photons: for example ${\cal U}_2(x,q^2)= (q^2/4\pi^2 \alpha_{em}) (\sigma_T+\sigma_L)$.

Let us focus on $ {\cal T}_2(x,q^2)$. (A similar analysis can also be carried out for ${\cal T}_1$.) The s-channel physical region corresponds to  $1<x^{-1}<\infty$, with ${\cal U}_2(x,q^2) = 2\pi \,{\rm Im}\, {\cal T}_2(x,q^2)\,$. As a real-analytic function of $x^{-1}$, ${\cal T}_2(x,q^2)$ is odd and has symmetric branch cuts for $1<|x|^{-1}<\infty$. We can re-express ${\cal T}_2$ through a dispersion integral in $x^{-1}$,  ${\cal T}_2(x, q^2)= \frac{2x}{\pi} \int_0^1 dx' \frac{ {\cal U}_2(x', q^2)}{x^2-x'^2}$. The full amplitude can then be expanded for $1<|x|$ as ${\cal T}_2(x, q^2)= (2/\pi) \sum_{n=1,2,\cdots} u(2n,q^2) \, x^{1-2n}\label{eq:DLC-MCB}$ where
\be
u(\ell, q^2)=  \int\limits_0^1 dx x^{\ell-2}  \, {\cal U}_2(x, q^2). \label{eq:f2Mellin}
\ee
Note that, initially, $u(\ell, q^2)$ is defined  for $\ell=2,4, \cdots$, corresponding to even moments $M_n(q)$ of ${\cal U}_2$.  As an integral 
over the ${\cal U}_2$, the imaginary part of ${\cal T}$ in the s-channel physical region, it also  defines an analytic function~\footnote{We assume that ${\cal U}_2(x,q)< O(x^{-1})$ at $x=0$, consistent with the requirement of energy-momentum conservation, i.e., the $u(2,q)$ integral is finite.   It follows, for the inverse transform, Eq. (\ref{eq:F2InvMellin}), $L_0$ can be chosen so that $2-\epsilon<L_0<2$, with $\epsilon$ infinitesimal. For a related discussion, see \cite{komargodski:2016gci}.} of $\ell$, regular for $2\leq {\rm Re}\, \ell$. 

Eq. (\ref{eq:f2Mellin}) also corresponds to the Mellin transform of ${\cal U}_2$ with respect to $x^{-1}$. It follows that, for $0<x<1$, ${\cal U}_2(x,q)$ can be recovered via an inverse Mellin transform, 
 \be
 {\cal U}_2(x,q^2) =  \int\limits^{L_0+i\infty}_{L_0-i\infty}\,  \frac{d \ell\, }{2\pi i} \, x^{1-\ell}  \, u(\ell, q^2)\, . \label{eq:F2InvMellin}
 \ee
 with $2-\epsilon <L_0<2$. With $u(\ell,q^2)$ bounded for ${\rm Re}\, \ell \rightarrow \infty$, the full amplitude   can be represented for  $-1<x<1$ as ${\cal T}_2(x, q)=-   \int^{L_0+i\infty}_{L_0-i\infty}\,  \frac{d \ell\, }{2\pi i}\frac{1+e^{-i\pi \ell}}{\sin \pi \ell}\,  x^{1-\ell} \,   u(\ell, q)\, . \label{eq:DLC-MCB-summed}$ Correspondingly, (\ref{eq:F2InvMellin}) defines a distribution with  ${\cal U}_2(x,q^2)=0$ for $1<|x|<\infty$. Eq. (\ref{eq:DLC-MCB}) can be considered as the starting point of a ``primitive" t-channel OPE. Therefore the Mellin-representation for  ${\cal T}_2(x, q^2)$ corresponds to a Sommerfeld-Watson re-summation introduced earlier,  with Eq. (\ref{eq:F2InvMellin}) providing the imaginary part in the s-channel scattering region, $0<x<1$.

\subsection{Reduction to \texorpdfstring{$d=2$}{d=2}}\label{sec:red22d}

Let us turn next  to a CFT description. For the 4-point correlator, the forward limit, $t=0$, corresponds to integrating $F^{(M)}(w,\sigma)$, Eq. (\ref{eq:newgroupexpansion}) over impact space. Because of conformal invariance, $b_\perp^2$ enters only through $\sigma$, it follows that, from Eq. (\ref{eq:sigma}), the amplitude  at $t=0$,  with fixed conformal virtualities, is a total derivative.  The contribution at $\vec{b}=\infty$ vanishes, and the total contribution becomes~\footnote{We also mention, for DIS, we have $W_s$ having symmetry $(-1)^{s-1}$. More generally, the reduction of one power of $s$ reverses the symmetry pattern for each.}
$
T(s,0; z, \bar z ) = 4 \pi\, (z\bar z)\,  s  \,  P_{(12)}(z) \,P_{(34)} (\bar z)\, W(w, \sigma_0) \,
$,
where  $W(w, \sigma_2) $   is a  2-d reduced  function of $w$ and $\sigma_2= \sigma(z,\bar z, \vec b=0) = \frac{z^2+\bar z^2}{2 z\bar z}$, 
\begin{align}
W(w, \sigma_2) &=  \int  \frac{d^{2} b_\perp}{2\pi  z\bar z}   \,  F^{(M)}(w,\sigma) \nn
&=\int\limits^\infty_{\sigma_2} d\sigma\, \,  F^{(M)}(w,\sigma)\, .  \label{eq:B}
\end{align}
Kinematically, this represents a reduction   in dimension from $d=4$  to $d=2$ and is analogous to the application of Eq.   (\ref{eq:lowering}).   It follows from Eq. (\ref{eq:newgroupexpansion}) that it also admits a Mellin-like representation
\begin{align}
W&(w,\sigma_2) =W_{0}(w,\sigma_2)  - \sum_\alpha\int\limits_{L_0 -i\infty}^{L_0 + i\infty}  \frac {d\ell}{2\pi i} \nn
&\times \frac {1+ e^{-i \pi \ell } }{\sin\pi \ell } a_\alpha^{(12),(34)}( \ell) K_{\alpha}(w,\sigma_0; \ell)  \, .\label{eq:RedConformalBlock}
\end{align}
where  $K_{\alpha}(w,\sigma_2; \ell)$ is a reduced Minkowski conformal block, 
\be
K_{\alpha}(w,\sigma_2; \ell) = \int\limits^\infty_{\sigma_2} d\sigma\, \,  G_\alpha(w,\sigma; \ell)\, .
  \label{eq:RedConformalBlock2}
\ee 

Since $a_\alpha^{(12),(34)}( \ell)$, $ \chi_{\alpha}(w,\sigma_2; \ell)$, and $
W_{0}(w,\sigma_2)$ are real, the imaginary part of $W(w,\sigma_2)$  is again given by a Mellin-like representation
\begin{align}
{\rm Im}& W(w,\sigma_2) \nn
&= \sum_\alpha\int\limits_{L_0 -i\infty}^{L_0 + i\infty}  \frac {d\ell}{2 i}  a_\alpha^{(12),(34)}( \ell )\chi_{\alpha}(w,\sigma_2; \ell)  \, .\label{eq:RedConformalBlock-Disc}
\end{align}

We stress that Eq. (\ref{eq:RedConformalBlock-Disc}) is a new feature for Minkowski  OPE. It occurs whenever one deals with  an inclusive cross section which is related to a discontinuity in the forward limit. From a CFT perspective, one is now working with Wightman functions. This analysis can be generalized to treating other more involved inclusive processes~\cite{Nally:2017nsp}.

In addition, since Eq. (\ref{eq:RedConformalBlock-Disc}) is related to a cross section, a positivity constraint applies. In the Regge limit of $w\rightarrow \infty$ and $\sigma_2$ large, we can keep  the leading order for $\chi_{\alpha}(w,\sigma_2; \ell)$. As in Eq. (\ref{eq:lowering}), it becomes $\chi_{\alpha}(w,\sigma_2; \ell)\simeq w^{\ell-1} h_0(\sigma_2, \Delta_\alpha)$, where, from  (\ref{eq:lowering}) for $d=4$,
\be
h_0(\sigma_2,\Delta )= \int\limits^\infty_{\sigma_2} d\sigma g_0(\sigma,\Delta,4)= \frac{e^{-(\Delta-2) \xi}}{\Delta-2}\, .
\ee

Focusing on the leading twist-two contribution to a total cross section, with 
 $\Delta_\alpha=\Delta_P(\ell)$, we find a {\bf positivity constraint} requiring
 \be
 \Delta_P(\ell)\geq 2\,. \label{eq:positivity}
\ee
From Eq. (\ref{eq:Delta-j-analyticity}), one has the promised upper bound
 \be
 \ell_{eff}\leq 2.
 \ee
 The extra factor $(\Delta_p(\ell)-2)^{-1}$  also changes the power of $\ln w$ in Eq. (\ref{eq:reducedF}) from $-3/2$ to $-1/2$. This is an enhancement which can also be attributed to the positivity  condition at $t=0$ mentioned above~\footnote{This also has an interesting phenomenological consequence for DIS distribution, \cite{Brower:2010wf}.}.
 
It is now clear that, in treating DIS as a CFT scattering process, several essential steps must be followed. To extract the DIS cross section, it is necessary (1) to approach the forward limit of $t=0$ and (2) to take  the imaginary part of a 4-point CFT amplitude. 

\subsection{BFKL-DGLAP Equation}\label{sec:BFKL-DGLAP}

$SO(2,2)$ invariance for QCD can be illustrated by a joint-integral-differential equation of BFKL-DGLAP~\cite{Balitsky:1978ic,Kuraev:1977fs,Fadin1975,Gribov1972,Altarelli1977,Dokshitzer1977} and its solution. These lead to a $\Delta-\ell$ spectral-curve from which the effective spin can be extracted. The BFKL program \cite{Balitsky:1978ic,Kuraev:1977fs} demonstrates the inter-relation between the scaling behavior in longitudinal boost: the effective spin and the anomalous dimensions control the scaling under dilatation for moments of DIS structure functions as we see below.

Let's us focus on the DIS structure $U_2$, introduced above. The BFKL-DGLAP integro-diferential equation corresponds to summing dominant contributions to the DIS cross section in the multiperipheral region. This summation is based on a series of approximations-``leading-log", ``$k_T$-factorization", etc.-and, in its integral form, it can be expressed as
\begin{align}
{\cal U}_2(x,q^2)  &\simeq  {\cal U}^{(1)}_2(x) \nn
&+ \int\limits_{x}^{1}dx'\int\limits^{\infty}_{0}\frac{dq'}{q'}  \widetilde {\cal R}(x, x', q^2, q'^2)  {\cal U}_2(x',q'^2)\label{eq:BFKL-DGLAP1}
\end{align}
We will refer to this as the BFKL-DGLAP equation. A corresponding (Bethe-Salpeter) integral equation can also be written for the full amplitude, $H$, symbolically expressed as
\be \label{eq:BFKL-full}
H=H^{(1)}+ {\cal R} \, \otimes \, H \,,
\ee  
with  ${\cal U}^{(1)}_2= {\rm Im}\, H^{(1)}$ and $\widetilde {\cal R}= {\rm Im}\, {\cal R}$. One advantage of working with ${\cal U}$ is the fact that the integration in Eq. (\ref{eq:BFKL-DGLAP1}) is over physical region only. The solution, formally expressed as ${\cal U}_2(x,q^2)=\sum_n{\cal U}_{2\,(n)}(x,q^2)$, corresponds to summing dominant contributions to the DIS cross section in the multiperipheral region~\footnote{We have  made use of gluon-dominance by dropping quark-contributions as well as other technical simplification in order to bring the equation into a manageable form. Therefore, Eq. (\ref{eq:BFKL-DGLAP1}) should be interpreted as a schematic representation. For more realistic discussion, see \cite{Balitsky:1978ic,Kuraev:1977fs,Fadin1975,Gribov1972,Altarelli1977,Dokshitzer1977,Kwiecinski:1997bt,Kwiecinski:1997ee}}. Each term in the sum, ${\cal U}_{2\,(n)}(x)$,  can be associated with the cross section for the production of n gluons.  This simplification becomes particularly useful when we discuss SYK-like 1-d CFT models in Sec. \ref{sec:SYK}.

Approximate boost and dilatation invariance are reflected by the fact that the kernel is a function of $x/x'$ and $q/q'$  for $x,x'<<1$ and $q_0<<q,q'$. If one extends  this to the whole physical range, the integral equation can be solved by a double Mellin-Fourier transform with respect to $x^{-1}$, as in Eq. (\ref{eq:F2InvMellin}), and $\eta =\log q$ respectively, leading to
\begin{align}
{\cal U}_2(x,q^2)&=\int\limits^{L_0+\infty}_{L_0-i\infty} \frac{d \ell}{2\pi i}  \,  x^{1-\ell} \int\limits_{-\infty}^{\infty} \frac{ d \nu}{2\pi}    e^{- i \nu \eta} \, f(\ell,\nu)\, . 
\end{align}
The simplest approximation has ${\cal R}(x, x', q^2, q'^2)\rightarrow {\cal R}_0(x/x')\simeq {\rm constant} $, leading to 
$
u(\ell, \nu) =\frac{1}{\nu^2+\epsilon^2}\frac{u_1(\ell)}{1- R_0(\ell)}, 
$
where  $R_0(\ell) =\lambda (\ell-1)^{-1}$.  With $u(\ell, q)$ having a pole at $\ell_{eff}=1+\lambda$, 
this leads to Regge behavior at small-$x$,
\be
{\cal U}_2(x,q^2) \sim  \, x^{1-\ell_{eff}}\, . \label{eq:BFKL-Regge2}
\ee
Furthermore, ${\cal U}_2$ is $q$-independent, corresponding to naive Bjorken scaling, with vanishing anomalous dimension, $\gamma_n=0$. In a more refined treatment  by taking into account $q^2$-dependence properly, one finds~\cite{Kwiecinski:1997bt,Kwiecinski:1997ee}
\be
u(\ell, \nu) \simeq \frac{r(\ell)}{\nu+i G(\ell,\nu)} +{\rm regular \, terms}  \label{eq:BFKL-DGLAP3}
\ee
Here we focus on the singularities in the lower-half $\nu$-plane, appropriate for $q$ large. At $\ell=n$,   $u(n,\nu)$ has a pole at $\nu =- i \gamma_n$, where  the anomalous dimension is obtained by solving: $
\gamma_n= G(n, i \gamma_n)
$.
There could multiple solutions and we will retain only the lowest solution for each $\ell$. This leads to  DGLAP-like evolution equation for $M_n(q)$ with $n=2,4,\cdots$,
\be
-\frac{d M_n(q)}{d \log q} \simeq  \gamma_n \, M_n(q)  \quad \rightarrow \quad   M_n(q) \sim q^{-\gamma_n}  \label{eq:f2DGLAP}
\ee
as $q\rightarrow \infty$. This characterizes the dilatation symmetry as realized in DIS. In particular, $\gamma_2=0$, due to energy-momentum conservation.

Since $\ell$ enters as a continuous parameter, it is possible to consider $\gamma(\ell)$ as an analytic function of $\ell$.  This defines a spectral curve, $\Delta(\ell) \equiv \gamma(\ell) - \ell -d/2$. If one shifts $\nu$ by $i\ell$, 
one can formally express the singular part of Eq. (\ref{eq:BFKL-DGLAP3}) as
$$
u(\ell, \nu) \simeq \frac{r'(\ell)}{\nu^2+\widetilde \Delta(\ell)^2}  
$$
where $\nu\rightarrow \nu -i\ell$ and $\widetilde \Delta = \Delta -2$.  Note that we have endowed $f(\ell, \nu)$ certain analyticity and symmetry structure, similar to that in Eq. (\ref{eq:trueCFT}). In particular, due to conformal invariance, after inversion to $\ell(\Delta)$, one has $\frac{d \ell(\Delta)}{d\Delta}=0$ at $\Delta=2$. It follows that the spectral curve $\Delta(\ell)$  has a square-root branch point at $\ell_{eff}$, which can be found by   solving
\be
\Delta (\ell_{eff})= 2.
\ee
Due to the presence of this singularity, one finds
\be
{\cal U}_2(x)(x,q) \simeq   \, x^{1-\ell_{eff}}/|\log x|^{1/2}\, . \label{eq:BFKL-Regge3}
\ee
In weak coupling, one generically has~\cite{Balitsky:1978ic,Kuraev:1977fs,Fadin1975}
\be
\ell_{eff}=1+O(\lambda). \label{eq:Pomeron-weak}
\ee
In contrast, at strong coupling, as shown in Sec. \ref{sec:SpectralCurve}, one finds, $\ell_{eff}=2-O(1/\sqrt \lambda)$. For both limits, Eq. (\ref{eq:BFKL-Regge3}) is the consequence of boost invariance for DIS.

\section{Hilbert Space Treatment for \texorpdfstring{$d=1$}{d=1} CFT}\label{sec:green}

\subsection{Minkowski Green's Functions: Spectral Analysis}\label{sec:spectral1}

Before discussing the case where functions are polynomially bounded, let us first take a closer look at our presentation of Eq. (\ref{eq:1D-ImConformalBlock-2}) for the Hilbert space of square-integrable functions, i.e. the space of functions with a standard inner product,
 \be
 \langle f | g\rangle = \int\limits_1^\infty dw f(w)^* g(w) \, .
 \ee
The differential operator $-{\cal D}_w$, Eq. (\ref{eq:Legendre2}), can be expressed directly in a positive self-adjoint form, $-\frac{d}{dw} (w^2-1)\frac{d}{dw}$, which follows from an effective 1-d action $S=\int_1^\infty dw [(w^2-1) f'(w)^2 + m^2 f(w)^2]$.  Self-adjointness also  requires  $-{\cal D}_w$ must act on functions bounded at $w=1$ and $w=\infty$.  More generally, a Green's function, $G(w,w')$, for $-{\cal D}_w+m^2$, 
\be
 [-\frac{d}{dw} (w^2-1)\frac{d}{dw} + m^2] G(w,w') = \delta(w-w'),
 \ee 
 can be found directly by the Wronskian method
 \footnote{${\cal D}_w$  is the same as ${\cal L}_{0,\sigma}$, Eq. (\ref{eq:Legendre}), evaluated for $d=3$, with solutions given by Legendre functions. A similar treatment  can also be carried out for   ${\cal L}_{0,\sigma}$,  $d\neq 3$, acting on reduced functions $\tilde f(w) = (w^2-1)^{(d-3)/2} f(w)$, leading to  appropriate associated Legendre functions. We will treat this more general case elsewhere.}.   The desired Green's function is simply  given by
 \be
G(w,w') = P_{\nu_+}(w_<)Q_{\nu_+}(w_>), \label{eq:wronskian}
 \ee
 where $\nu_+=-1/2 + \sqrt{m^2+1/4}$ and, as usual,  $w_< =(w,w')_{\rm min}$ and $w_>=(w,w')_{\rm max}$.   $P_\nu(w)$ and $Q_\nu(w)$  are Legendre functions of the first and second kind discussed earlier.

It is also instructive to arrive at the same answer by a spectral analysis to illustrate the non-compactness involved. Consider first  $m^2=0$. The eigenvalue problem can be expressed as 
\be
\left[ -\frac{d}{dw} (w^2-1)\frac{d}{dw} \right] P(w) = \lambda P(w),
\ee
with $\lambda>0$. Eigenfunctions at $w\sim +\infty$ are oscillatory in $\xi=\cosh^{-1} w$, $P(w)\sim e^{i\pm k \xi}$, with wave-number $k=\sqrt {\lambda -1/4}$. With eigenfunctions also bounded at $w=1$,  the spectrum for $-{\cal D}_w$ is positive and continuous, with $\lambda=k^2 +1/4 $,  $0<k <\infty$.   The corresponding eigenfunctions are Legendre functions, $P_{-1/2+i k}(w)=P_{-1/2-i k}(w)$. With index taking on values $-1/2 +ik$, these are also known as toroidal or ring  functions. 
 
It is worth noting that our analysis is comparable to that done in\cite{Polchinski:2016xgd,Maldacena:2016hyu} in treating SKY model, but yet differs in a significant detail. In \cite{Maldacena:2016hyu}, the Hilbert space deals with functions, in terms of variable $\tau$, Eq. (\ref{eq:tau1}),  defined over $(0,2)$. (The range is  extended to $(-\infty,\infty)$ by symmetry.)  The range $(0,2)$ in $\tau$ corresponds to $0<w<\infty$.  In our treatment, by restricting $w$ to the range $(1,\infty)$, the spectrum for ${\cal D}_w$ is strictly positive and continuous. There is no accompanying discrete spectrum involved.  Our ability to take advantage of this simplification is part due to our ability to deal with the absorptive part, ${\rm Im}\, \Gamma(w)$, for a scattering process.

The eigenfunctions  $P_{-1/2+i k}(w)=P_{-1/2-i k}(w)$ satisfy  orthonormal and completeness conditions,  
\begin{align}
 \int\limits_1^\infty dw& \,  P_{-1/2-i k}(w) P_{-1/2+i k'}(w) \nn
 &=  \,\frac{1}{k\tanh \pi k}\, \delta(k-k'), \label{eq:ON-w}\\ 
\text{and} \,\, \int\limits_0^\infty \,dk  \,&{ k\tanh \pi k} \,P_{-1/2-i k}(w) P_{-1/2+i k}(w') \nn
& = \delta(w-w').  \label{eq:Completeness-w}
 \end{align}
  It follows that the desired propagator, with $m^2\neq 0$, is
\begin{align}
G(w,& w') = \frac{1}{2} \int\limits_{-\infty}^{\infty} \,  { dk  }\,\,{ k\,\tanh \pi k}\, \nn
&\times \frac{P_{-1/2+ik}(w) P_{-1/2 +ik}(w')}{k^2 +1/4 + m^2 }\, .
\end{align}
where we have extended the integration  over $-\infty<k<\infty$, with  $P_{-1/2-i k}(w)=P_{-1/2+i k}(w)$. By replacing $P_{-1/2+ik}(w')$ by the identity in Eq. (\ref{eq:PQidentity}), it separates the above representation into two integrals, one involving $  Q_{-1/2+ik}(w')$ and another $Q_{-1/2-ik}(w')$. We can now analytically  continue each  integral into the complex $k$-plane. Consider the case $1<w<w'<\infty$. Since $Q_{-1/2-ik}(w')$ vanishes as $w'\,^{ik-1/2}$ as ${\rm Im} k\rightarrow \infty$, it dominates over $P_{-1/2+ik}(w)$ and the contour can be closed in the upper-half plane, picking up a pole contribution at $k= i\sqrt{1/4+m^2}$.  For the term involving $Q_{-1/2+ik}(w')$, the contour can be closed in the lower half plane, yielding an identical contribution. This can be repeated for $1<w'<w<\infty$. Together they lead to the same result 
by the direct computation using a Wronskian approach, Eq. (\ref{eq:wronskian}), as expected.

As a direct application of Eqs. (\ref{eq:ON-w}) and (\ref{eq:Completeness-w}), every  real, square-integrable function, $F(w)$,  defined over $1<w<\infty$, can be expressed as
\be
F(w) = \int\limits_{-\infty}^{\infty} \frac{dk }{2\pi } \,\, k \, f(k )\,\, P_{-1/2 + i k}(w)\, ,\label{eq:spectralrep1}
\ee
The transformed function $f(k)$ is real and antisymmetric, $f(-k)  =- f(k)$,
\be
f(k) =  \pi \tanh  k   \int\limits_1^\infty d w \,\,  F(w)\, P_{-1/2 + i k}(w) \, .\,  \label{eq:transform}
\ee

\subsection{Mellin-Like Representation and Polynomial Boundedness}\label{sec:spectral2}

We now address the important case where $F(w)$ is not square-integrable but polynomially bounded. This can be handled by either working with a reduced function, for example $\tilde F(w)=w^{-L} F(w)$, or adopting a deformed representation for Eq. (\ref{eq:spectralrep1}).  We will adopt the latter approach, which can be seen to correspond to a Sommerfeld-Watson resummation from the OPE context. 

We begin by first  re-expressing Eq. (\ref{eq:spectralrep1}) by a change of variable from $k$ to $\ell=\ell_0 + i k$, with $\tilde f(\ell)=f(k)$. The choice of this constant $\ell_0$ is arbitrary. We shall choose $\ell_0=1/2$ so that the integral path in Eq. (\ref{eq:spectralrep1}) corresponds to ${\rm Re}\,\tilde \ell=0$, with $\tilde \ell=\ell -1/2$, for $d=1$~\footnote{From Eq. (\ref{eq:spectralrep1}), a more natural choice is $\ell_0=-1/2$, which would lead to an expressions more familiar in form to a $d=3$ partial wave expansion. Our choice corresponds to a shift from $\ell$ to $\ell-1$.}.

For a square-integrable function $F(w) \sim O(w^{-1/2-\epsilon})$, the transform $ \tilde f(\ell)$ is analytic in the strip $1/2 -\epsilon < {\rm Re} \ell < 1/2+ \epsilon$.  It is convenient, using Eq. (\ref{eq:PQidentity}),  to separate $\tilde f(\ell)$ into two pieces, $ \tilde f(\ell)=  \tilde f_+(\ell) - \tilde f_-(\ell)$, and $ \tilde f_-(\ell)= \tilde f_+(-\ell+1)$,  where
\begin{align}
\tilde f_+(\ell) &=   \,\int\limits_1^\infty \,d w \,  F(w)\,  Q_{\ell-1}(w) \,,\, \text{and} \nn
 \tilde f_-(\ell) &=   \,\int\limits_1^\infty \,d w \,  F(w)\,  Q_{-\ell}(w) \,. 
\end{align}
With $P_{\ell}(w)=P_{-\ell+1}(w)$, the contribution from $\tilde f_+$ and $\tilde f_-$ are equal, leading to a new representation involving $\tilde f_+(\ell)$ only,
\be
F(w) = \int\limits_{1/2 -i \infty}^{1/2 +i\infty} \frac{d\ell}{2\pi i} (2\ell+1) \, \tilde f_+(\ell )\; {P_{\ell-1}(w) }\, ,\label{eq:spectralrep2}
\ee

Since $Q_{\ell}(w)\sim w^{-\ell-1}$, it follows that $ \tilde f_+(\ell) $ is analytic in the right-half $\ell$-plane,  $ 1/2<{\rm Re}\, \ell$,  and  $ \tilde f_-(\ell) $ is analytic in the left half-plane, $ {\rm Re}\, \ell<1/2$.  Applying the identity in Eq. (\ref{eq:PQidentity}) to $P_{\ell-1}$ above, we see the term coming from $Q_{\ell}(w)$ can be dropped in closing the contour to the right, leading to 
\be
F(w) =-  \int\limits_{1/2- i\infty}^{1/2 +i\infty} \frac{d\ell (2\ell-1)}{2\pi +1}     \, \tilde f_+(\ell )\,  \frac{\tan \ell \pi}{\pi} \, Q_{-\ell }(w)\, ,\label{eq:spectralrep3}
\ee
where the factor ${\tan \ell \pi}$ above plays the same role of $c_\ell$ in Eq. (\ref{eq:1dMCB}), rendering the integrand finite at positive integral values for ${\rm Re}\, \ell>1/2$.   

Let us now turn to functions which grow with $w$. To deal with functions which grow with $w$ as a power, it is possible  to enlarge the Hilbert space \cite{Toller3,Herman}, and, for the class of functions which are polynomially bounded $F(w)=O(w^{L_0})$, the region of analyticity for $\tilde f_+(\ell)$ gets pushed out to the right. In other words, $L_0  < {\rm Re} \ell <\infty$.  It is possible to define $f_+(\ell) $ as an analytic function by 
\be
f_+(\ell) \equiv     \,\int\limits_1^\infty \,d w \,  Q_{\ell-1}(w) \, F(w), 
\ee
initially for $L_0<{\rm Re} \ell$, and then analytically continue in the the region to the left of ${\rm Re} \ell  =L_0$.  The function $F(w)$ can be recovered by
\be
F(w) =-  \int\limits_{L_0-i\infty}^{L_0+i\infty} \frac{d\ell}{2\pi i} (2\ell-1) \,  f_+(\ell )\,  \frac{\tan \ell \pi}{\pi}\, Q_{-\ell}(w)\, .\label{eq:spectralrep4}
\ee
This is precisely what we have arrived at earlier via Minkowski OPE analysis, (\ref{eq:1D-ImConformalBlock}).  As also mentioned earlier, an equivalent representation, which we will make use of in Sec. \ref{sec:SYK},  is Eq. (\ref{eq:1D-ImConformalBlock-2}).
If $ f_+(\ell )$ contains a singularity at $\ell_{eff}$ where $1/2<\ell_{eff}<L$, e.g., a pole, by pulling the contour in (\ref{eq:spectralrep4}) to the left passing to pole, one finds  $F(w)$ diverges at $w\rightarrow \infty$ as
\be
F(w) =O(w^{\ell_{eff}-1})\, .
\ee

\subsection{\texorpdfstring{$AdS_2/CFT_1$}{AdS2/CFT1} Duality} \label{sec:syk-back}

In recent years, much work has been done to elucidate the duality between some string theory in $AdS_2$ and CFT in $d=1$.  Pure Einstein gravity in two dimensions has no propagating degrees of freedom, but if other fields are included in the theory, there can be interesting dynamics.  Most of the interesting work has been aimed at resolving issues with the \emph{black hole information loss paradox}. (For reviews see \cite{Mathur:2009hf,Chen:2014jwq,Unruh:2017uaw}) Maldacena \cite{Maldacena:2001kr} pointed out that d dimensional theories with eternal black holes in Anti-de Sitter space can have interesting consequences for the information loss paradox; the past boundary could be decomposed into two copies of the boundary CFT and initial states could be thought of as a thermal ensemble of two CFT states. The object of interest is often called a \emph{thermofield double} (TFD) state
\be \label{eq:tfd}
\ket{TFD} = Z^{-1/2} \sum_n e^{\frac{-\beta E_n}{2}}\ket{E_n}_L^{CFT}\otimes \ket{E_n}_R^{CFT}\, ,
\ee
where the two copies are called left and right.  Importantly, although separate conformal transformations can be performed that leave the L/R density of states, $\sqrt{\rho} = Z^{-1/2} \sum_n e^{\frac{-\beta E_n}{2}}\ket{E_n}^{_{^{^{CFT}}}}\otimes \bra{E_n}^{_{^{^{CFT}}}}$, invariant, it was pointed out \cite{Czech:2012be} that these transformations can affect the properties of entangled states as they fall into a black hole.

Of critical importance is to understand how information sent in from the boundary becomes ``smeared" across the horizon as quanta falls into the black hole.  Classical dynamics describes such a system as chaotic; a sensitive dependence upon initial conditions describes a situation in which final state information is ``smeared" on a mathematically dense phase space.  To this end, scrambling of infalling quanta can be defined via a maximal Lyapunov exponent, $\lambda_L$.  This can be determined by considering an out-of-time correlator~\footnote{It was pointed out in \cite{Swingle:2016var} that this correlator can be considered a quantum variant on the Loschmidt echo.  \cite{Swingle:2016var} then goes on to propose a cold-atom qubit set up that could be used to experimentally such correlations.} 
\be\label{eq:correl}
\braket{W_R^{\dagger}(t)V_L^{\dagger}(0)W_L(t)V_R(0)}_{\beta} \sim 1 - \tilde{\alpha} e^{\lambda_L t} + \mathcal{O}(\tilde{\alpha})
\ee 
for small $\tilde{\alpha}$.  Here the index $\beta$ indicates that it is a thermal correlation function, and $\tilde{\alpha}$ is a constant that encodes information.\footnote{$\tilde{\alpha}$ should be a function of the entropy. A common definition is for $\tilde{\alpha}\equiv 1/B$, Where $B$ is the number of bits of information. For holographic theories, $\tilde{\alpha}\sim 1/N^2$.} The scrambling time can then be defined as the time when the exponential becomes order one, 
\be \label{eq:scramble}
t_* = \frac{1}{\lambda_L} \text{ln}(S) 
\ee 
for an entropy S.  

The correlator Eq. (\ref{eq:correl}) \footnote{Often in the literature a similar correlator is defined with different operator ordering and operators living on different initial state CFTs.  These various 4 point correlators are all related via analytic continuation by continuing various operators by $\beta/2$ around the imaginary time thermal boundary $S_{\beta}$. An example can be seen in Eq. (\ref{eq:analytic-cont}).} probes chaotic behavior \footnote{It should be noted that this requires a chaotic system.  For example, in integrable models, as in the $d=2$ Ising model, the behavior of Eq. (\ref{eq:correl}) will not be seen.\cite{Roberts:2014ifa}}. It can be calculated from the gravitational theory \cite{Shenker:2013pqa,Shenker:2014cwa}.  A key element to the calculation is that the operator $W(t)$ can be seen to create a shock wave \cite{DRAY1985173,Dray1985}. This shock wave can be interpreted as a boosting scattering particles by exp$(2\pi t/\beta)$ to arbitrary high energies \footnote{In principle, for late enough times, this can boost particles to the string or Planck scales. }. The result is that Eq. (\ref{eq:correl}) can be described by high energy elastic eikonal scattering at fixed impact parameter b, similar to that described in \cite{Cornalba:2006xk,Cornalba:2006xm,Cornalba:2007fs,Cornalba:2008qf,Cornalba:2007zb,Brower:2007qh} and shown in Eq. (\ref{eq:T-chi}).  

Finally we emphasize that while the thermofield double approach, Eq. (\ref{eq:tfd}), to calculating Eq. (\ref{eq:correl}) has been convenient in gravitational literature, it is not the only approach that can be used.  Consider a $d=1$ CFT\footnote{A similar procedure exists for $d=2$ CFTs as described in \cite{Roberts:2014ifa}.  Here a separate holomorphic and antiholomorphic conformal transformation exist for the two degrees of freedom.}. At finite temperature, a thermal correlator can be related to a vacuum correlator via a conformal transformation of the form $f(t)=$exp$(2\pi t /\beta)$. The invariance of conformal correlation functions then leads to 
\begin{align} \label{eq:conftrans}
\braket{\mathcal{O}_1(t_1)\mathcal{O}_2(t_2)...} = \left|\frac{df}{dt}\right|_{t=t_1}^{\Delta_1}\left|\frac{df}{dt}\right|_{t=t_2}^{\Delta_2}...\nn
\times \braket{\mathcal{O}_1(f(t_1))\mathcal{O}_2(f(t_2))...}_{\beta} \, .
\end{align}

\paragraph{SYK Theory} Most of the discussion in this work applies generally to CFTs of arbitrary dimension. While these results are more general, without a specific theory the details of particular dynamic behavior can be hard to suss out.  For integrable theories the canonical non-trivial example is that of the well known duality between string theory on $AdS_5\times S^5$ and $\mathcal{N}=4\, SYM$, which is conjectured to be integrable \footnote{This is backed up by a large library of literature.  In the large N limit, some sectors of the theory have been shown to be exactly integrable.  For a comprehensive review see \cite{Beisert:2010jr}.}.  For chaotic systems with black holes, holographic examples have been harder to come by.

In this vein, much recent attention has been given to the Sachdev-Ye-Kitaev (SYK) model, first proposed by Kitaev in a series of talks \cite{kitaev1,kitaev2,Sachdev:1992fk}. The details of this theory were quickly expanded upon in \cite{Polchinski:2016xgd,Maldacena:2016hyu,Jevicki:2016ito,Jevicki:2016bwu,Haldar2017,Murugan:2017eto,Das:2017wae,Das:2017hrt,Das:2017pif,Kitaev:2017awl,Narayan:2017hvh,Altland:2017eao,Forste:2017apw}. The boundary theory of this model is a many body system of Majorana fermions with an all-to-all four point interaction $\mathcal{J}$.  In the low temperature limit\footnote{An effective coupling $\lambda = \beta \mathcal{J}$ interpolations between a holographic $\lambda >>1$ limit where the theory is nearly conformal and a thermal $\lambda <<1$ limit where the conformal symmetry is broken.}  the system is approximately conformal. The full 2-pt function of the theory can be found using a Schwinger-Dyson equation.  The four point function can found using a Bethe-Salpeter equation which involves a ladder like exchange process as in Eq. (\ref{eq:SD-W-1}). In this theory, calculation of Eq. (\ref{eq:correl}) leads to a Lyapunov exponent $\lambda_L=2\pi/\beta$ with corrections coming from Regge string effects, longitudinal string spreading, and non-linear interactions.  This calculation can be done from the perspective of the bulk string theory using the approach of \cite{Shenker:2014cwa}.  The bulk process is described by an eikonal scattering who's dominant contribution can be traced to the BPST Pomeron.  In both cases the $\Delta(\ell)$ spectral curve plays an important role as the conformal weight of the dominant OPE contribution for the conformal theory and as the Virasoro operator dimension for the bulk theory. 

We should take a moment to emphasize that both of the above approaches in the literature follow one path: define the correlation function in the euclidean limit, calculate the correlator, then \emph{carefully} analytically continue to the Minkowski region by adding an imaginary piece to the Euclidean time \footnote{The type of analytic continuation was first spelled out in great detail in \cite{Cornalba:2006xk,Cornalba:2006xm,Cornalba:2007fs,Cornalba:2008qf}.}. For a finite temperature conformal transformation $f(t)=$exp$(2\pi t /\beta)$ as in Eq. (\ref{eq:conftrans}) this process can be outlined as 
\be \label{eq:analytic-cont}
\begin{aligned}
&\text{correlator:}\\ 
&\braket{D(t_4)C(t_3)B(t_2)A(t_1)} \\
&\text{imaginary time order:}\\
&\langle D(f(t_4=t)C(f(t_3=0)) \\
& \qquad \qquad B(f(t_2=t))A(f(t_1=0))\rangle \\
&\text{Lorentz correlator:} \\  
&\langle D(f(4i\epsilon)f(t))C(f(3i\epsilon)f(0)) \\
&\qquad \qquad B(f(2i\epsilon)f(t))A(f(i\epsilon)f(0))\rangle \, .
\end{aligned}
\ee
In the end we arrive at the time ordered process we are interested in in this paper (1,3)$\rightarrow$(2,4) as in Eq. (\ref{eq:A(X)}).  The advantage of our approach is that, by taking advantage of boundary conditions, one can directly write down the Minkowski solution without having to carefully do the analytic continuation.

\bibliographystyle{utphys}
\bibliography{conformalc}

\end{document}